\documentclass[options]{JHEP3}
\usepackage{epsfig} 
\usepackage{amsmath}
\usepackage{amsfonts}
\usepackage{graphicx}

\newcommand{\insertplot}[5]{\begin{figure}
 \hfill\hbox to 0.05in{\vbox to #5in{\vfill
 \inputplot{#1}{#4}{#5}}\hfill}
 \hfill\vspace{-.1in}
 \caption{#2}\label{#3}
 \end{figure}}
 \newcommand{\inputplot}[3]{
 \special{ps: plotfile #1}
}
\newcounter{fig}   
\newcommand{\vphi}{\varphi}

\newcommand{\zz}{\theta\theta}
\newcommand{\beq}{\begin{equation}}
\newcommand{\eeq}{\end{equation}}
\newcommand{\beqs}{\begin{eqnarray}}
\newcommand{\eeqs}{\end{eqnarray}}

\numberwithin{equation}{section}
\newcommand{\be}{\begin{equation}}
\newcommand{\ee}{\end{equation}}
\newcommand{\bea}{\begin{eqnarray}}
\newcommand{\eea}{\end{eqnarray}}

\abstract{
We argue that the Weyl coordinates and the
 rod-structure employed to construct 
static axisymmetric solutions in higher dimensional Einstein gravity
can be generalized to the Einstein-Gauss-Bonnet theory.
As a concrete application of the general formalism,
we present numerical evidence for the existence of static black ring solutions
in  Einstein-Gauss-Bonnet theory in five spacetime dimensions.
They approach asymptotically the Minkowski background and
are supported against collapse by a conical singularity in the form of a disk.
An interesting feature of these solutions is that the
Gauss-Bonnet term reduces the conical excess 
of the static black rings.
Analogous to the Einstein-Gauss-Bonnet black strings, 
for a given mass the static black rings
 exist up to a maximal value 
of the Gauss-Bonnet coupling constant {$\alpha'$.
Moreover, in the limit of large ring radius,
the suitably rescaled black ring maximal value of $\alpha'$
and the black string maximal value of $\alpha'$ agree.}
}  
\preprint{hep-th/yymmddd}

\title{Generalized Weyl solutions in $d=5$ Einstein-Gauss-Bonnet theory: 
the  static black ring  } 
  
 \author{
 {\large Burkhard Kleihaus}, {\large Jutta Kunz} 
and {\large Eugen Radu}  
\\ 
\\
{\small Institut f\"ur Physik, Universit\"at Oldenburg, Postfach 2503
D-26111 Oldenburg, Germany}
}
\maketitle

\begin{document}

\section{ Introduction}

In recent years it has been realized that higher dimensions $d>4$ allow
for a rich landscape of black hole solutions 
that do not have four dimensional counterparts.
The vacuum black ring solution of Emparan and Reall \cite{Emparan:2001wn,Emparan:2001wk} in $d=5$ Einstein gravity
is perhaps the best known example of such a
configuration.  
The  black ring has a horizon with topology $S^2\times S^1$,
while the Myers-Perry black hole \cite{Myers:1986un} has a  horizon topology $S^3$. 
This solution provided also
 the first concrete piece of evidence that
in higher dimensional gravity, the no-hair theorems of $3+1$ dimensions do not apply.
For example, in a $4+1$ dimensional asymptotically flat spacetime
 with a given ADM mass and angular momentum, the geometry need not necessarily
 be that of the Myers-Perry black hole.

The $d=5$  Emparan and Reall  black ring solution 
has been generalized in various directions, 
including configurations with abelian matter fields
\cite{Elvang:2003yy}-\cite{Chng:2008sr}.
Physically interesting solutions describing superposed black objects 
(black saturns \cite{Elvang:2007rd}, 
bicycling black rings \cite{Elvang:2007hs}, \cite{Izumi:2007qx},
and concentric rings \cite{Iguchi:2007is}, \cite{Evslin:2007fv}, 
\cite{Gauntlett:2004wh}) were also constructed. 
However, in the static limit, all known $d=5$ asymptotically flat
solutions with a nonspherical horizon topology possess
a conical singularity or other pathologies  \cite{Emparan:2001wk},
\cite{Kunduri:2004da}, \cite{Chng:2008sr}.

All these results concern the case of $d=5$ Einstein gravity theory and its 
various extensions with abelian matter fields.
However, in five dimensions, the most general theory of gravity leading to
second order field equations for the metric is the so-called 
Einstein-Gauss-Bonnet (EGB) theory, which contains quadratic powers of 
the curvature. 
The  Gauss-Bonnet (GB) term appears as the first curvature stringy
correction to general relativity~\cite{1,Myers:1987yn}, when assuming
that the tension of a string is
large as compared to the energy scale of other variables.
Inclusion of this term in the action leads to a variety of 
new features  (see \cite{Garraffo:2008hu}, \cite{Charmousis:2008kc}
for a review of the  higher order gravity theories).

Although the generalization of the spherically symmetric Schwarzschild solution
in EGB theory has been known for quite a long time \cite{Deser}, 
the issue of axially symmetric solutions with a GB term is basically unexplored.
In particular we do not know if the $d\geq 5$ black holes with a nonspherical
topology of the horizon continue to exist when including
stringy correction to the action. 
The main obstacle here seems to be that the Weyl formalism,
which has proven so useful in the case of Einstein gravity,
allowing for the discovery of a plethora of interesting exact solutions,
has no straightforward extension in the presence of a GB term.

In the absence of exact solutions, a natural way to approach this issue 
is to construct such configurations numerically.
This paper aims at a first step in this direction,
since we propose a framework for a special class of $d=5$ static configurations
with three commuting Killing vectors. In the absence of 
a GB term, this framework reduces to that used in  \cite{Emparan:2001wk}
to construct generalized Weyl solutions.
Here we argue that some basic properties 
there are still valid in the presence of a 
GB term, in particular the rod structure of the solutions.

As the simplest example of a
$d=5$ black object with a nonstandard topology of the event horizon,
we present numerical evidence
for the existence of static black rings in EGB theory.
These solutions are found within a nonperturbative approach, 
by directly solving the second order field equations
with suitable boundary conditions.
These black rings share most of the features of
the Einstein gravity solution in  \cite{Emparan:2001wk}. 
Although the inclusion of the  
GB term in the action reduces the conical excess, 
these configurations still possess an angular deficit.
Moreover,  for a given value of the mass,
similar to the case of a $d=5$ EGB black string \cite{Kobayashi:2004hq}, the static black ring  
solutions exist up to a maximal value of the GB coupling constant.

The plan of the paper is the following.  
In the next Section we present a brief review of the Weyl formalism 
in Einstein gravity and argue that the coordinate system and the rod structure 
used there can be employed to construct EGB solutions as well.
Section 3 consists of a discussion of the 
Schwarzschild black hole 
and the uniform black string in EGB theory.
There we present evidence that these configurations 
can also be viewed as generalized Weyl
solutions within the framework of Section 2.
Section 4 contains the main results of this work consisting 
of a systematic study of the static black rings in EGB theory.
We give our conclusions and remarks in the final section. 
There we report also our results on charged generalizations 
of the static black rings present in EGB-Maxwell theory. 
In Appendix A we present some details on the EGB equations.
Appendix B  contains a discussion of some technical aspects involved
in the numerical construction of the EGB static black rings. 
This includes a new coordinate system
which has proven more suitable for the numerical study of the 
black ring solutions.

\section{The general formalism}

\subsection{The Einstein-Gauss-Bonnet theory}

We consider the EGB action in five space-time dimensions
\begin{equation}
I = \frac{1}{16\pi G} 
   \int_\mathcal{M}d^5x \sqrt{-g}\left[ R + \alpha' L_{\rm GB}\right]
   \ , 
\label{action}
\end{equation}
where $G$ is the five dimensional Newton constant 
and $\alpha'$ is the GB coefficient with dimension $(length)^2$. 
In string theory,
the GB coefficient is positive,
and this is the only case considered in this 
work\footnote{Also, a negative value of $\alpha'$ leads to a number of 
 pathological features of the theory, see $e.g.$ \cite{Garraffo:2008hu}.}.
$R$ denotes the Ricci scalar and 
\begin{equation}
L_{\rm GB}  = R^2 - 4 R_{\mu\nu}R^{\mu\nu} 
            + R_{\mu\nu\rho\sigma}R^{\mu\nu\rho\sigma} 
\label{GBterm}
\end{equation}
the Gauss-Bonnet term with Ricci tensor $R_{\mu\nu}$ and 
Riemann tensor $R_{\mu\nu\rho\sigma}$.

The variation of the action (\ref{action}) with respect to the
metric tensor yields the EGB equations	    
\begin{equation}
E_{\mu\nu}=G_{\mu\nu} +\alpha' H_{\mu\nu}=0 \ , 
\label{EGBeqs}
\end{equation}
where
\begin{eqnarray}
G_{\mu\nu} & = & R_{\mu\nu} -\frac{1}{2} g_{\mu\nu} R \ , 
\nonumber \\
H_{\mu\nu} & = & 2\left[R R_{\mu\nu} -2 R_{\mu\rho}R^\rho_\nu
                        -2 R_{\mu\rho\nu\sigma}R^{\rho\sigma}
			+R_{\mu\rho\sigma\lambda}R_\nu^{\ \rho\sigma\lambda}
		   \right]
		   -\frac{1}{2}g_{\mu\nu}L_{\rm GB}	.
\nonumber 
\end{eqnarray}
For a well-defined variational principle, one has to supplement the 
action (\ref{action}) with the Gibbons-Hawking surface term \cite{GibbonsHawking1}
\begin{equation}
I_{b}^{(E)}=
-\frac{1}{8\pi G}
\int_{\partial \mathcal{M}}d^{4}x\sqrt{-\gamma }K~,
\label{Ib1}
\end{equation}
and its counterpart for Gauss-Bonnet gravity  \cite{Myers:1987yn} 
\begin{equation}
I_{b}^{(GB)}=
-\frac{\alpha'}{4\pi G}
\int_{\partial \mathcal{M}}d^{4}x\sqrt{-\gamma }%
 \left( J-2{\rm G}_{ab} K^{ab}\right)  ~,
\label{Ib2}
\end{equation}
where $\gamma_{ab}$ is the induced metric on the boundary,  
$K$ is the trace of the extrinsic curvature of the boundary,
${\rm G}_{ab}$ is the Einstein tensor of the metric $\gamma _{ab}$ 
and $J$ is the trace of the tensor
\begin{equation}
J_{ab}=\frac{1}{3}%
(2KK_{ac}K_{b}^{c}+K_{cd}K^{cd}K_{ab}-2K_{ac}K^{cd}K_{db}-K^{2}K_{ab})~.
\label{Jab}
\end{equation}

\subsection{The Weyl solutions in $d=5$ Einstein gravity and the rod structure}
 
Following the approach in
\cite{Emparan:2001wk},
we consider asymptotically flat, five-dimensional
static and axisymmetric vacuum spacetimes
with three commuting Killing vector fields $V_{(i)}$
($i=1,2,3$).
The commutativity of Killing vectors $[V_{(i)}, V_{(j)}]=0$
enables us to find a coordinate system such that
$V_{(i)}=\partial/\partial x^i$ and
the metric is independent of the coordinates $x^i$.
In what follows, we shall put
$x^1=t,\ x^2=\psi,$ and $x^3=\varphi$.
Then $(\partial/\partial x^1)$ is the Killing vector field
associated with time translation and
$(\partial/\partial x^2),(\partial/\partial x^{3})$
denote the spacelike Killing vector fields with closed orbits.

Here we invoke the particularization for $d=5$
of the general  theorem 2.1 in Ref. \cite{Emparan:2001wk}:

{\it Let
$V_{(i)}, i=1,2,3$, be three-commuting
Killing vector fields such that
\begin{enumerate}
\item
$V_{(1)}^{[\mu_1}V_{(2)}^{\mu _2} 
V_{(3)}^{\mu _{3}}D^\nu V_{(i)}^{\rho ]}=0$
holds at least at one point of the spacetime
for a given $i=1,2,3$.

\item
$V_{(i)}^\nu R_\nu ^{[\rho}
V_{(1)}^{\mu _1}V_{(2)}^{\mu _2} 
V_{(3)}^{\mu _{3}]}=0 $ holds for all
$i=1,2,3.$
\end{enumerate}
Then the two-planes
orthogonal to the Killing vector fields
$V_{(i)}, i=1, 2,3,$ are integrable.}

The first condition holds because we have assumed axisymmetry, while 
the second one is automatically satisfied
as long as we restrict
ourselves to the vacuum solutions of Einstein equations.

As a result, the metric can be written
in the canonical form~\cite{Harmark:2004rm} as
\begin{eqnarray}
\label{metric-canonical}
ds^2=e^{2\nu(\rho,z)}(d\rho^2+dz^2)+e^{2U_2(\rho,z)} d\psi^2+e^{2U_3(\rho,z)} d\varphi^2-e^{2U_1(\rho,z)}dt^2,
\end{eqnarray}
where $0\leq \rho<\infty,$ $-\infty< z<\infty$.
Here it is most convenient to choose
the three functions $U_i$ as to satisfy the condition
\begin{eqnarray}
\label{cons}
\sum_i U_i=\log \rho.
\label{eq:det}
\end{eqnarray}
This is compatible with the vacuum Einstein equations
$G_{ij}=0$ ($i=1,2,3$), which for the choice (\ref{cons}) reduce  to 
\begin{eqnarray}
\label{eq-U}
 \frac{\partial^2 U_i}{\partial \rho^2} + \frac{1}{\rho}\frac{\partial
 U_i}{\partial \rho} + \frac{\partial^2 U_i}{\partial z^2} = 0,
\end{eqnarray}
(the Einstein tensor for the metric ansatz (\ref{metric-canonical}) is presented in Appendix A).
One can see that (\ref{eq-U}) is just Laplace's equation in a 
(fictitious) three-dimensional flat space with
metric 
$
 ds^2 = d\rho^2 + \rho^2 d\theta^2 + dz^2.
$

 From the other components of the Einstein
equations $G_{\rho}^\rho-G_{z}^z=0$ and $G_{\rho}^z=0$,
we obtain the equations which determine the function $\nu(\rho,z)$
for a given solution of the  equation (\ref{eq-U})
\begin{eqnarray}
\label{eq-nu}
\nu'=-\frac{1}{2\rho}+\frac{\rho}{2}
\left( 
U_1'^2+U_2'^2+U_3'^2-\dot U_1^2-\dot U_2'^2-\dot U_3^2
\right),
~~
\dot \nu=\rho(\dot U_1'+\dot U_2'+\dot U_3'),
\end{eqnarray}
where a  prime denotes the derivative with respect to 
$\rho$ and a dot denotes the derivative with respect to $z$.
Solutions with the ansatz (\ref{metric-canonical}) and with $U_1,U_2,U_3$ and $\nu$
satisfying the equations (\ref{eq-U}), (\ref{eq-nu})
are usually called generalized Weyl solutions.

Although the Einstein equations take a simple form in terms of  ($U_i,\nu$), 
for the purposes of this paper it is more convenient to work with a set a functions
$f_i$ defined as follows
\begin{eqnarray}
\label{fi}
e^{2\nu(\rho,z)}=f_1(\rho,z),~~e^{2U_2(\rho,z)} =f_2(\rho,z),
~~e^{2U_3(\rho,z)} =f_3(\rho,z),~~e^{2U_1(\rho,z)} =f_0(\rho,z).
\end{eqnarray} 
This leads to a line element
\begin{eqnarray}
\label{metric} 
ds^2=-f_0(\rho,z)dt^2+f_1(\rho,z)(d\rho^2+dz^2)+f_2(\rho,z)d\psi^2+f_3(\rho,z)d\varphi^2,
 \end{eqnarray}
which was used in our study of the EGB static black ring solutions. 
 
 In this paper we are mainly interested in configurations approaching asymptotically the
 five dimensional  Minkowski spacetime, this being also the simplest solution of the equations (\ref{eq-U}), \ref{eq-nu}).
In this case, the metric functions $f_i$ have the following expression:
 \begin{eqnarray}
\label{Mink}
 f_0(\rho,z)=1,~~
 f_1(\rho,z)=\frac{1}{2\sqrt{\rho^2+z^2}},~~
f_2(\rho,z)= \sqrt{\rho^2+z^2}+z,~~~
f_3(\rho,z)= \sqrt{\rho^2+z^2}-z.~~~~{~~}
\end{eqnarray}
The usual form of the flat spacetime metric in the Hopf coordinates 
 \begin{eqnarray}
\label{mink-1}
 ds^2=-dt^2+dr^2+r^2 (d\theta^2+\cos^2 \theta d\psi^2+\sin^2 \theta d\varphi^2  ),
\end{eqnarray}
is found from (\ref{metric}), (\ref{Mink}) via the coordinate transformation
 \begin{eqnarray}
\label{coord-tr}
\rho=\frac{1}{2}r^2 \sin 2 \theta,~~z=\frac{1}{2}r^2 \cos 2 \theta,
\end{eqnarray}
with $0\leq r<\infty$, $0\leq \theta  \leq \pi/2$.

The  equations (\ref{eq-U}), (\ref{eq-nu}) possess a variety of 
physically interesting solutions.
They can be uniquely characterized 
by the boundary conditions on the $z-$axis, known as the {\it rod-structure} \cite{Emparan:2001wk},
\cite{Harmark:2004rm}, \cite{Hollands:2007aj}.
In pure Einstein gravity, the physically relevant solutions for $U_i$  
can
also be thought of as Newtonian potentials produced by thin rods 
of zero thickness with linear mass density $1/2$, placed
on the axis of symmetry
in the auxiliary three-dimensional flat space. 
Then the constraint (\ref{cons}) states that these sources must add up
to give an infinite rod.

In this approach, the $z-$axis is divided into $N$ intervals (called rods of the 
solution), $[-\infty, z_1]$,
$[z_1,z_2]$,$\dots$, $[z_{N-1},\infty]$.
As proven in \cite{Harmark:2004rm},  in order to avoid curvature naked 
singularities at $\rho=0$, it is a 
necessary condition that only one of the functions
$f_0(0,z)$, $f_2(0,z)$, $f_3(0,z)$ becomes zero for a given rod,
except for isolated points between the intervals.

 For the static case discussed here, a horizon corresponds to 
 a timelike rod where $f_0(0,z)=0$ while $\lim_{\rho \to 0}f_0(\rho,z)/\rho^2>0$.
 There are also spacelike rods corresponding to compact directions  specified by the
 conditions $f_{a}(0,z)=0$, $\lim_{\rho\to 0}f_{a}(\rho,z)/\rho^2>0$, with $a=2,3$.
A semi-infinite spacelike rod corresponds to an axis of rotation, the associated coordinate being
a rotation angle.
For example, the Minkowski spacetime (\ref{mink-1}) corresponds to two
semi-infinite rods $[-\infty,0]$ and $[0,\infty]$.
Demanding regularity of the solutions at $\rho=0$
imposes a periodicity $2\pi$ for both $\psi$ and $\varphi$.
(However, when several  $\psi$- or $\varphi$-rods are present, 
it may be impossible to satisfy simultaneously all the periodicity
conditions).

One of the main advantages of this approach is that
the topology of the horizon is automatically imposed by the rod structure. 
This provides a simple way to construct a variety of solutions with 
nontrivial topology of the horizon (including multi-black objects). 
Since (\ref{eq-U}) is linear, one can superpose different solutions 
for the same potential $U_i$.
The nonlinear nature of the Einstein gravity manifests itself through the equation (\ref{eq-nu}) 
for the metric
function $\nu$.

\subsection{The static, axially symmetric Einstein-Gauss-Bonnet configurations}
\subsubsection{The equations}
One of the main purposes of this work is argue that
the ansatz (\ref{metric-canonical}) and the associated rod structure
 can be used to construct  physically relevant solutions in EGB theory.
{\it A priori}, it is not clear that this metric ansatz is valid also in this case, since
 the second assumption in the general theorem mentioned above
 ($i.e.$ $V_{(i)}^\nu R_\nu ^{[\rho}
V_{(1)}^{\mu _1}V_{(2)}^{\mu _2} 
V_{(3)}^{\mu _{3}]}=0 $)
does not hold in general for EGB theory.
Thus the situation here is similar to the case 
of caged black holes  \cite{Sorkin:2003ka}, \cite{Kudoh:2003ki}
 or nonuniform black strings 
\cite{Wiseman:2002zc}, \cite{Kleihaus:2006ee}
 in Kaluza-Klein Einstein gravity,
where the validity of the metric ansatz could be proven only {\it a posteriori},
after solving the field equations. 

The equations for the functions $f_1,~f_2,~f_3$ 
and $f_0$ are found by using a suitable combination of the EGB equations,
$E_t^t =0,~E_\rho^\rho+E_z^z =0$, $E_{\psi}^{\psi} =0$,
and  $E_{\varphi}^{\varphi} =0$.
Details on these equations and the explicit form of the tensors
$G_\mu^\nu$ and $H_\mu^\nu$ are presented in Appendix A. 

The remaining equations $E_z^\rho =0,~E_\rho^\rho-E_z^z  =0$
yield two constraints. Following \cite{Wiseman:2002zc}, we note that
setting $E^t_t =E^{\varphi}_{\varphi} =E^\rho_\rho+E^z_z=0$
in $\nabla_\mu E^{\mu \rho} =0$ and $\nabla_\mu E^{\mu z}=0$, 
we obtain the Cauchy-Riemann relations
\begin{eqnarray}
\partial_\rho\left(\sqrt{-g} E^\rho_z \right) +
\partial_\rho\left(  \sqrt{-g} \frac{1}{2}(E^\rho_\rho-E^z_z) \right)
= 0 ,~~
 \partial_\rho\left(\sqrt{-g} E^\rho_z \right)
-\partial_z\left(  \sqrt{-g} \frac{1}{2}(E^\rho_\rho-E^z_z) \right)
~= 0 .~~~{~~~} 
\end{eqnarray}
Thus the weighted constraints satisfy Laplace equations, and the constraints 
are fulfilled, when one of them is satisfied on the boundary and the other 
at a single point
\cite{Wiseman:2002zc}. 

Due to the GB contributions, the second order equations for the functions $f_i$
are much more complicated than in the case of Einstein gravity
and do not reduce to the simple Laplace equation.
Thus the functions $U_i$ are no longer harmonic.
Also, one can verify that the central property (\ref{cons}) of the Einstein gravity Weyl-solutions
does not hold in the presence of a GB term and thus one cannot set 
$f_0 f_2 f_3=\rho^2$.
 
Therefore finding closed form solutions within this approach looks unlikely.
However, the solutions can be constructed numerically, by solving boundary value problems.
A major advantage of the ansatz (\ref{metric-canonical}) is that
the $(\rho,z)$ coordinates have a rectangular boundary in which all boundaries coincide with the coordinate lines 
and thus are suitable for numerics\footnote{This is not the case of the ring coordinates
used in most of the studies on  black ring solutions.
The spatial infinity corresponds there to a single point.}.

\subsubsection{The rod structure}
The central point in this approach is that the rod structure, as explained 
above for the case of Einstein gravity,
can be used also for solutions of the EGB 
theory\footnote{However, note that the interpretation of a rod
as corresponding to a zero thickness  source  with linear mass density $1/2$, placed
on the axis of symmetry
in a auxiliary three-dimensional flat space  is no longer valid in EGB theory.}.
This would fix the boundary conditions  along the
$z-$axis for the functions $f_i$ and thus the topology of the horizon.

Here one starts by noticing that the following generic form of a solution near the $z-$axis
is compatible with the EGB
equations:
\begin{eqnarray}
\label{rods}
 f_i(\rho,z)=f_{i0}(z)+\rho^2f_{i2}(z)+\dots,
\end{eqnarray}
where the functions $f_{ik}(z)$ are 
solutions of a complicated set 
of nonlinear second order ordinary differential equations.
Then, similar to the case of Einstein gravity,
the $z-$axis is divided into $N$ intervals--the rods of the solution.
Except for isolated points between the rods, one assumes that
only one of the functions
$f_0(0,z)$, $f_2(0,z)$, $f_3(0,z)$ becomes zero for a given rod, while 
the remaining functions stay finite at $\rho = 0$ in general.
(In fact, if more than one of these functions is going to zero  for a given $z$ inside a rod,
one can prove following the arguments in \cite{Harmark:2004rm}, that
there is a curvature singularity at that point.)
Again, one imposes the condition that the 
$N$ intervals must add up
to give an infinite rod.

For example, for a rod in the $\psi$-direction,
one finds the following expansion of the metric functions as $\rho\to 0$:
\begin{eqnarray}
\label{nrod1}
&&f_0(\rho,z)=f_{20}(z)+\rho^2f_{22}(z)+\dots,~~f_1(\rho,z)=f_{10}(z)+\rho^2f_{12}(z)+\dots,
\\
\nonumber
&&f_2(\rho,z)= \rho^2f_{22}(z)+\rho^4f_{24}(z)+\dots,~~f_3(\rho,z)=f_{30}(z)+\rho^2f_{32}(z)+\dots~.
\end{eqnarray}
The important feature here is that
the constraint equation $E_\rho^z=0$ implies  $f_{10}(z)/f_{22}(z)=c_1$, $i.e.$
a well-defined periodicity for the  coordinate $\psi$.

Therefore,
in order to cure the conical singularity at the rod, the coordinate $\psi$
should have a periodicity $\Delta \psi=2\pi \sqrt{c_1}$.
A different periodicity of $\psi$ implies the occurrence of a   
conical singularity 
(this is the case if there are are several different $\psi$-rods). 

Similar results holds also for a rod in the $\varphi$-direction ($i.e.$ when interchanging $f_2$ and $f_3$),
the periodicity of $\varphi$ there being again fixed by the constraint equation $E_\rho^z=0$, 
$i.e.$ $\lim_{\rho\to 0}\rho^2 f_1/f_3=c_2$.

%
\newpage
 {\small \hspace*{0.4cm}{\it Schwarzschild-Gauss-Bonnet black hole} \hspace{2.49cm} {\it black ring}}
\begin{figure}[ht]
\hbox to\linewidth{\hss%
	\resizebox{6cm}{4cm}{\includegraphics{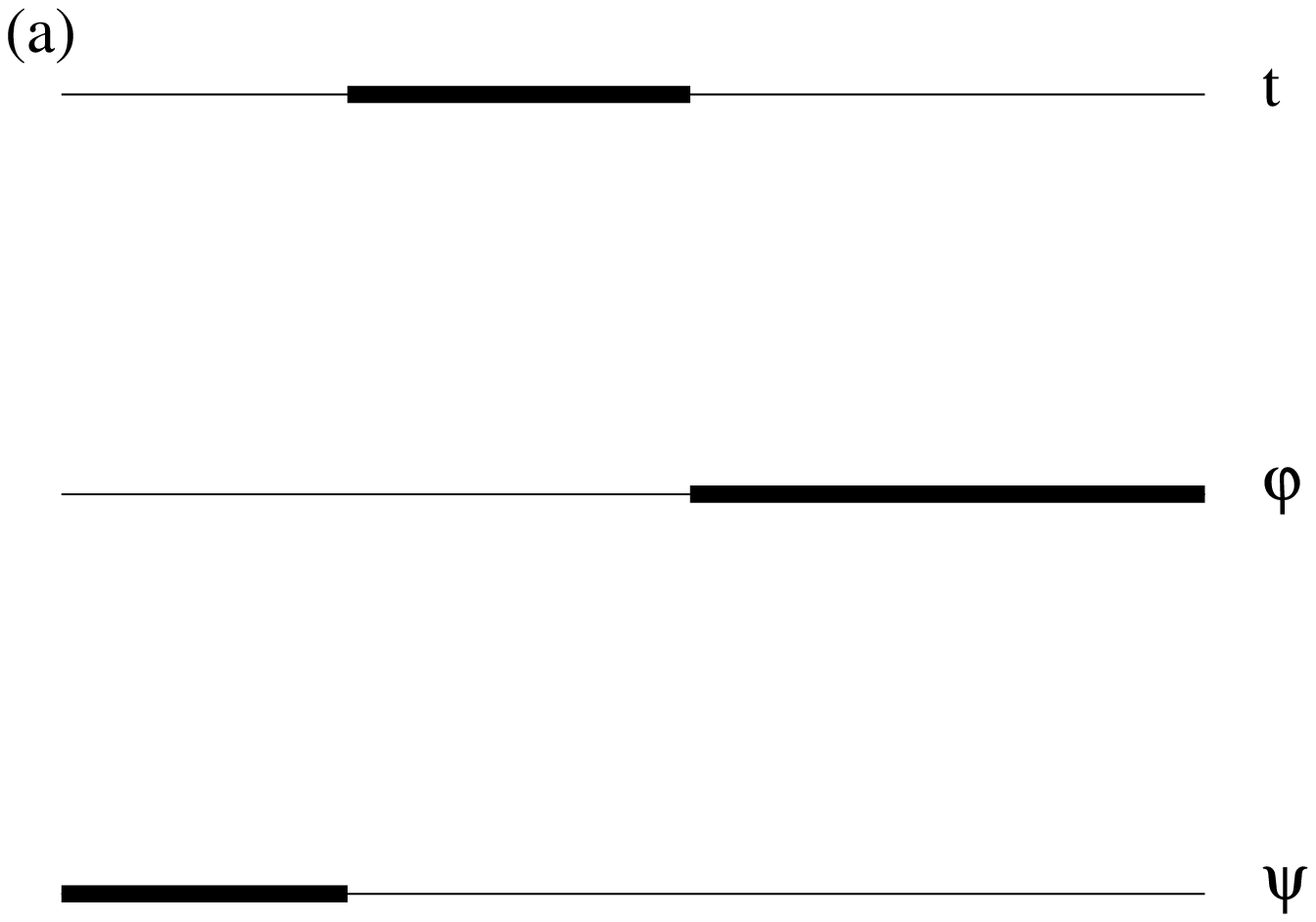}}
\hspace{15mm}%
        \resizebox{6cm}{4cm}{\includegraphics{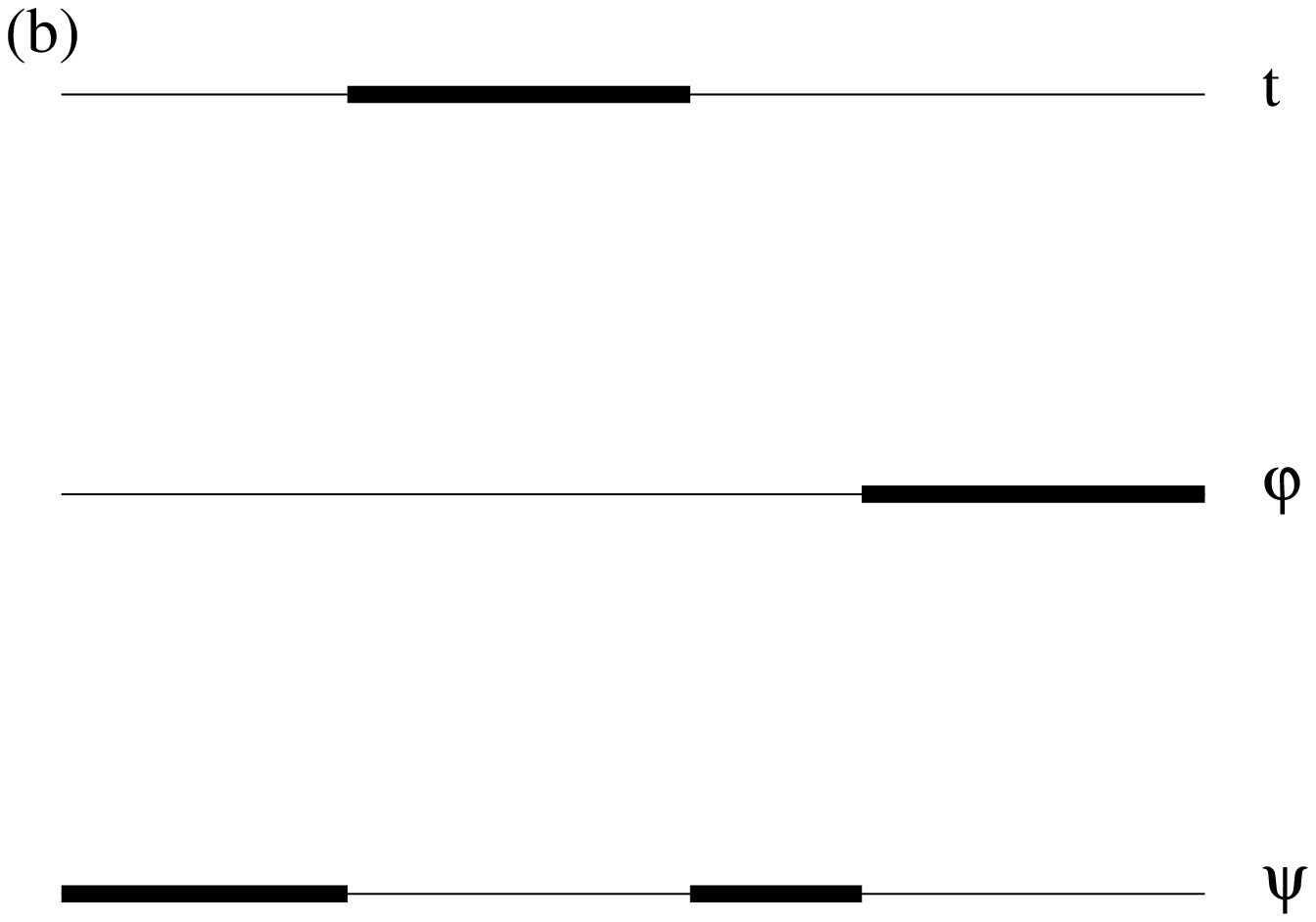}}	
\hss}
\label{rod-sch-ring}
 \end{figure}

\vspace*{0.8cm}
 {\small \hspace*{1.5cm}{\it black string} \hspace{5.9cm} {\it black Saturn} }
\begin{figure}[ht]
\hbox to\linewidth{\hss%
	\resizebox{6cm}{4cm}{\includegraphics{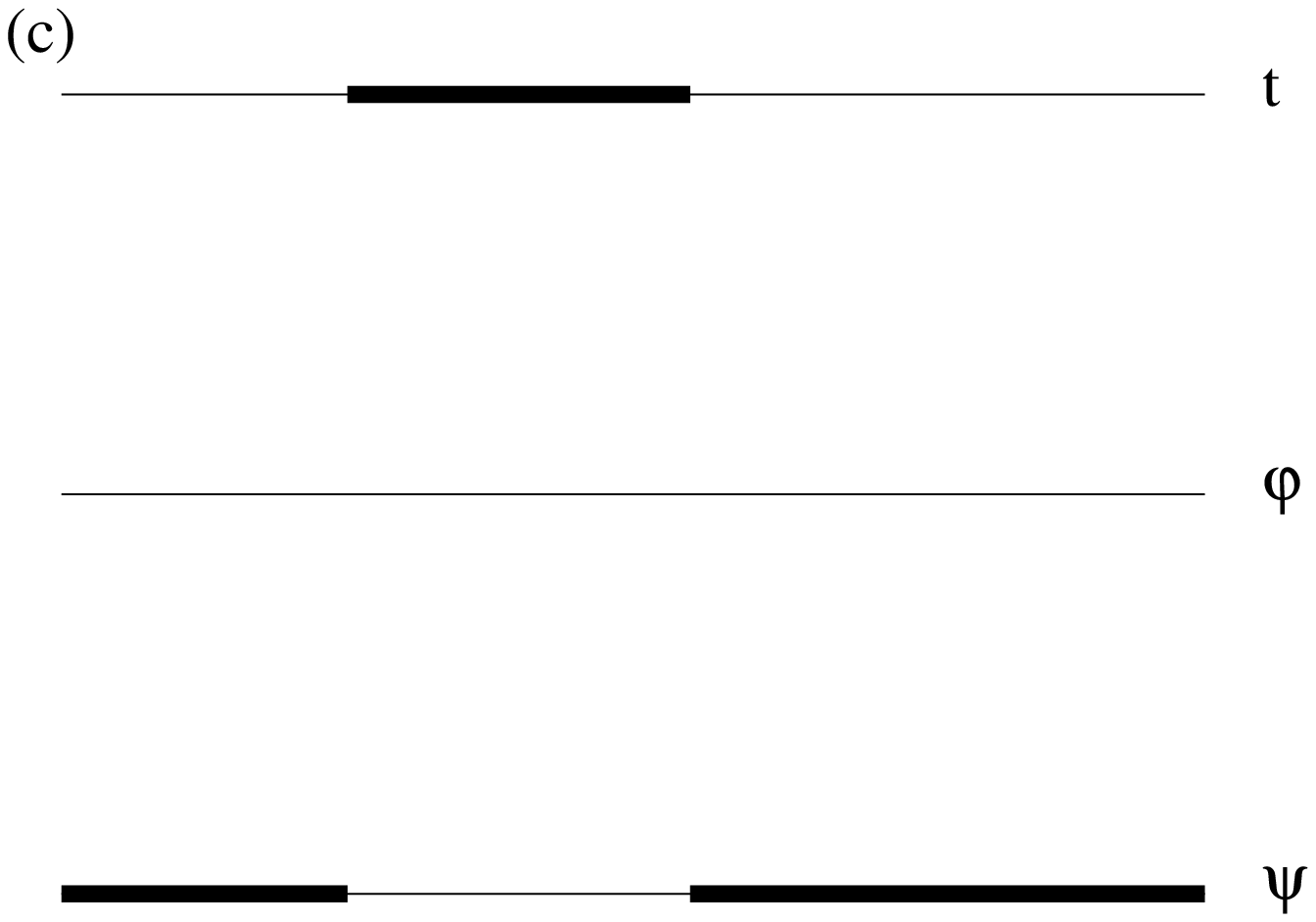}}
\hspace{15mm}%
        \resizebox{6cm}{4cm}{\includegraphics{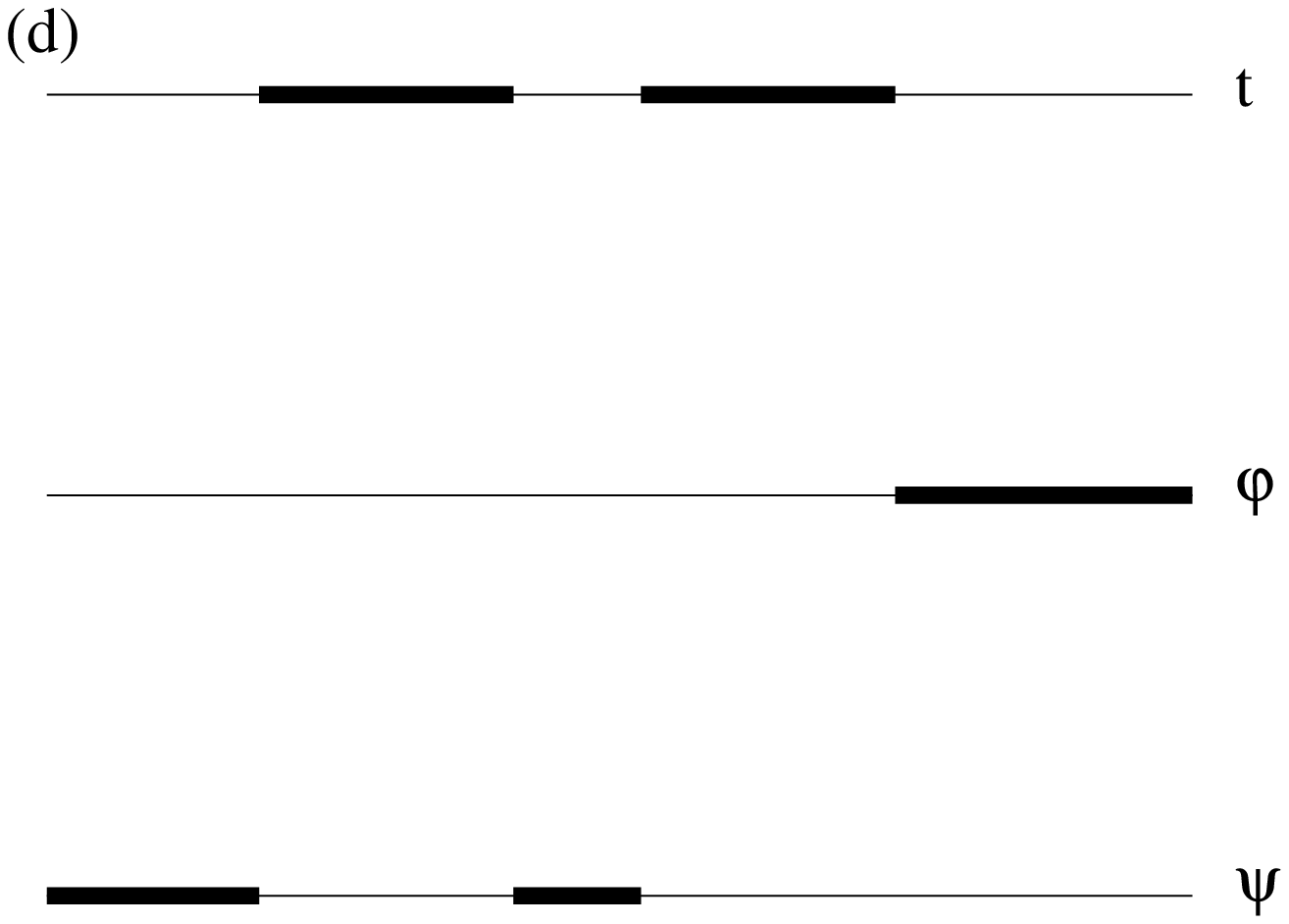}}	
\hss}
\label{rod-string-saturn}
\end{figure}
\\
{\small {\bf Figure 1.}
The  rod structure of the solutions is shown for several EGB
 solutions. 
The thin lines denote the $z-$axis and the thick lines denote the rods. 
}
\vspace{0.7cm}
\\
 A finite timelike rod corresponds to an event horizon, where\footnote{$f_{ik}(z)$ here 
should not be confused with those in (\ref{nrod1}).} 
\begin{eqnarray}
\label{nrod2}
&&f_0(\rho,z)=\rho^2f_{02}(z)+\rho^4f_{04}(z)+\dots,~~f_1(\rho,z)=f_{10}(z)+\rho^2f_{12}(z)+\dots,
\\
\nonumber
&&f_2(\rho,z)= f_{20}(z)+\rho^2f_{22}(z)+\dots,~~f_3(\rho,z)=f_{30}(z)+\rho^2f_{32}(z)+\dots,
\end{eqnarray}
 with $\lim_{\rho\to 0}\rho^2 f_1/f_0=c_3$, which fixes the Hawking temperature of solutions. 
 
Thus, depending on the physical situation we consider,
 the boundary conditions along the $z-$axis are fixed by the above relations.
The obvious boundary conditions for large $\rho,z$ is 
that the functions $f_i$ approach the Minkowski background
functions (\ref{Mink}).

Similar to the case of Einstein gravity, 
the topology of the horizon is fixed by the boundary conditions satisfied by $f_2$ and $f_3$
at the ends of the corresponding (finite) timelike rod\footnote{A timelike rod extending to 
infinity corresponds to an acceleration horizon.}.

For example, if either end of this rod continues with rods
of different angular directions, then the event horizon
has an $S^3$ topology (see Figure 1a).
A black ring corresponds to $f_2$ or $f_3$ vanishing at both ends of 
the  finite timelike rod associated with the horizon (see Figure 1b).
One can consider as well a black Saturn 
combining both types of black objects above, 
with two different horizons
(see Figure 1d).
Moreover, if both $\psi-$ and $\varphi-$rods extend to infinity, 
then the spacetime is asymptotically flat.
Solutions in Kaluza-Klein theory 
(for example a black string without a rod on the
$\psi-$ or $\phi-$direction (see Figure 1c)) can also be considered.

It is tempting to conjecture that, similar to 
the case of Einstein gravity \cite{Hollands:2007aj},
a $d=5$ EGB solution within the ansatz (\ref{metric}), 
is uniquely specified by its rod structure. 

\subsubsection{The physical quantities}

The general results in the literature \cite{Deser:2002jk} 
show that, similar to the case of Einstein gravity, the mass $M$ of an asymptotically flat
EGB 
solution can be read from the asymptotic expression for the metric component $g_{tt}$  
\begin{eqnarray}
\label{gtt}
-g_{tt}=f_0\sim 1-\frac{4 G M}{3\pi \sqrt{\rho^2+z^2}}+\dots~.
\end{eqnarray}
Supposing we have an event horizon for $z_1\leq z\leq z_2$,
the horizon metric is given by\footnote{If there are several horizons, then one should write
such an expansion for each of them.}
\begin{eqnarray}
d\sigma^2=f_1(0,z)dz^2+f_2(0,z)d\psi^2+f_3(0,z)d\varphi^2.
\end{eqnarray}
Two quantities associated with the event horizon are
the event horizon area $A_H$ and
the Hawking temperature. For the metric ansatz (\ref{metric}) these are given by
 \begin{eqnarray}
\label{AHTH}
A_H=\Delta \psi \Delta \varphi\int_{z_1}^{z_2}dz \sqrt{f_1(0,z)f_2(0,z)f_3(0,z)}, ~~~
T_H=\frac{1}{2\pi}\lim_{\rho\to 0} \sqrt{\frac{f_{0}(\rho,z)}{\rho^2 f_{1}(\rho,z)}}.
\end{eqnarray}
For solutions in EGB theory it is also of interest to compute the Ricci scalar 
of the horizon
 \begin{eqnarray}
\label{Rh}
& R_{\Sigma_h}=\frac{1}{2f_1(0,z)}
\bigg(
\frac{\dot f_1(0,z)\dot f_2(0,z)}{f_1(0,z)f_2(0,z)}
+\frac{\dot f_1(0,z)\dot f_3(0,z)}{f_1(0,z)f_3(0,z)}
-\frac{\dot f_2(0,z)\dot f_3(0,z)}{f_2(0,z)f_3(0,z)}
+\frac{\dot f_2^2(0,z)}{f_2^2(0,z)}+\frac{\dot f_3^2(0,z)}{f_3^2(0,z)}
-2(\frac{\ddot f_2(0,z)}{f_2(0,z)}+\frac{\ddot f_3(0,z)}{f_3(0,z)})
\bigg).~~~{~~~~}
\end{eqnarray}
%
Considering now the case of a space-like $\psi-$rod for $z_3\leq z\leq z_4$,
one writes the line element on this three-dimensional surface ${\Sigma_\delta}$
\begin{eqnarray}
d\sigma^2=f_1(0,z)dz^2+ f_3(0,z)d\varphi^2- f_0(0,z)dt^2.
\end{eqnarray}
The first quantity of interest is the proper
length of the rod
 \begin{eqnarray}
\label{L}
 L=\int_{z_3}^{z_4}dz\sqrt{f_1(0,z)},
\end{eqnarray}
(note that for a finite rod, $L$ differs from the coordinate distance $\Delta z=z_4-z_3$).

The solutions we are interested in  may possess a conical singularity  
along some region of the symmetry axis.
To define a
conical singularity for a rotational axis with angle $\psi$ one computes
the proper circumference $C$ around the axis and its proper radius $R$ and
defines:
\begin{eqnarray}
\alpha&=&\frac{dC}{dR}\bigg|_{R=0}=\lim_{\rho\rightarrow 0}\frac{\sqrt{%
g_{\psi\psi}}\Delta\psi}{\int_0^{\rho}\sqrt{g_{\rho\rho}}d\rho}%
=\lim_{\rho\rightarrow 0}\frac{\partial_{\rho}\sqrt{g_{\psi\psi}}%
\Delta\psi}{\sqrt{g_{\rho\rho}}},
\end{eqnarray}
where $\Delta\psi$ is the period of $\psi$. The presence of a conical
singularity is now expressed\footnote{Note that, in some sence, 
fixing $\delta$ is the analogue of computing the Hawking temperature on the Euclidean section.} by means of:
\begin{eqnarray}
\label{delta}
\delta&=&2\pi-\alpha=2\pi\left(1-\lim_{\rho\rightarrow 0} \sqrt{\frac{f_{2}(\rho,z)}{\rho^2f_{1}(\rho,z)}}\right),
\end{eqnarray}
such that $\delta>0$ corresponds to a conical deficit,
while $\delta<0$ corresponds to a conical excess.
A conical deficit  can be interpreted as a string stretched along  on a certain segment of the $z-$axis, while a conical  
excess is a strut pushing  apart the rods connected to that segment (in fact, for $d=5$, the struts and strings
 are two dimensional surfaces). 
Similar to Einstein gravity,  a constant rescalings of
$\psi$  can be used to eliminate possible conical singularities on a
given segment, but in general, once this is fixed, there will remain
conical singularities at other $\psi$-segments.  

For $\delta<0$, we have found it convenient to introduce the quantity
\begin{eqnarray}
\label{rel-delta}
\bar \delta=\frac{\delta/(2\pi)}{1-\delta/(2\pi)},
\end{eqnarray}
which has a finite range and measures the 
'relative angular excess'.

Of interest here is to compute the spacetime area spanned by the $\psi$-rod.
This is done by going to the Euclidean section $t\to i \tau$ and evaluating the quantity
\begin{eqnarray}
\label{area}
Area=\beta \Delta \varphi \int_{z_3}^{z_4}dz\sqrt{f_0(0,z)f_1(0,z) f_3(0,z)},
\end{eqnarray}
where $\beta=1/T_H$
is the periodicity of the Euclidean time.
For completeness, we give here also the expression for the 
Ricci scalar on the $\psi-$rod,
 \begin{eqnarray}
\label{Rpsi}
&  R_{\Sigma_\delta}=\frac{1}{2f_1(0,z)}
\bigg(
\frac{\dot f_1(0,z)\dot f_0(0,z)}{f_1(0,z)f_0(0,z)}
+\frac{\dot f_1(0,z)\dot f_3(0,z)}{f_1(0,z)f_3(0,z)}
-\frac{\dot f_0(0,z)\dot f_3(0,z)}{f_0(0,z)f_3(0,z)}
+\frac{\dot f_0^2(0,z)}{f_0^2(0,z)}+\frac{\dot f_3^2(0,z)}{f_3^2(0,z)}
-2(\frac{\ddot f_0(0,z)}{f_0(0,z)}+\frac{\ddot f_3(0,z)}{f_3(0,z)})
\bigg).~~~{~~~~}
\end{eqnarray}
Of course, similar expressions can be written when considering instead a $\varphi$-rod.

\subsubsection{Remarks on the free energy and thermodynamics}
The discussion in this subsection applies to asymptotically flat solutions 
(although it can easily be generalized to the Kaluza-Klein case).

The gravitational thermodynamics of the EGB black objects can be formulated via the path integral approach
\cite{GibbonsHawking1,Hawking:ig}.
In what follows it is important to use the observation that one can write
\begin{eqnarray}
\label{total-action1} 
&&R_t^t\sqrt{-g}=-\frac{1}{2}
\left(
\partial_\rho(\sqrt{\frac{f_2f_3}{f_0}}  f_0')+
\partial_z(\sqrt{\frac{f_2f_3}{f_0}}  \dot f_0)
\right),
\\
&&(H_t^t+\frac{1}{2}L_{GB})\sqrt{-g}= \frac{1}{2} \left(\partial_\rho T_{\rho}+\partial_z T_{z}\right),
\end{eqnarray} 
where
\begin{eqnarray}
\nonumber
&&T_\rho= \sqrt{\frac{f_2f_3}{f_0}}
\bigg(
\frac{f_0'}{f_1}(\frac{f_2'f_3'}{f_2f_3}-\frac{\dot f_2 \dot f_3 }{f_2f_3})
+\frac{\dot f_0}{f_1}
( \frac{\dot f_2 f_0'}{f_0f_2}+\frac{\dot f_3 f_0' }{f_0f_3}+\frac{\dot f_2 f_3'}{f_2f_3}+\frac{\dot f_3 f_2' }{f_2f_3}  )
+\frac{f_0'}{f_1^2f_2}( f_1'f_2'+\dot f_1 \dot f_2)
\\
\nonumber
&&{~~~~~~~~~~~~~~}
+\frac{\dot f_0}{f_1^2 f_2}(\dot f_2 f_1'-\dot f_1 f_2')
+\frac{\dot f_0}{f_1^2 f_3}(\dot f_3 f_1'-\dot f_1 f_3')
+ \frac{f_0'}{f_1^2f_3}( f_1'f_3'+\dot f_1 \dot f_3)
\\
\label{Tr} 
&&{~~~~~~~~~~~~~~}
-\frac{\dot f_0^2}{f_0f_1}(\frac{f_3'}{f_3}+\frac{f_2'}{f_2})
-\frac{2\dot f_0'}{f_1}(\frac{\dot f_2 }{f_2}+\frac{\dot f_3 }{f_3})
+\frac{2\ddot f_0}{f_1}(\frac{f_2'}{f_2}+\frac{f_3'}{f_3})
\bigg) ,
\\
&&T_z= \sqrt{\frac{f_2f_3}{f_0}}
\bigg(
\frac{f_0'}{f_1f_2f_3}(\dot f_2 f_3'+\dot f_3 f_2') 
+\frac{\dot f_0}{f_1f_2 f_3}(\dot f_2 \dot f_3-f_2'f_3')
+\frac{\dot f_0 f_0'}{f_0f_1}(\frac{f_3'}{f_3}+\frac{f_2'}{f_2})
\nonumber
\\
\label{TZ} 
\nonumber
&&{~~~~~~~~~~~~~~}
-\frac{f_0'^2}{f_0f_1}(\frac{\dot f_2}{f_2}+\frac{\dot f_3}{f_3})
+\frac{f_0'}{f_1^2f_2}(\dot f_1 f_2'-\dot f_2 f_1')
+\frac{\dot f_0}{f_1^2f_2}( f_1'f_2'+\dot f_1 \dot f_2)
\\
\nonumber
&&{~~~~~~~~~~~~~~}
+\frac{f_0'}{f_1^2f_3}(\dot f_1 f_3'-\dot f_3 f_1')
+\frac{\dot f_0}{f_1^2f_3}( f_1'f_3'+\dot f_1 \dot f_3)
-\frac{2\dot f_0'}{f_1}(\frac{f_2'}{f_2}+\frac{f_3'}{f_3})
+\frac{2f_0''}{f_0}(\frac{\dot f_2}{f_2}+\frac{\dot f_3}{f_3})
\bigg).
 \end{eqnarray} 
  When computing the classical bulk action evaluated on the equations of motion,
one replaces the $R + {\alpha'} L_{GB}$ volume term with
$2(R_t^t+ {\alpha'} (H_t^t+L_{GB}/2))$ and make use of (\ref{total-action1}) to express
it as a difference of two boundary integrals.
The boundary integral  at infinity should be evaluated  together with the contributions
from $I_{b}^{(E)}$ and $I_{b}^{(GB)}$.
 As usual, this quantity is divergent.
 To regularize it, one has to subtract the contribution of the Minkowski background  for {\it both}
Einstein and Gauss-Bonnet boundary 
terms\footnote{Note that for asymptotically flat solutions,
the Gibbons-Hawking boundary term
(\ref{Ib1}) gives a divergent contribution to $I_0$,
while the GB boundary term reduces to a constant factor.
It would be interesting to generalize the quasilocal formalism 
and the renormalized boundary stress-tensor  
from Einstein gravity to EGB theory.
This will avoid the requirement to choose a background for the  solutions.
For example, we have found that 
the usual counterterm  used in \cite{Astefanesei:2005ad} for the 
black rings in Einstein theory 
regularizes also the mass and action of some 
(asymptotically flat-) EGB solutions,
including the black rings discussed in Section 4. 
However, this approach implies the existence
of a constant term $\sim \beta \alpha'$ in the action, 
originating in the contribution of the GB boundary term (\ref{Ib2}). }.
 
A direct computation implies the following expression for the 
tree level action of a  single black object  in $d=5$ EGB theory 
(the extension to multi-black objects is straightforward)
\begin{eqnarray}
\label{I0} 
I_0=\beta (M-\frac{1}{4G}T_H(A_H +\alpha' A_1)),
\end{eqnarray} 
with
\begin{eqnarray}
\label{I0-c} 
A_1= 2\int_{\Sigma_h} d^{3}x \sqrt{\tilde h}  R_{\Sigma_h} ,
\end{eqnarray} 
 where $\tilde h=\sqrt{f_1f_2f_3}$ is the determinant of the induced metric on the horizon 
 and $ R_{\Sigma_h}$ is the event horizon curvature as given by (\ref{Rh}).

The above results hold for the case of configurations 
with a regular $z-$axis ($i.e.$ the periodicity
of both angles $\psi$ and $\varphi$ is $2\pi$ everywhere). 
It is interesting to extend this analysis to the case of EGB solutions
with conical singularities 
(this is the case of the black rings discussed in the Section 4).
For simplicity, we shall consider here the case of  
a single singular section of the $z-$axis
associated with a $\psi-$rod 
(the generalization to other cases is straightforward).
Then the conical singularity will add an extra contribution to
the total tree level  Euclidean action of the system, which leads to a
more complicated thermodynamics 
of the system (see $e.g.$ \cite{Costa:2000kf}
for a related discussion for the $d=4$ Israel-Kahn solution).
This contribution can be evaluated by using the relations \cite{Fursaev:1995ef}
\begin{eqnarray}
\label{comp1} 
 \frac{1}{2}\int_{\Sigma_\delta} d^5 x \sqrt{ g} R= Area~\delta,~~~
 \frac{1}{2}\int_{\Sigma_\delta} d^5 x \sqrt{ g}L_{\rm GB} = Area_1~\delta,
 \end{eqnarray}
where $Area$ is the space-time area of the surface 
spanned by the conical singularity and 
  \begin{eqnarray}
\label{comp2} 
 Area_1=2\int_{\Sigma_\delta} d^3 x \sqrt{\bar h} R_{\Sigma_\delta} , 
 \end{eqnarray}
with $ \bar h=\sqrt{f_1f_0f_3}$ the determinant 
of the induced metric on the surface ${\Sigma_\delta}$
and $ R_{\Sigma_\delta}$ the Ricci scalar on $\Sigma_\delta$ 
as given by (\ref{Rpsi}).

Thus the total action in the presence of a conical singularity becomes
   \begin{eqnarray}
\label{tot-act} 
I=\beta \big(M-\frac{1}{4G}T_H(A_H +\alpha' A_1) \big) -\frac{1}{8\pi G } (Area+\alpha' Area_1) ~\delta.
 \end{eqnarray}
 The free energy of the solutions is identified as $F =T_H I={\cal  M}-T_H S$, where ${\cal  M}$
is the mass which enters the thermodynamics and $S$ is the entropy.
 The entropy 
of the EGB black hole solutions  {\it without}
conical singularities can be written as an integral over the event horizon \cite{Wald:1993nt}:
 \begin{eqnarray}
 \label{S-Noether} 
 S=\frac{1}{4G}\int_{\Sigma_h} d^{3}x \sqrt{\tilde h}(1+ 2\alpha'  R_{\Sigma_h}),
  \end{eqnarray}
which is the sum of one quarter of the event horizon area plus a
Gauss-Bonnet correction.
In this case, the mass ${\cal  M}$ computed from the first law of thermodynamics
is equal to the mass $M$ computed at infinity.

It would be interesting 
to perform a similar computation for solutions with conical singularities. 
Similar to the case of 
Einstein gravity \cite{carlos}, a new  extensive parameter 
${\cal A}=(Area+\alpha' Area_1)/\beta$  
associated with the rod containing the conical
singularity will appears here, 
while the free energy becomes a function of both $ T_H $ and $ {\cal A}$.
Also, the first law of  thermodynamics  
contains an extra-work term ${\cal T} d{\cal A}$ 
(with ${\cal T}=-\delta/8\pi G$ the tension associated with ${\cal A}$).
Then the entropy $S$ and the thermodynamical mass ${\cal M}$ 
of the physical system are given by
\begin{eqnarray}
\label{r1} 
 S=-\frac{\partial F}{\partial T_H}\bigg|_{{\cal A}},~~
~{\cal M}=F+T_HS.
 \end{eqnarray}
The thermodynamical stability of the EGB solutions can be studied in the usual way. For example, one defines
the specific heat of a black object $C = T_H \left( \frac{\partial S}{\partial T_H} \right) $, 
with  $C>0$ for thermodynamically stable solutions.

\section{The known static solutions in $d=5$ EGB theory}

\subsection{The Schwarzschild black hole in EGB theory}

\subsubsection{The solution in Schwarzschild coordinates}

The black hole solutions of EGB gravity have been studied begining with the
work of \cite{Deser} more than 20 years ago.
Most of the solutions in the literature restricted to static, spherically symmetric
configurations and a  Schwarzschild coordinate system.
A suitable metric ansatz in this case is 
 \begin{eqnarray}
\label{general-metric-form}
ds^2=\frac{dr^2}{N(r)}+r^2(d\theta^2+\sin^2\theta d\varphi^2+\cos^2\theta d\psi^2)-N(r)\sigma^2(r)dt^2,
\end{eqnarray}
and the expressions for $N(r)$ and $\sigma(r)$ depend 
on the matter content of the theory.

The Hawking temperature of a generic black hole (\ref{general-metric-form}) is $T_H=N'(r_h)\sigma(r_h)/(4 \pi)$, where a prime denotes 
the derivative with respect to the radial coordinate and
 $r_h$ is the largest positive root of $N(r)$, typically associated to the outer horizon of a black hole.
 Also, the event horizon area of a black hole is $A_H=V_3r_h^3$ (with $V_3=2 \pi^2$ the area of the three-sphere).

The complexity of the EGB theory, 
basically due to higher order terms in the curvature tensor,
makes the task of finding exact solutions very difficult.
To the best of our knowledge, the only $d=5$ static, spherically
symmetric EGB solutions known in closed form
are the generalizations of  
(electro-)vacuum Einstein gravity configurations\footnote{Exact 
solutions describing cosmic strings in $d=5$ EGB theory were found  in 
\cite{AzregAinou:1996eb}. 
Obviously, these solutions cannot be described within the metric 
ansatz  (\ref{general-metric-form}).}.
The Schwarzschild black hole in EGB theory has $\sigma(r)=1$, 
the metric function
$N(r)$ being given by\footnote{The usual form for $N(r)$ in the literature
is in terms of $r_h^2=m-2\alpha'$, 
which corresponds to fixing the mass of the solutions. 
However, the expression (\ref{N1}) fits better with the purposes of this work.}
 \begin{eqnarray}
\label{N1}
N(r)=1+\frac{ r^2}{ 4\alpha'}
 \bigg (
  1-\sqrt{1+ \frac{8\alpha'(r_h'^2+2\alpha')}{r^{4}} }
 \bigg).
\end{eqnarray}
The constant $r_h$ in (\ref{N1}) corresponds to the event horizon radius. 
One should note that the EGB Schwarzschild solution exists for 
all $r_h>0$ and $\alpha' \geq 0$.
As $r\to r_h$ one finds
\begin{eqnarray}
\label{N3}
N(r)=\frac{2r_h}{r_h^2+4\alpha'}(r-r_h)+O(r-r_h)^2.
\end{eqnarray}

At short distances, the EGB Schwarzschild solution and  
its Einstein gravity counterpart are
substantially different due to the effect of the GB term.
An interesting feature is that the EGB Schwarzschild metric  
turns out to be finite at the origin, 
since $N(r)\to 1-\sqrt{(r_h^2+2\alpha')/(2\alpha')}$ as $r\to 0$.
However, at large distances ($r^2\gg \alpha'$)
the EGB black hole behaves like the Schwarzschild solution.
Once the event horizon radius $r_h$ is fixed 
(as is done in the numerical construction of the solutions), 
the parameter $\alpha'$ enters the $1/r^2$ term in the asymptotics,
\begin{eqnarray}
\label{N2}
N(r)=1-\frac{r_h^2+2\alpha'^2}{r^2}+\frac{2\alpha'(r_h^2+2\alpha'^2)^2}{r^6}+O(1/r^{10}).
\end{eqnarray}
Also, as $\alpha'\to 0$, one recovers to leading order the Schwarzschild black hole expression
\begin{eqnarray}
\label{N4}
N(r)=1-\frac{r_h^2}{r^2}+(-\frac{2}{r^2}+\frac{2r_h^4}{r^6})\alpha'+O(\alpha')^2.
\end{eqnarray}
The Hawking temperature and the mass of these solutions are given by
\begin{eqnarray}
\label{SGB-TH}
 T_H=\frac{r_h}{2\pi(r_h^2+4\alpha')},~~M=\frac{3V_3}{16\pi G}(r_h^2+2\alpha').
\end{eqnarray} 
The above expression for $M$ shows the existence of a mass gap: 
black holes exist only for $M>3V_3 \alpha'/(8\pi G)$.
Thus the mass spectrum of the spherically symmetric EGB black holes 
is bounded from below.

A straightforward computation leads to the following expression 
for the entropy of the EGB Schwarzschild black holes
\begin{eqnarray}
\label{BH-entropy}
 S=S_0+S_{c}~~~{\rm with}~~~S_0=\frac{A_H}{4G}~,
~~~~S_c=3\alpha'  \frac{V_3r_h}{G}  .
\end{eqnarray}  
One can easily verify that the first law of thermodynamics 
$dM = T_H dS$ also holds.

Without entering into details, 
we mention the existence of some substantial 
differences between the thermodynamics
of the EGB Schwarzschild solutions and their Einstein gravity counterparts.
If the black holes are large enough $r_h\gg \sqrt{\alpha'}$,
then they behave like their Schwarzschild-Tangerlini counterparts.
A different picture is found for small values of $r_h$, 
since $T_H\simeq r_h/(8\pi \alpha')$ in that case.
Therefore, the specific heat changes its sign 
at length scales of order $r_h\sim\sqrt{\alpha'}$.
This implies the existence of a branch 
of five dimensional EGB black holes which is thermodynamically stable
(see the Ref. \cite{Garraffo:2008hu} for a review of these aspects).

\subsubsection{The solution in Weyl-type coordinates}
To bring the generic metric (\ref{general-metric-form}) to Weyl form (\ref{metric}), one 
considers the coordinate transformation
\begin{eqnarray}
\label{coord-transf-SGB1}
\rho=\frac{1}{2}r_0^2\sinh G(r) \sin 2\theta,~~z=\frac{1}{2}r_0^2\cosh G(r) \cos 2\theta,
 \end{eqnarray}
 where $r_0$ is defined by the asymptotic expansion of $N(r)=1-(r_0/r)^2+\dots$ ($i.e.$ $r_0^2=r_h^2+2\alpha'$).
The function $G(r)$ above is defined as
 \begin{eqnarray}
\label{integral}
G(r)=2\int_{r_h}^r \frac{dx}{x \sqrt{N(x)} }
  \end{eqnarray}
(with $G(r_h)=0$).
Then the Schwarzschild coordinate $r$ is expressed in terms of 
Weyl coordinates $\rho,z$ as
 \begin{eqnarray}
\label{coord-transf-SGB2}
r(\rho,z)=G^{-1}(\operatorname{arsinh}\sqrt{X(\rho,z)}),
  \end{eqnarray}
   where
 \begin{eqnarray}
\label{coord-transf-SGB3}
 X(\rho,z)=
 \frac{1}{2}
 \left(
 \frac{4}{r_0^2} (\rho^2+z^2)+
 \sqrt{\frac{16 \rho^2}{r_0^2}+(1-\frac{4(\rho^2+z^2)}{r_0^2})^2}
 \right).
  \end{eqnarray}
  
\begin{figure}[ht]
\hbox to\linewidth{\hss%
	\resizebox{8cm}{6cm}{\includegraphics{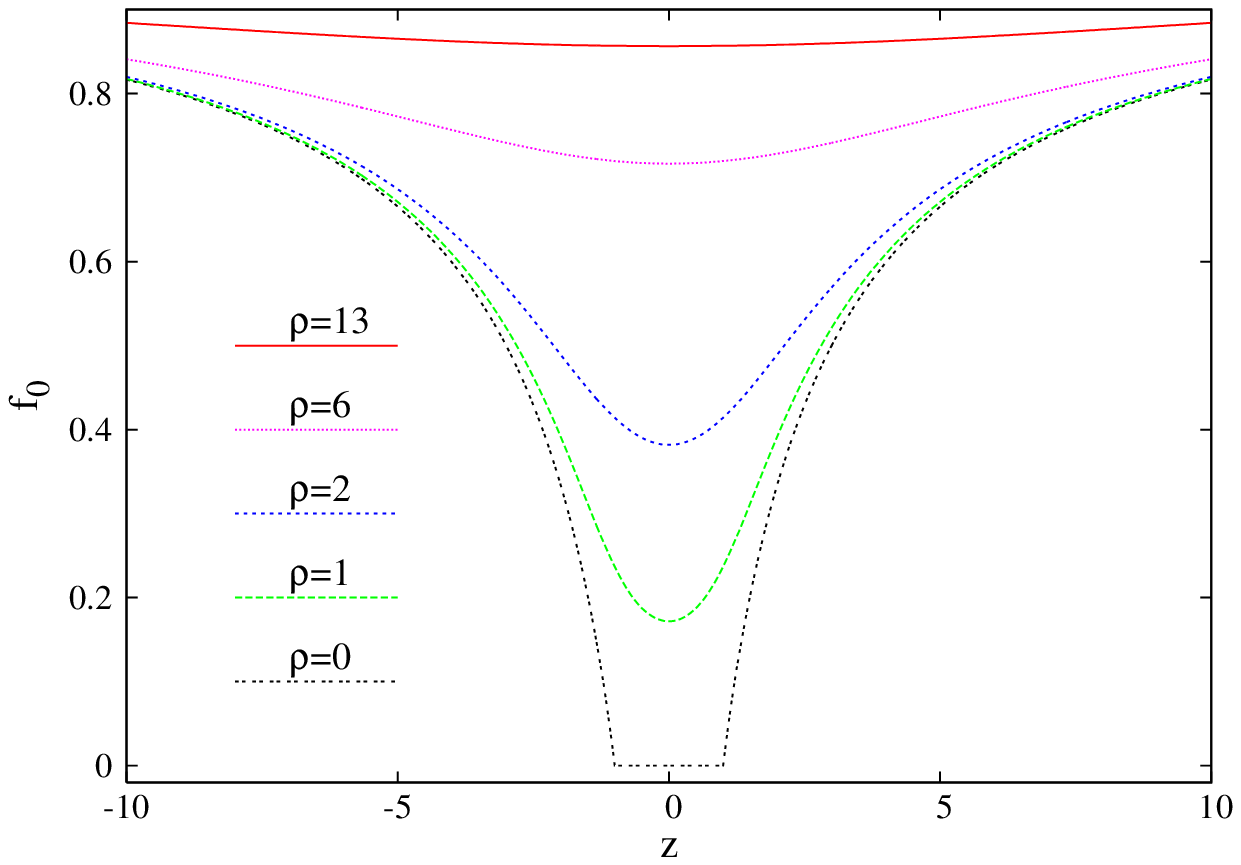}}
\hspace{5mm}%
        \resizebox{8cm}{6cm}{\includegraphics{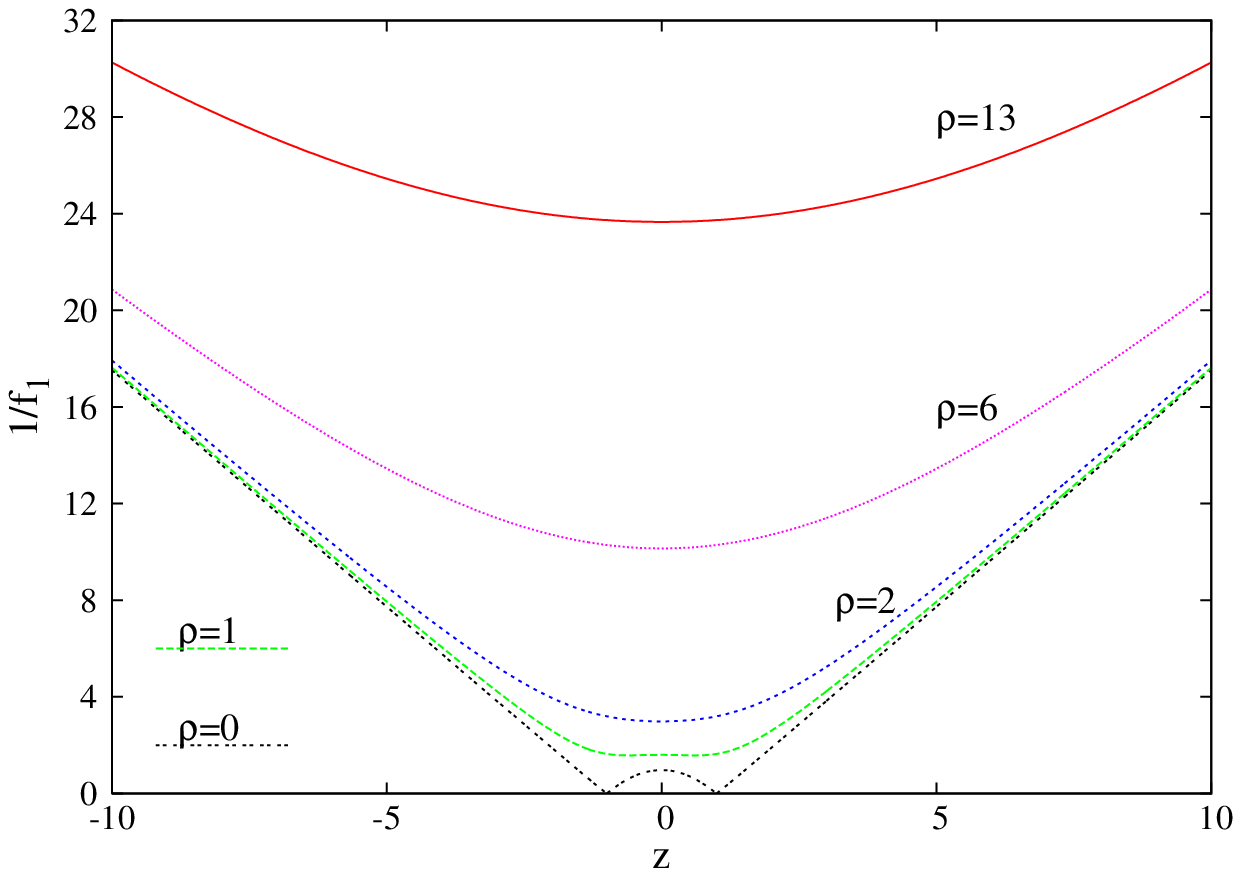}}	
\hss}
\end{figure}
\begin{figure}[ht]
\hbox to\linewidth{\hss%
	\resizebox{8cm}{6cm}{\includegraphics{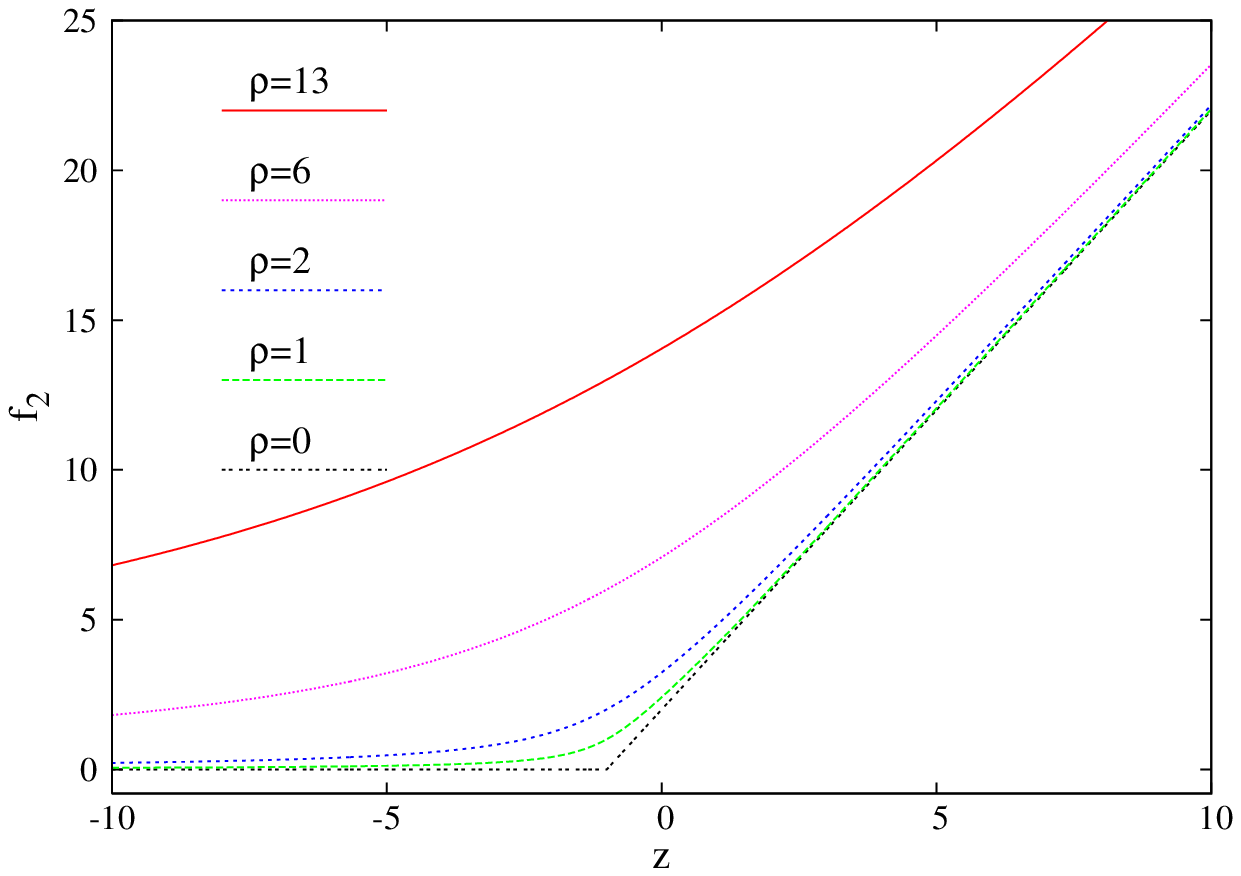}}
\hspace{5mm}%
        \resizebox{8cm}{6cm}{\includegraphics{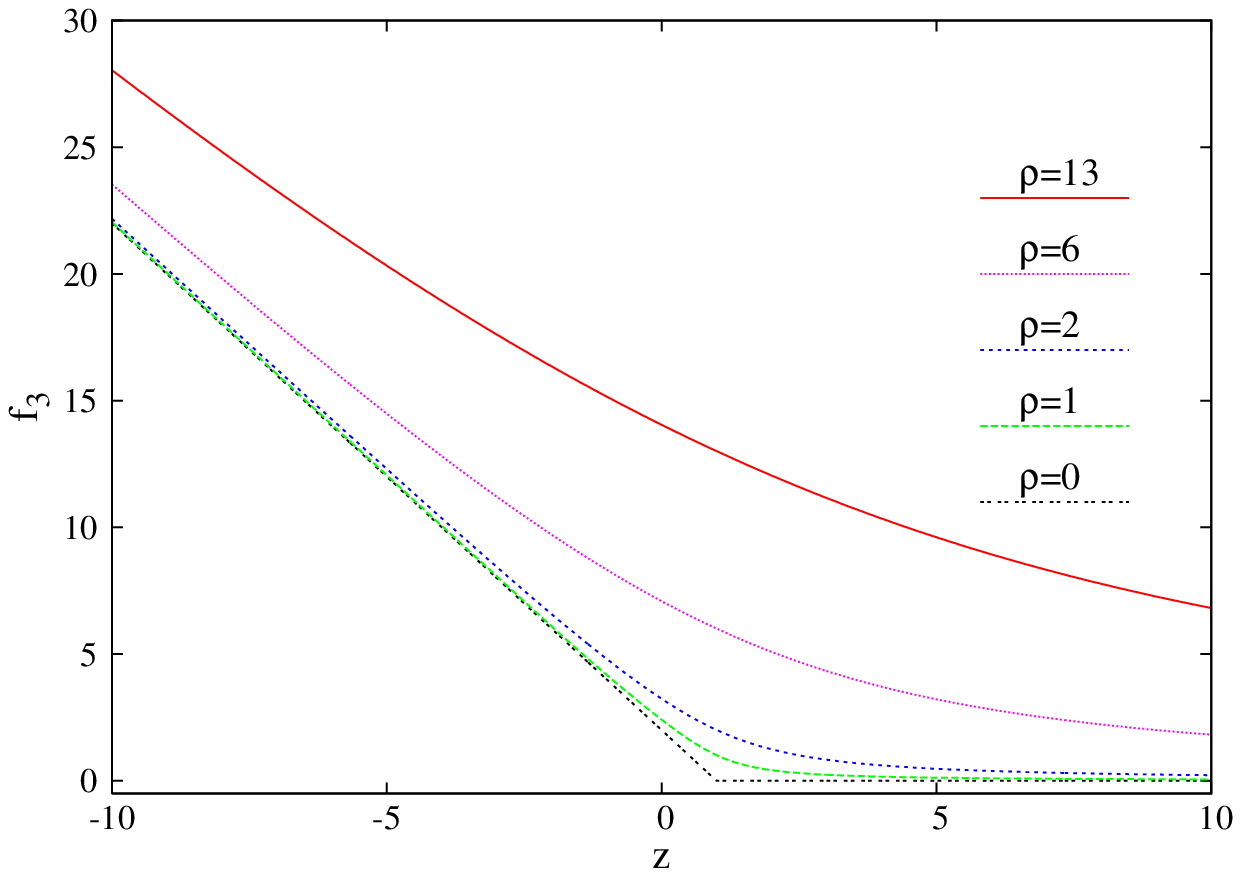}}	
\hss}
\label{Schw1}
\end{figure}
\vspace{0.3cm}
{\small {\bf Figure 2.}
The metric functions $f_i$ 
of an EGB Schwarzschild black hole
are shown versus $z$ for several values of $\rho$.
The relevant parameters here are
  $a=1$, $\alpha'=0.01$.
  }
 \\
 \\
A straightforward but cumbersome computation 
leads to the following expression for the
metric functions  for the parametrization (\ref{metric}) 
\begin{eqnarray}
\label{metric-weyl}
&&f_0(\rho,z)= F(r (\rho,z)) \sigma^2(r(\rho,z)),~~~
 f_1(\rho,z)= \frac{r(\rho,z)P^2(\rho,z)}{1+P^4(\rho,z)},
 \\
 \nonumber
&&f_2(\rho,z)=\frac{1}{2}r^2(\rho,z) (1+\frac{1}{\sqrt{P^2(\rho,z)+1}}),~~
  f_3(\rho,z)=\frac{1}{2}r^2(\rho,z) (1-\frac{1}{\sqrt{P^2(\rho,z)+1}}),
\end{eqnarray}
 with
   \begin{eqnarray}
\label{coord-transf-SGB4}
 P(\rho,z)=\frac{\rho}{z}\sqrt{\frac{X(\rho,z)+1}{X(\rho,z)}}.
  \end{eqnarray}
Unfortunately, the integral (\ref{integral}) cannot be computed in closed form
for the known solutions of EGB theory (except for the asymptotic expressions as $r\to r_h$ and $r\to \infty$).
However, the  expressions (\ref{integral}), (\ref{metric-weyl}) can be evaluated numerically.

For any value of $\alpha'$,
the rod structure of the EGB Schwarzschild black hole
as resulting from (\ref{metric-weyl})
consists of a semi-infinite space-like rod $[-\infty,-a]$
(with $f_2(0,z)=0$ there),
a finite time-like rod $[-a, a]$ ($f_0(0,z)=0$) and a  
semi-infinite space-like rod $[a,\infty]$ (with vanishing $f_3(0,z)$)
in the $\varphi$-direction (with $a=r_0^2/4$).
Thus the topology of the horizon is $S^3$ as required (see Figure 1a).
A plot of the  metric functions $f_i$ exhibiting 
this rod structure for a typical
EGB Schwarzschild solution is shown in Figure 2.
In principle, most of the physically relevant 
properties of the EGB Schwarzschild black hole 
can also be rederived within the metric ansatz (\ref{metric}).
However, the required computation is much more difficult for that coordinate system.
 
\subsection{The static uniform black string}
It is of interest to briefly review the situation for a different 
type of black object in EGB theory
 which can also be studied within the ansatz (\ref{metric}).
In Einstein gravity, one can construct uniform $d=5$ black string solutions
by adding a flat direction to any  $d=4$ vacuum black hole. 
(These solutions can still be written in the $d=5$ Weyl form (\ref{metric}), the new direction $\psi$
being trivial.)
However, it is straightforward to check that this simple construction does
not work in the presence of a GB term in the 
action\footnote{It is interesting to notice that
this is valid also for solutions with a
cosmological constant.},
 unless the solutions are conformally flat.  
Although no exact solutions describing $d=5$ black strings
in EGB theory are known so far, 
these configurations were studied numerically in Ref. \cite{Kobayashi:2004hq}.
These solutions can be constructed within a metric ansatz related to (\ref{general-metric-form})
 \begin{eqnarray}
\label{UBS}
 ds^2=\frac{dr^2}{N(r)}
 +r^2(d\theta^2+\sin^2 \theta d\psi^2)-N(r)\sigma^2(r)dt^2+b(r)d\varphi^2,
\end{eqnarray}
(note that $0\leq \theta\leq \pi$ in this case while the periodicity of $\varphi$
is not fixed {\it a priori}).
The new feature here as compared to the case of Einstein gravity 
is that the metric component
$g_{\varphi\varphi}$ differs from one.

The event horizon of a uniform black string  is located
at $r=r_h>0$,
where the following approximate form of the metric functions holds
 \begin{eqnarray}
\label{UBS-eh}
&&N(r)=N_1(r-r_h)+O(r-r_h)^2,
~~\sigma(r)=\sigma_h+ \sigma_1(r-r_h)+O(r-r_h)^2,
\\
\nonumber
&&b(r)=b_h+b_1(r-r_h)+O(r-r_h)^2,
\end{eqnarray}
where
 \begin{eqnarray} 
 \nonumber
 &&
 ~~
 b_1=\frac{2b_h(r_h-N_1r_h^2+4N_1 \alpha'+2N_1^2r_h\alpha')}
 {N_1(r_h^2+4\alpha')(2N_1\alpha'-r_h)},
 \\
 && \sigma_1= \sigma_h
 \bigg(
 -r_h^2(-1+N_1r_h)(1+6N_1r_h)
 +2N_1 r_h(4+3N_1r_h(3+4N_1r_h))\alpha'
 \\
  \nonumber
 &&{~~~~~~~~} -8N_1^2(1+3N_1r_h(-3+N_1r_h))\alpha'^2
 -112N_1^4\alpha'^3
 \bigg) 
 \bigg( 6N_1^2 r_h(r_h^2+4\alpha')(-2\alpha'+(r_h-2N_1 \alpha')^2)\bigg)^{-1}~,
\end{eqnarray}
while
 \begin{eqnarray} 
N_1=\frac{r_h-\sqrt{r_h^2-8\alpha'}}{4\alpha'}.
\end{eqnarray}

\begin{figure}[ht]
\hbox to\linewidth{\hss%
	\resizebox{8cm}{6cm}{\includegraphics{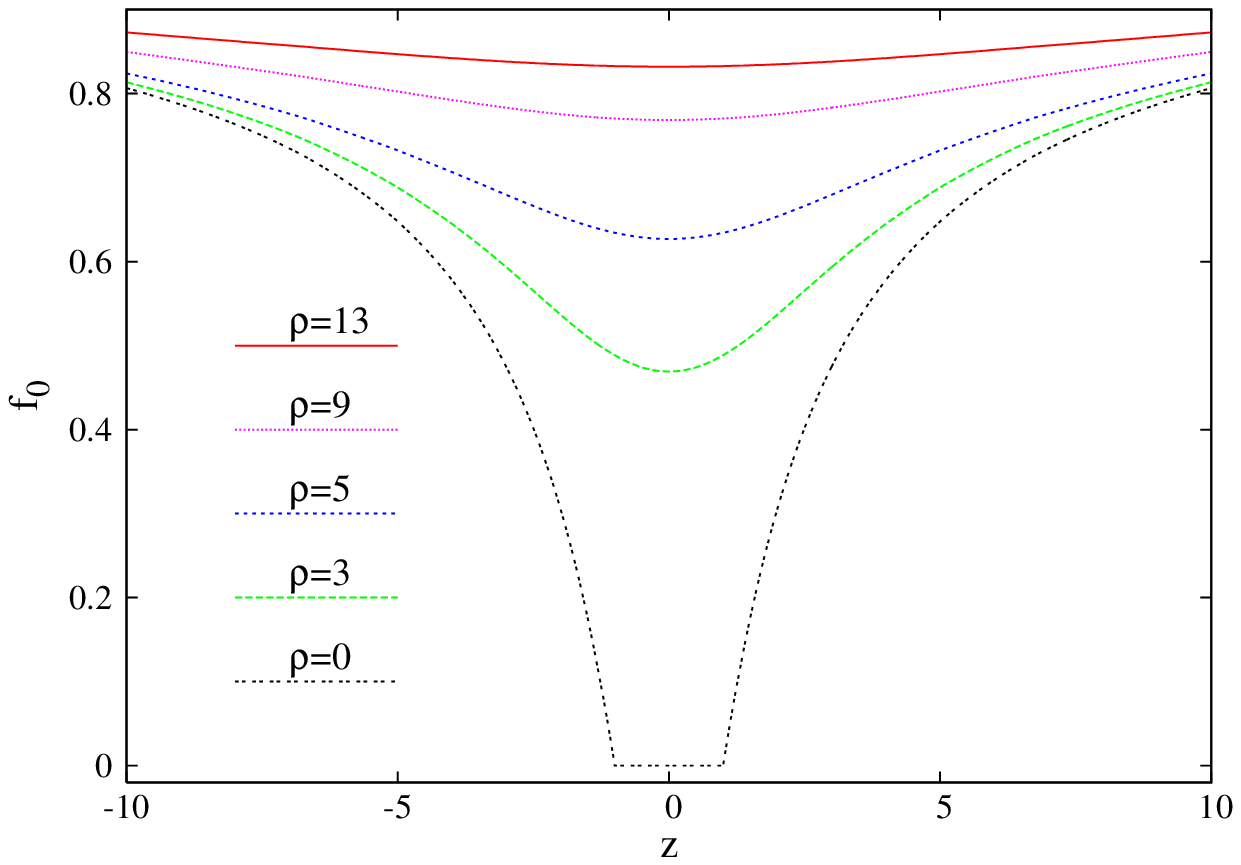}}
\hspace{5mm}%
        \resizebox{8cm}{6cm}{\includegraphics{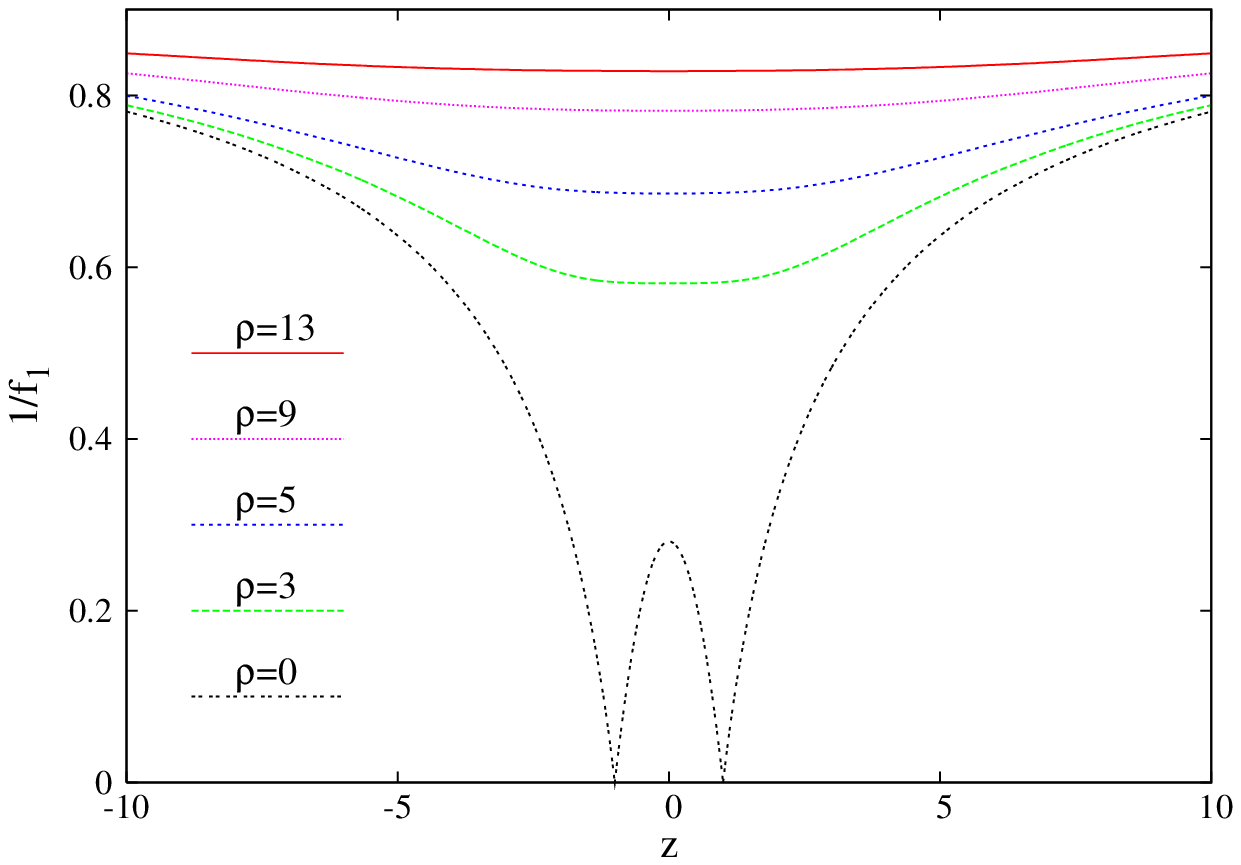}}	
\hss}
\end{figure}
\begin{figure}[ht]
\hbox to\linewidth{\hss%
	\resizebox{8cm}{6cm}{\includegraphics{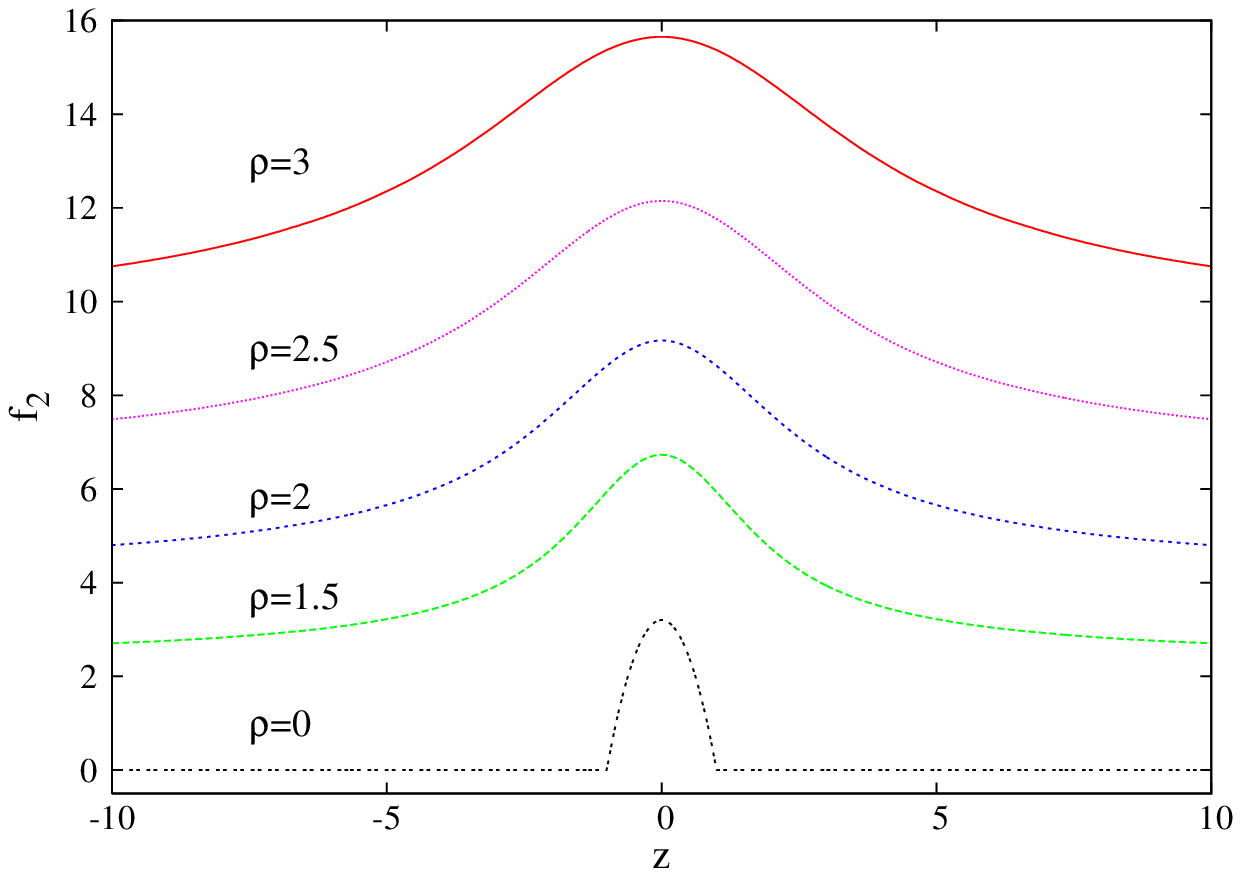}}
\hspace{5mm}%
        \resizebox{8cm}{6cm}{\includegraphics{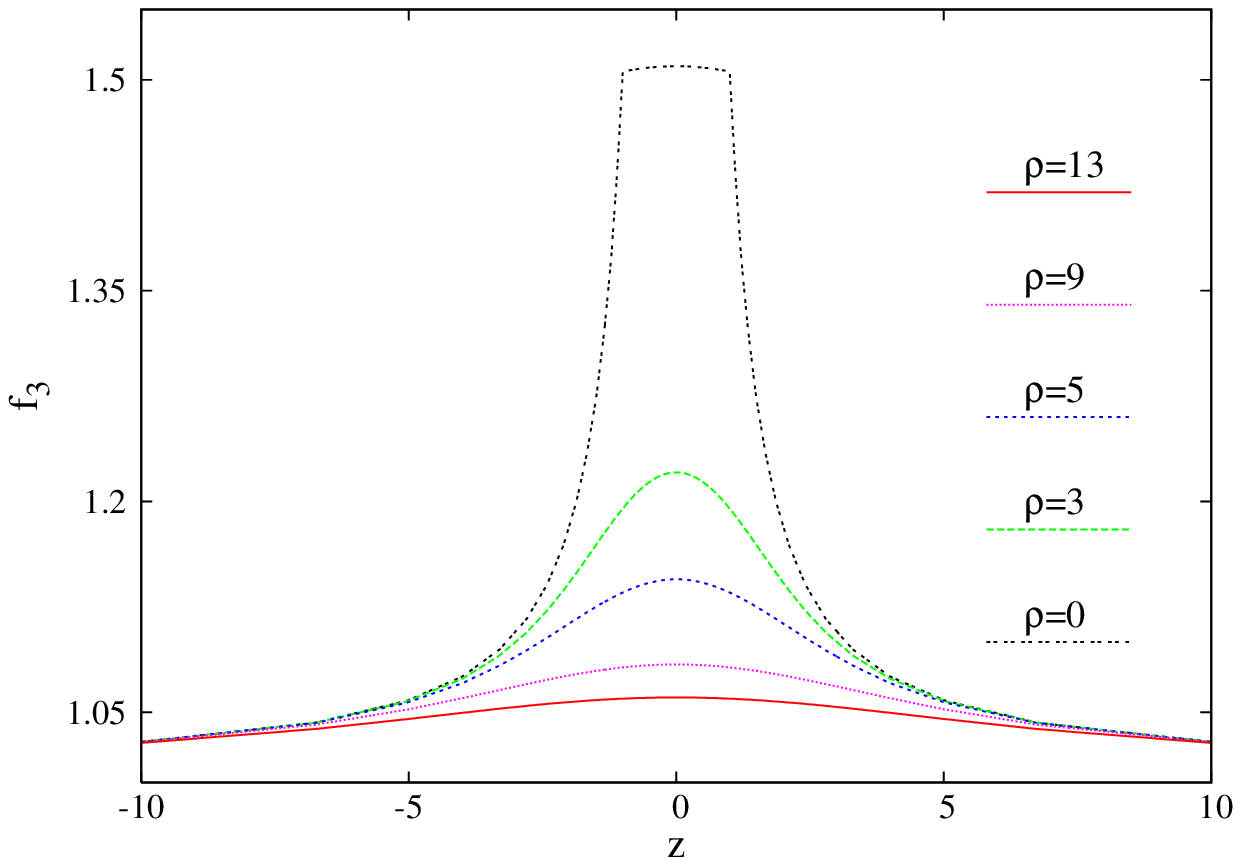}}	
\hss} 
\label{Schw2}
\end{figure}
\vspace{0.5cm}
{\small {\bf Figure 3.}
The metric functions $f_i$ 
of a typical uniform black string solution in EGB theory
are shown versus $z$ for several values of $\rho$.
The relevant parameters here are
  $a=1$, $\alpha'=0.175$.
  }
 \\
 \\
The free parameters in the near horizon expansion are $b_h>0$ and $\sigma_h>0$.

The above relations imply the existence of a minimum horizon size for
a given value of the GB coupling constant
 \begin{eqnarray}
\label{UBS-min}
r_h\geq \sqrt{~8\alpha'}.
\end{eqnarray}
Since, as discussed in Ref. \cite{Kobayashi:2004hq}, the horizon radius is 
decreasing monotonically with the mass of the solutions,
the relation (\ref{UBS-min}) shows again the existence 
of a minimal value of the mass for a given 
GB coupling constant $\alpha'$.

 At infinity, the background approached by a black string is the four-dimensional spacetime times the $\varphi-$direction,
 $ds^2= dr^2 
 +r^2(d\theta^2+\sin^2 \theta d\psi^2)- dt^2+ d\varphi^2$.
The solution  as $r\to \infty$ is written in terms of two parameters $c_t,c_z$:
  \begin{eqnarray}
\nonumber
&&N(r)=1+\frac{c_z-c_t}{r}+\frac{c_tc_z}{4r^2}+O(1/r^3),~~
\sigma(r)=1-\frac{c_z}{2r}+\frac{3c_z(c_z-c_t)}{8r^2}+O(1/r^3),
\\
\label{UBS-inf}
&&
b(r)=1+\frac{c_z}{r}+\frac{c_tc_z}{r^2}+O(1/r^3).
\end{eqnarray}

Similar to the case of Einstein gravity,
the EGB black string solutions possess two global charges
-- the mass $M$ and the tension ${\cal T}$,
associated with the Killing vectors $\partial/\partial t$ and $\partial/\partial \varphi$, respectively.
These global charges are fixed by the constants $c_t$, $c_z$ in the asymptotic expansion (\ref{UBS-inf}):  
\begin{eqnarray}
\label{2} 
M=\frac{\Delta \varphi}{4 \pi G}(2c_t-c_z),
~~{\mathcal T}=\frac{ 1}{4 \pi G}(c_t-2c_z).
\end{eqnarray}
Without entering into details, we mention that 
the  black string solutions can also be recovered within the metric ansatz
(\ref{metric}). The coordinate transformation between this ansatz and (\ref{UBS})
can be worked out in a similar way to 
(\ref{coord-transf-SGB1})-(\ref{metric-weyl}) (note however that 
$\rho=\frac{1}{2}r_0\sinh \bar G(r) \sin\theta,$ and $z=\frac{1}{2}r_0\cosh \bar G(r) \cos\theta$ in this case).
These solutions have a  semi-infinite space-like rod $[-\infty,-a]$
along the $\psi-$direction,
a finite time-like rod $[-a , a ]$ corrresponding to the event horizon and a second
semi-infinite space-like rod $[a ,\infty]$  again 
in the $\psi$-direction (thus there is no rod along the $\varphi-$direction, see Figure 1c).
This can also be seen in Figure 3, where the metric functions $f_i$ of a tyical EGB uniform black string solution 
are shown as a function of $z$ for several values of $\rho$ (one can notice the nontrivial shape of the
metric function $f_3$).

However, the EGB black string solutions  are much more difficult to study in the coordinate system (\ref{metric}),
since in this case one deals with partial differential equations.
In particular, it is much more difficult to prove analytically the existence of a maximal value of
$\alpha'$ for a given length of the finite timelike rod\footnote{However, the coordinate system (\ref{metric})
makes possible to attempt a numerical construction 
of more complicated Kaluza-Klein solutions in EGB theory, $e.g.$
multi-black strings or configurations with bubbles.}.

\section{The static EGB black rings}

The physical intuition (supported by the results in the previous Section) suggests that all known solutions in Einstein gravity
admit generalizations in EGB theory.
While it is rather inconvenient  
to use the metric ansatz (\ref{metric}) for the study of 
EGB Schwarzschild black holes and EGB black strings, 
this is not the case for more complicated solutions with a nonspherical
topology of the horizon.
In fact, in our opinion, the simplest way to construct 
EGB generalizations of such objects 
is within the metric ansatz (\ref{metric}),
by imposing the same rod structure as in the absence of  the GB term.

In this Section we present numerical evidence for the existence 
of static black rings,
as the simplest example of an asymptotically flat 
black object with a nonspherical
topology of the horizon in  EGB  theory.

\subsection{The static black rings in Einstein gravity}

The rod structure of a static black ring solution  in Einstein gravity
is exhibited in Figure 1b. It
consists of a semi-infinite space-like rod $[-\infty,z_1]$ 
in the $\psi$-direction (thus $f_2(0,z)=0$ there),
a finite time-like rod $[z_1, z_2]$ ($f_0(0,z)=0$),
a second (and finite) space-like rod $[ z_2,z_3]$ in the $\psi$-direction, 
where $f_2(0,z)=0$ again, and a semi-infinite space-like rod $[z_3,\infty]$ 
($f_3(0,z)=0$) in the $\varphi$-direction (and $z_1<z_2<z_3)$.

The  metric functions $f_i$ of the static black 
ring\footnote{Note that the function $f_1(0,z)$ behaves as $1/|z-z_i|$ as
$z\to z_i$.} are given  by \cite{Emparan:2001wk},\cite{Harmark:2004rm}
\begin{eqnarray}
\nonumber
&&f_0 =\frac{R_2+\xi_2}{R_1+\xi_1},~~
f_1 = \frac{(R_1+\xi_1+R_2-\xi_2)((1-c)R_1+(1+c)R_2+2c R_3)}{8(1+c )R_1R_2R_3},
\\
\label{BR5d}
&&f_2 =\frac{(R_2-\xi_2)(R_3+\xi_3)}{R_1-\xi_1},~~f_3 =R_3-\xi_3~,
\end{eqnarray}
  where
\begin{eqnarray}
\label{rel1}
\xi_i=z-z_i,~~
R_i=\sqrt{\rho^2+\xi_i^2}
~~~~
{\rm and~}~~~~
z_1=-a,~~z_2=a,~~z_3= b,
\end{eqnarray}
$a$ and $ b$ being two positive constants, with $c=a/b<1$.
{Roughly speaking, $a$ fixes the size of the horizon, while $b$ provides
a measure of the radius of the ring's $S^1$.}

Since the orbits of $\psi$ shrink to zero at $-a$ and $a$ while 
those of $\varphi$ do not vanish anywhere there, the topology of the horizon 
is $S^2\times S^1$, see Figure 1b (although the $S^2$ is distorted away from perfect sphericity).

The mass, event horizon area  and Hawking temperature of this solution are:
\begin{eqnarray}
M^{(E)}=\frac{3aV_3}{4\pi G},~~
A_H^{(E)}=8a^2V_3\sqrt{\frac{2}{a+b}},~~
T^{(E)}_H=\frac{1}{4 \pi a}\sqrt{\frac{a+b}{2}}.
\end{eqnarray}
Although the static black ring solution is asymptotically 
flat\footnote{The Einstein gravity static black ring solution in
 \cite{Emparan:2001wk} admits an alternative interpretation as 
              a ring sitting on the rim of a membrane that extends to 
	      infinity. (This is found by requiring 
              that the periodicity of $\psi$ is $2\pi$ on the 
	      finite $\psi$-rod.) However, the asymptotic metric is a deficit 
	      membrane in this case.}, 
it contains a conical singularity
for the  finite $\psi$-rod, since $\delta$ as defined by (\ref{delta}) is nonzero:
\begin{eqnarray}
\delta=2\pi \left(1-\sqrt{\frac{b+a}{b-a}}\right).
\end{eqnarray}
One can easily see that this is a negative quantity, $\delta<0$,
which implies the existence of a two-dimensional disk-like deficit membrane (with negative deficit)
that prevent the configuration from collapsing.

Another quantity of interest is the   area of the spacetime spanned by the conical singularity  and  the proper length
 of the finite $\psi$-rod, which are computed according to (\ref{area}), (\ref{L}):
 \begin{eqnarray}
\label{rel2}
Area= \beta \frac{2\pi (b-a)^{3/2}}{\sqrt{a+b}} ,~~L=\sqrt{2}\sqrt{b-a}~E(n),
\end{eqnarray}
where
$n=(b-a)/(b+a)$, $E(n)$ being the complete elliptic integral of the second kind.

In terms of the dimensionless parameter $a/b$,
one may think of a static black ring as interpolating
between two limits.
As $a/b \to 1$, the finite $\psi-$rod vanishes
and the Schwarzschild  metric is approached,
with  $\delta\to -\infty$.
As $a/b\to 0$, the second $\psi-$rod extends to infinity and the 
solution becomes, after a suitable rescaling\footnote{For the line element used in this work, this rescaling is
$r\to \sqrt{2b}\bar r$,
$z\to \sqrt{2b}\bar z$,
$\varphi\to \bar \varphi/\sqrt{2b}$ together with $a\to \sqrt{2b} \bar a$.}, 
a black string, $i.e.$ the 
four dimensional Schwarzschild black hole uplifted to five dimensions.  
 
\subsection{The static black rings in EGB theory}
\subsubsection{The ansatz and the physically relevant quantities}

The EGB generalizations of the Emparan-Reall black rings are found by solving 
the EGB equations  for the metric ansatz (\ref{metric}). 

The boundary conditions satisfied by the EGB black ring metric functions 
are similar to those in Einstein gravity.
At $\rho=0$, the function $f_0$ vanishes for $-a\leq  z\leq a$ ($i.e.$ on the horizon), 
$f_2$ is zero for $-\infty\leq z\leq -a$ and  $ a\leq  z\leq b$, 
while $f_3$ vanishes for $z\geq b$.
As a result, along the  horizon the orbits of  $\psi$ shrink to zero at $z = -a$ and $z = a$, while the
orbits of $\varphi$ do not shrink to zero anywhere. Thus the topology of the horizon is $S^2 \times S^1$.
From (\ref{nrod1}), (\ref{nrod2}), one can see that,
for a given rod with one of the functions vanishing, $f_a=0$, the other $f_i$ satisfy Neumann-type boundary conditions,
$\partial_\rho f_i|_{\rho=0}=0$ (with $i\neq a$).
At infinity, we require that the functions $f_i$
approach the Minkowski form (\ref{Mink}). 
 
In practice, we 
have found it convenient to take
\begin{eqnarray}
\label{ans1}
f_i=f_i^{0}F_i ,
\end{eqnarray}
where $f_i^{0}$ are background functions, given by the metric functions of the Einstein gravity
black ring solution (\ref{BR5d}). 
The advantage of this approach is that the $f_i$ will automatically satisfy the
desired rod structure.
Moreover, this choice `absorbes' the divergencies of the functions $f_2$ and $f_3$ as $r\to \infty$
coming from the imposed asymptotic behaviour.  

The equations satisfied by the $F_i$ can easily be derived from the general set of EGB equations\footnote{Note, however,
that the field equations become much more complicated in terms of 
the $F_i$, with the number of terms increasing drastically.}.
As for the boundary conditions, the relations (\ref{nrod1}), (\ref{nrod2})
together with the expressions (\ref{BR5d}) of the background functions $f_i^{0}$
imply
\begin{eqnarray}
\label{bc-psi}
\nonumber
 \partial_\rho F_i|_{\rho=0}=0,~~~~{\rm for~~}-\infty<z<\infty,
 \end{eqnarray}
 and $F_i=1$ as $\rho\to \infty$ or $z\to \pm \infty$.

The constraint equation $E_\rho^z=0$ results in
 $F_2/F_1=const.$ on the $\psi$-rods.
Now, to be consistent with the assumption of asymptotic flatness, one finds $const.=1$
for $-\infty<z\leq -a$. The value of this ratio for the second rod with $a<z\leq b$
is obtained only as a result of the numerical solution.
A similar reasoning implies $F_3/F_1=1$ on the $\varphi$-rod ($b\leq z\leq \infty$).

All of the physically relevant quantities except for the mass 
are encoded in the values of
the  functions $F_i, f_i^0$ at $\rho=0$.
The
event horizon area of static black rings in EGB theory is given by
\begin{eqnarray}
A_H=\Delta \psi 2\pi \int_{-a}^a dz \sqrt{f_1 f_2f_3 }=
\Delta \psi   4 \pi
a \sqrt{\frac{2}{a+b}}  \int_{-a}^a dz~\sqrt{(b-z)F_1 F_2F_3 },
\end{eqnarray}
where  $\Delta \psi$ is the periodicity of the 
angular coordinate $\psi$ on the horizon.

The Hawking temperature can be computed from the surface gravity or by requiring regularity 
on the Euclidean section 
\begin{eqnarray}
T_H=\frac{1}{4 \pi a}\sqrt{\frac{a+b}{2}}\sqrt{\frac{F_0}{F_1}}~,
\end{eqnarray}
where the constraint equation $E_\rho^z$ guarantees that the ratio $F_0/F_1$ is constant on the event horizon.

At infinity, the five dimensional Minkowski background is approached, with $\Delta \psi=2 \pi$ there.
From (\ref{gtt}), the mass $M$ of the solutions can be read from the subleading term in the asymptotic
 expression for $f_0=f_0^{0}F_0$. 
To obtain a Smarr-like relation 
(which is useful in numerics, see Appendix B.2), 
we consider the EGB equation
$E_t^t=G_t^t+ \alpha'H_t^t = 0$ 
in the form
$R_t^t= 1/2 R - \alpha'H_t^t $ and integrate 
over a spacelike hypersurface.
The volume 
integral on the left hand side of the equation
reduces to surface integrals at the horizon and at infinity,
which can be evaluated 
in terms of the mass, area and Hawking
temperature,
\begin{equation}
\int{R_t^t\sqrt{-g} }d^4x = 
\frac{1}{2}\left(2\pi T_{\rm H} A_{\rm H}-\frac{16 \pi }{3}GM \right) \ . 
\end{equation}
Substituting this expression on the left hand side and solving for the mass
yields
\begin{equation}
16 \pi GM =  
6 \pi T_{\rm H} A_{\rm H} + I_{\alpha'} =
6 \pi T_{\rm H} A_{\rm H}  
-3 \int{\left\{ R - 2\alpha'H_t^t\right\}}\sqrt{-g} d^4x .
\label{smarr}
\end{equation}
In the limit $\alpha' \rightarrow 0$ the integral
$I_{\alpha'}$ vanishes and the relation reduces
to the usual Smarr relation. 

To further characterize the properties of the horizon, we introduce the
minimal and maximal $S^1$ horizon radii, $R_{\rm min}$ and $R_{\rm max}$,
defined via
\begin{eqnarray}
R_{\rm min} &=& \frac{1}{2\pi} \int_0^{2\pi} 
\left. \sqrt{g_{\vphi\vphi}}\right|_{\rho=0,z = a} 
d\vphi 
= \sqrt{2(b-a)F_3(0,a)},
\\
R_{\rm max} &=& \frac{1}{2\pi} \int_0^{2\pi} 
\left. \sqrt{g_{\vphi\vphi}}\right|_{\rho=0,z = -a} 
d\vphi 
= \sqrt{2(b+a)F_3(0,-a)}.
\label{rminmax}
\end{eqnarray}

Turning now to the finite $\psi-$rod, we note that
all solutions we have found have $F_2/F_1\neq 1-2a/(b+a)$
here.
{Thus, from (\ref{delta}),
the coordinate $\psi$ possesses a conical excess 
for $a\leq z \leq b$, 
which is}\footnote{The coordinate $\psi$ can of course be rescaled
such that its periodicity is $2\pi$ on the finite $\psi-$rod.
Then the interpretation of the solutions is somehow different, since they
would describe a static ring sitting on the rim 
of a deficit membrane that extends to infinity (in which case $\delta>0$). 
Since for this choice the spacetime is not asymptotically flat, 
we have prefered to consider the case of a conical singularity 
localized in the bulk.}
\begin{eqnarray}
\delta=2\pi \left(1-\sqrt{\frac{b+a}{b-a}}\sqrt{\frac{F_2}{F_1}} \right).
\end{eqnarray}
The proper length of the finite $\psi$-rod is given by
\begin{eqnarray}
L=\int_a^b dz \sqrt{f_1(0,z)}=\sqrt{\frac{b-a}{2(b+a)}}\int_a^b dz \sqrt{\frac{a+z}{(z-a)(b-z)}} \sqrt{F_1(0,z)}.
\end{eqnarray}
Of interest is also the expression for the space-time area spanned by the conical singularity,
\begin{eqnarray}
Area=2\pi \beta  \int_a^b dz \sqrt{f_1(0,z)f_3(0,z)f_0(0,z)}
=2\pi \beta \sqrt{\frac{b-a}{b+a}} \int_a^b dz \sqrt{F_1(0,z)F_3(0,z)F_0(0,z)}~.~~~{~~~~}
\end{eqnarray}
The GB correction $Area_1/\beta$ to the parameter $\cal A$ 
which enters the thermodynamics  can also be computed
from (\ref{comp2}), (\ref{BR5d}).
is $A=Area/\beta.$ 
 
\subsubsection{The numerical results}
 In the absence of analytical methods to construct EGB  black rings,
 a numerical approach of this problem seems to be a reasonable task.
We have solved the resulting set of four coupled nonlinear
elliptic partial differential equations numerically,
subject to the above boundary conditions.
Details on the numerical methods used and on
a new coordinate system
better suited for the numerical study of these solutions
are presented in Appendix B.

The problem has two dimensionless parameters, which we have chosen to be
$b/a$ and $\alpha'/a$. 
We note that solutions are equivalent under scaling
$$(a,b,\alpha') \to (\lambda^2 a,\lambda^2 b, \lambda^2 \alpha') . $$
Mass, area, temperature etc. scale according to their dimensions
$$(M, A_{\rm H}, T_{\rm H}, \dots) \to
(\lambda^2 M, \lambda^3 A_{\rm H}, \lambda^{-1} T_{\rm H}, \dots)$$
With $\lambda^2= 1/a$ we obtain the dimensionless quantities
$$ \hat{M} = M/a \ ,  \hat{A}_{\rm H} = A_{\rm H}/a^{3/2} \ ,
  \hat{T}_{\rm H} =a^{1/2} T_{\rm H}\ , \dots $$

Starting from black rings of Einstein gravity,
we have generated branches of EGB black rings
by increasing the GB coupling constant $\alpha$ from zero,
while keeping the parameters $a$ and $b$ fixed.
Typical profiles of the solutions are presented  
 in Figures 4 and 5.
We note that the functions $F_i$ are smooth outside of the $z-$axis,
showing no sign of a singular behaviour.

 \newpage
\setlength{\unitlength}{1cm}
\begin{picture}(15,21)
\put(-1,0){\epsfig{file=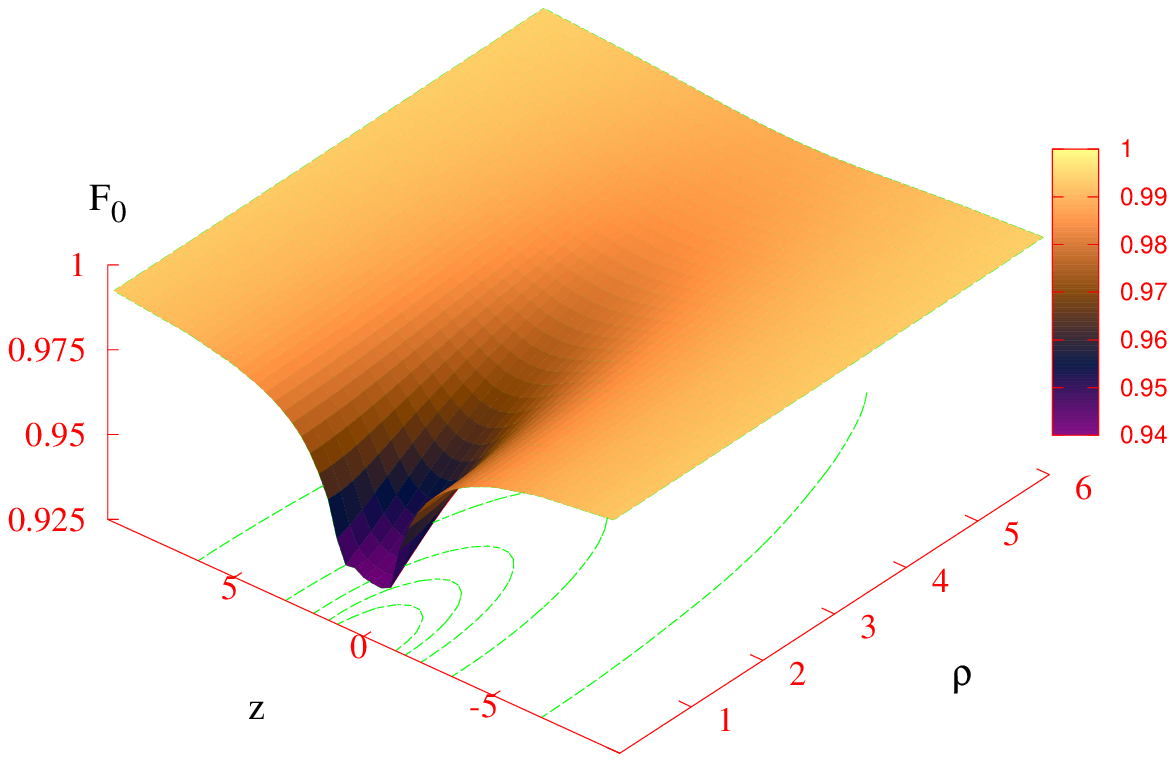,width=7.5cm}}
\put(7,0){\epsfig{file=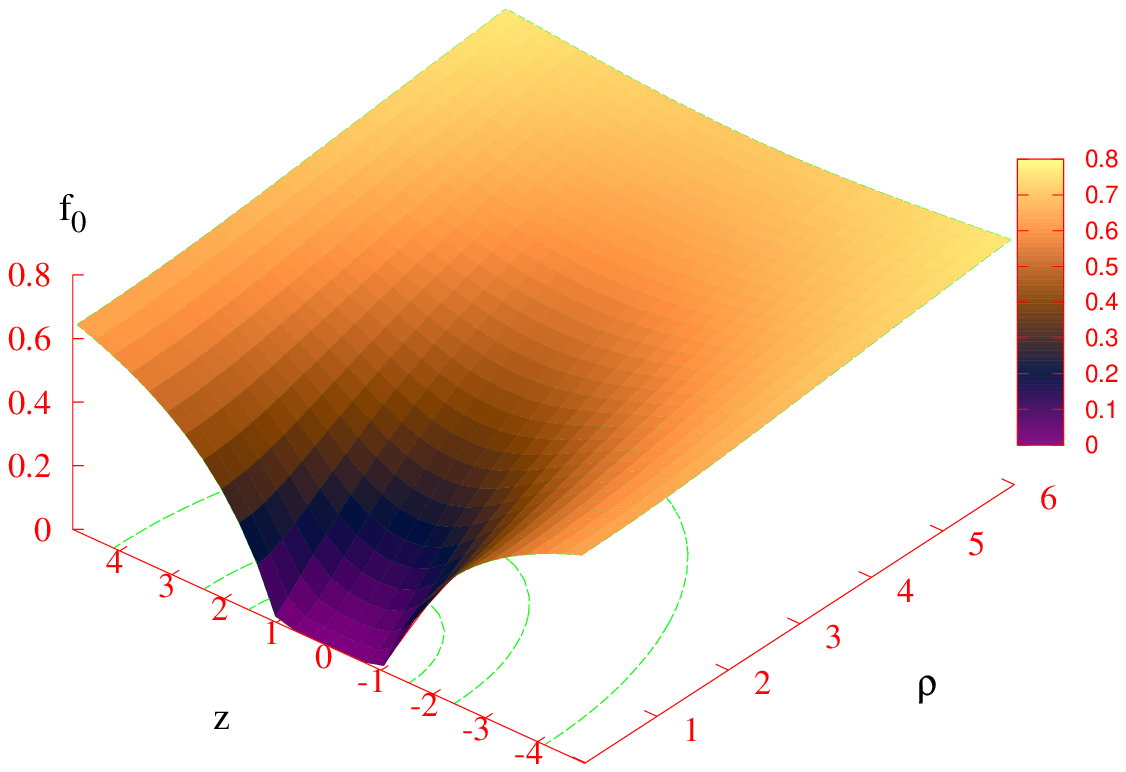,width=7.5cm}}
\put(-1,6){\epsfig{file=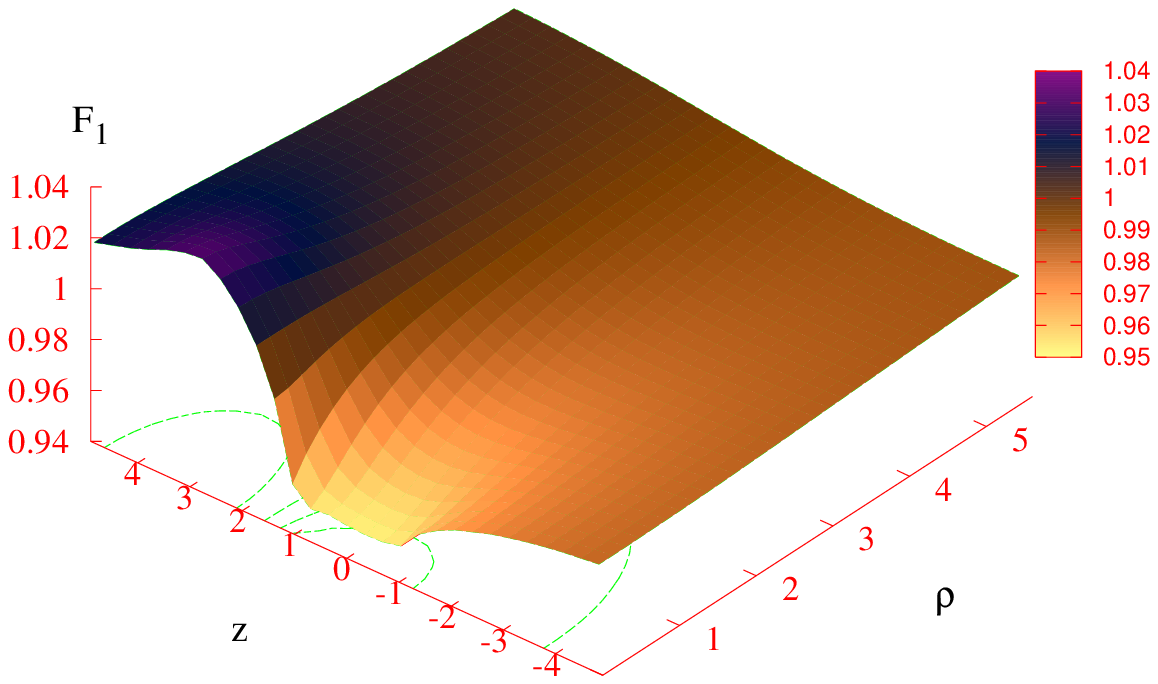,width=7.5cm}}
\put(7,6){\epsfig{file=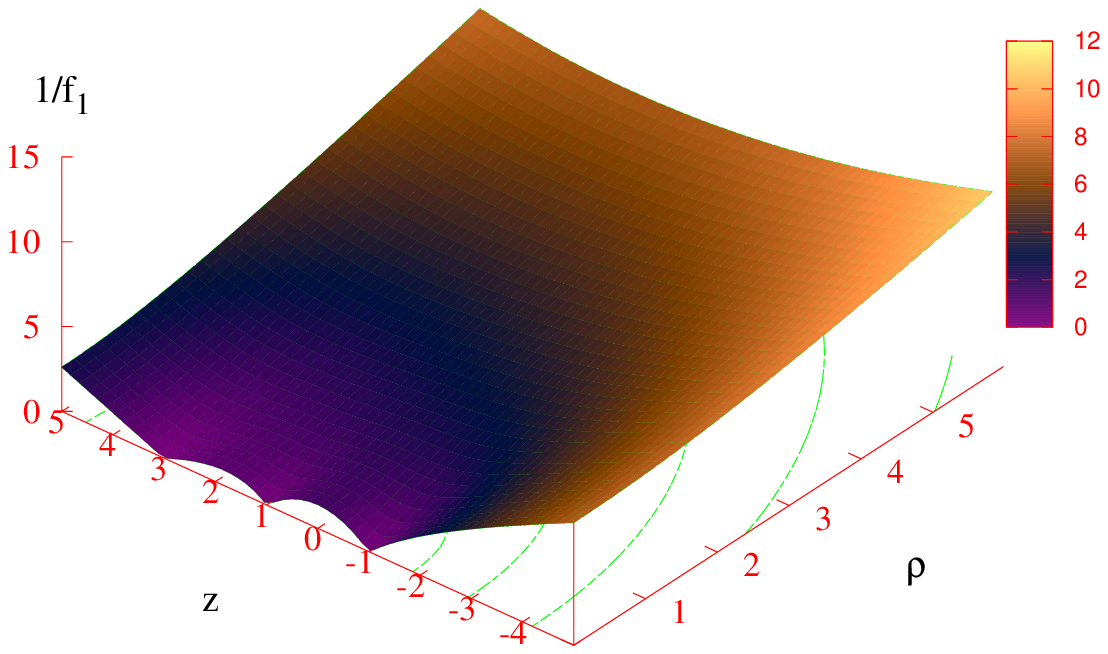,width=7.5cm}}
\put(-1,12){\epsfig{file=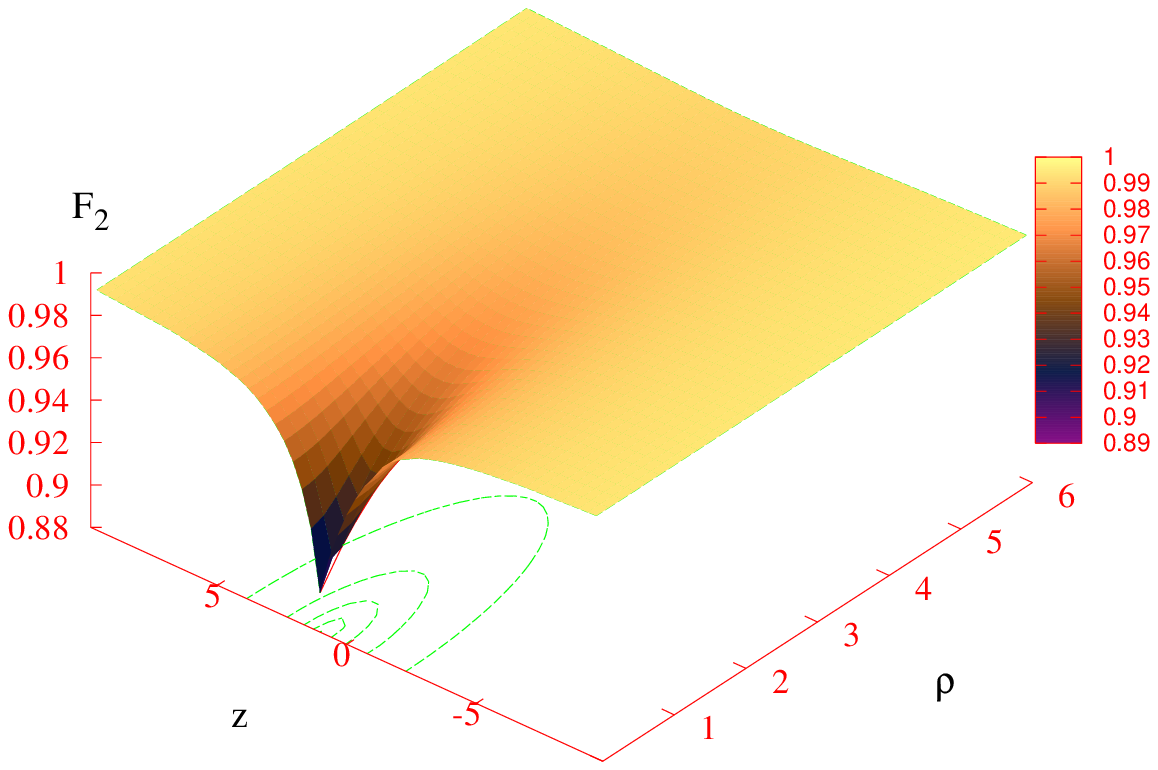,width=7.5cm}}
\put(7,12){\epsfig{file=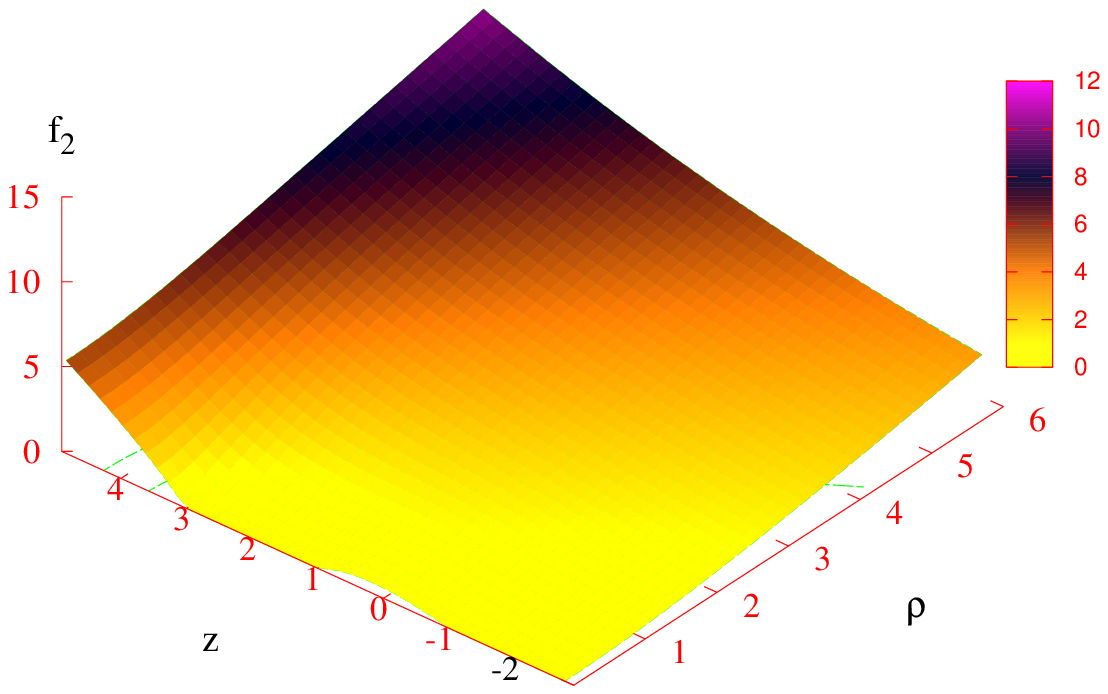,width=7.5cm}}
\put(-1,18){\epsfig{file=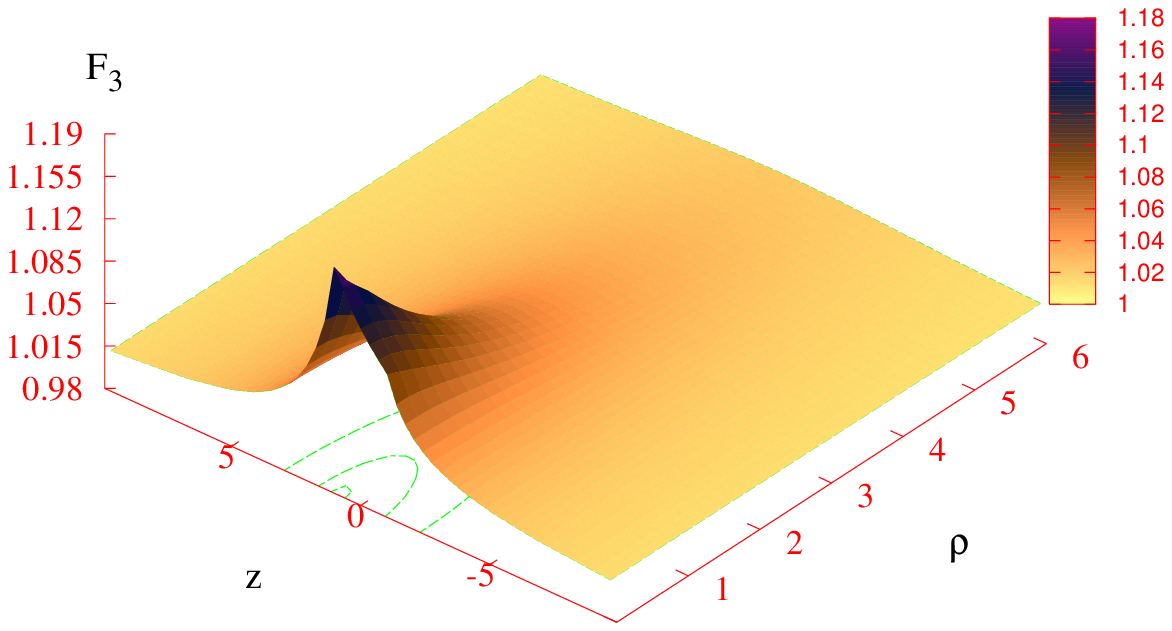,width=7.5cm}}
\put(7,18){\epsfig{file=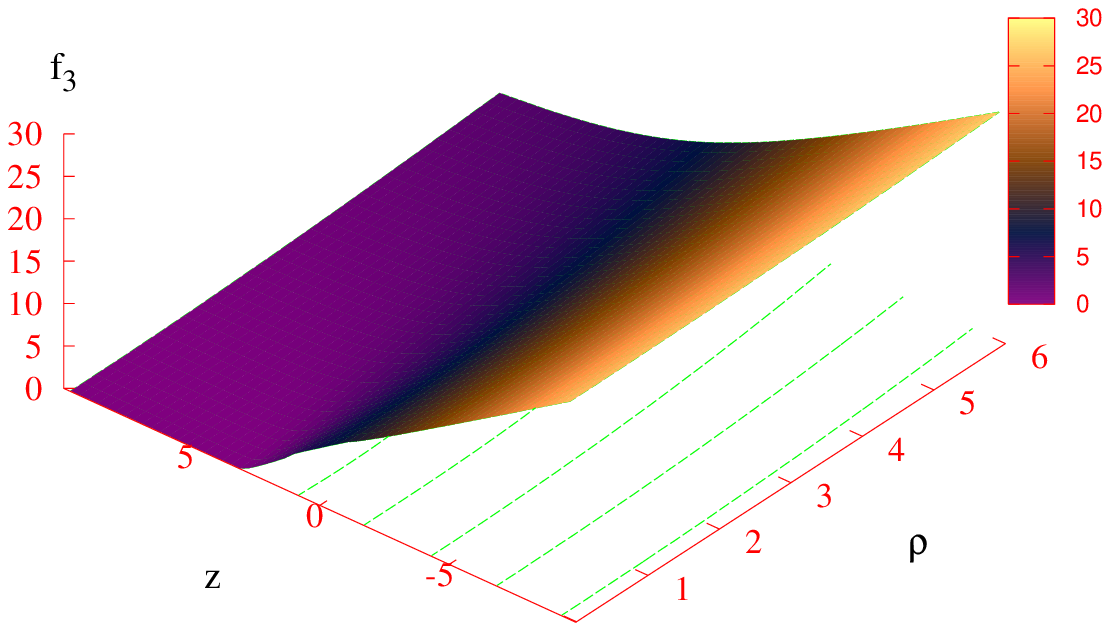,width=7.5cm}} 
\end{picture}
\\
{\small {\bf Figure 4.}
 The profiles of the functions $F_i$ employed in the numerical calculations and of the metric functions $f_i$ 
are shown 
for a typical EGB black ring solution with $a=1$, $b=3$, $\alpha'=0.0125$.
} 
 
 \newpage
\setlength{\unitlength}{1cm}
\begin{picture}(15,21)
\put(-1,0){\epsfig{file=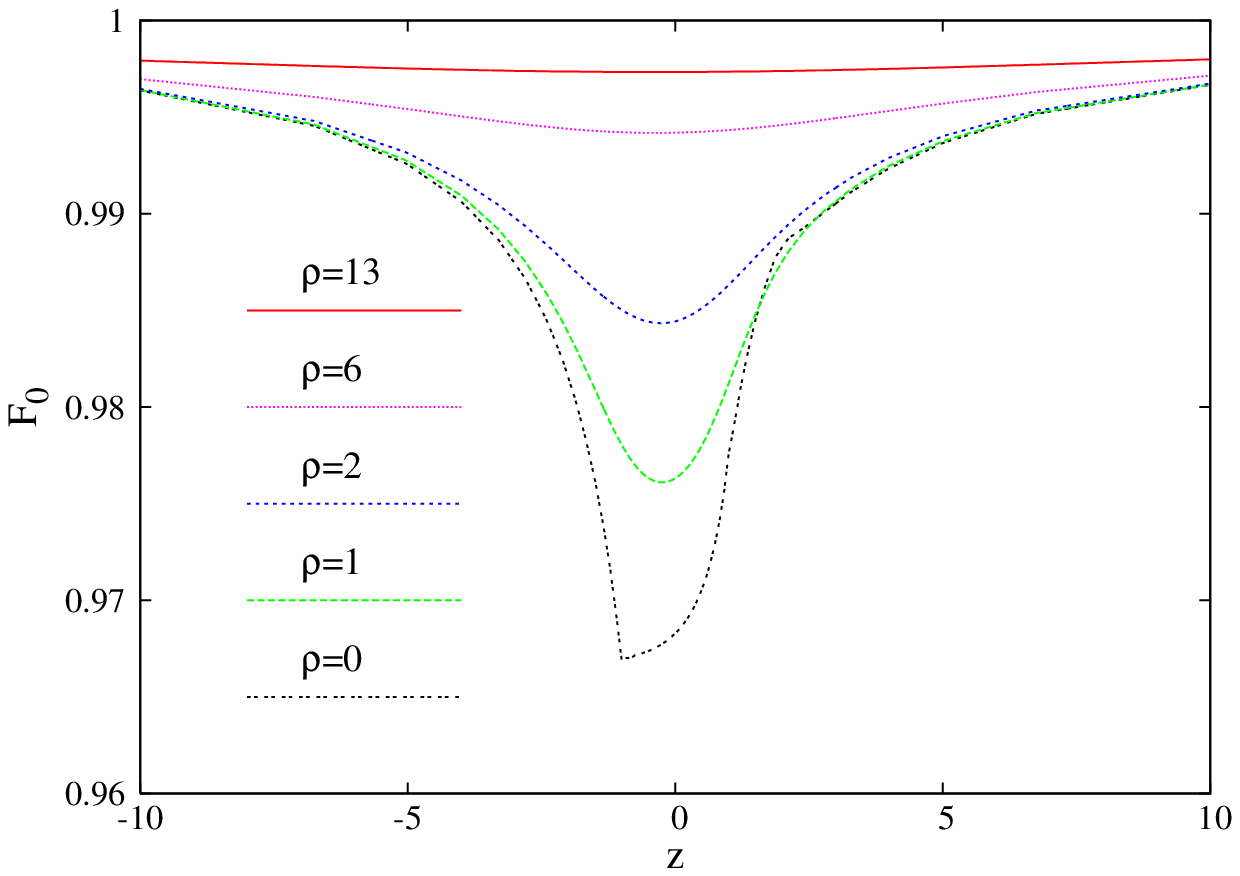,width=7.5cm}}
\put(7,0){\epsfig{file=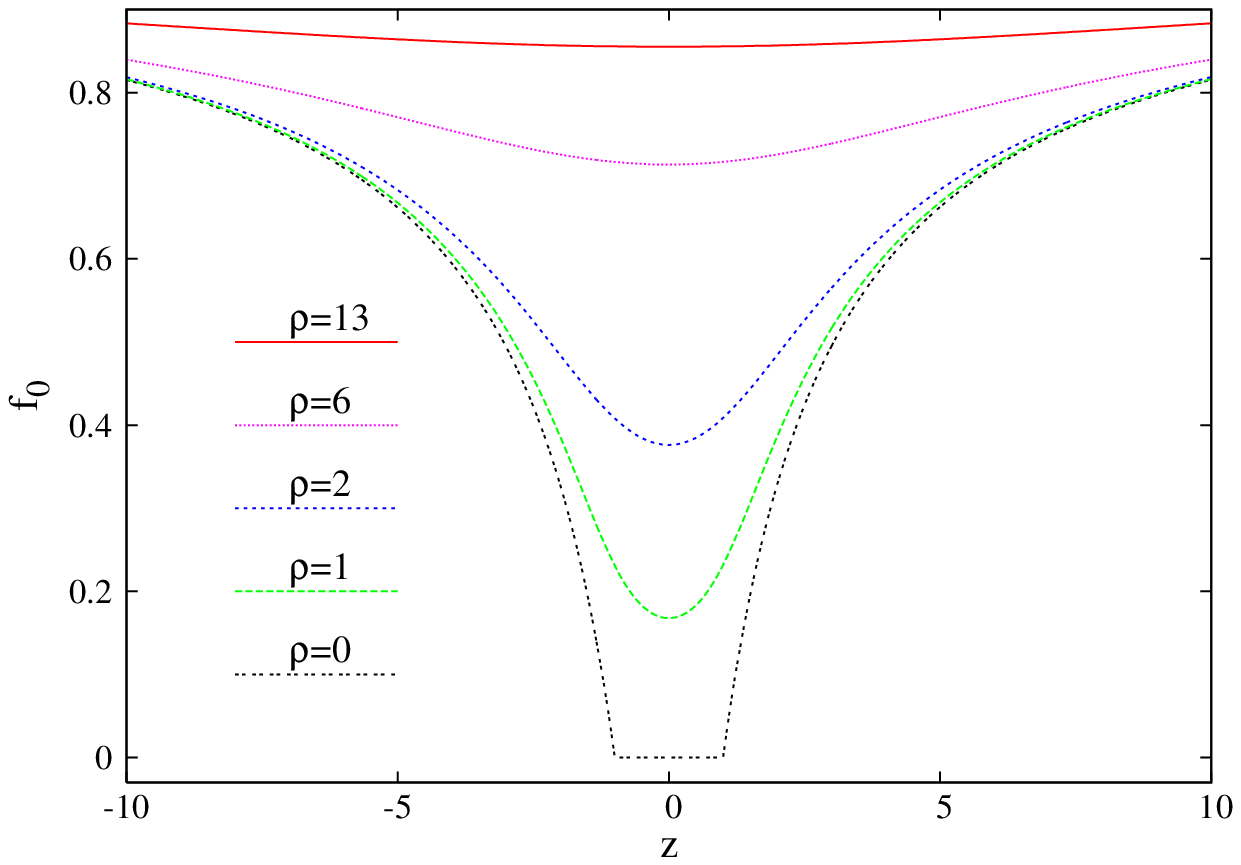,width=7.5cm}}
\put(-1,6){\epsfig{file=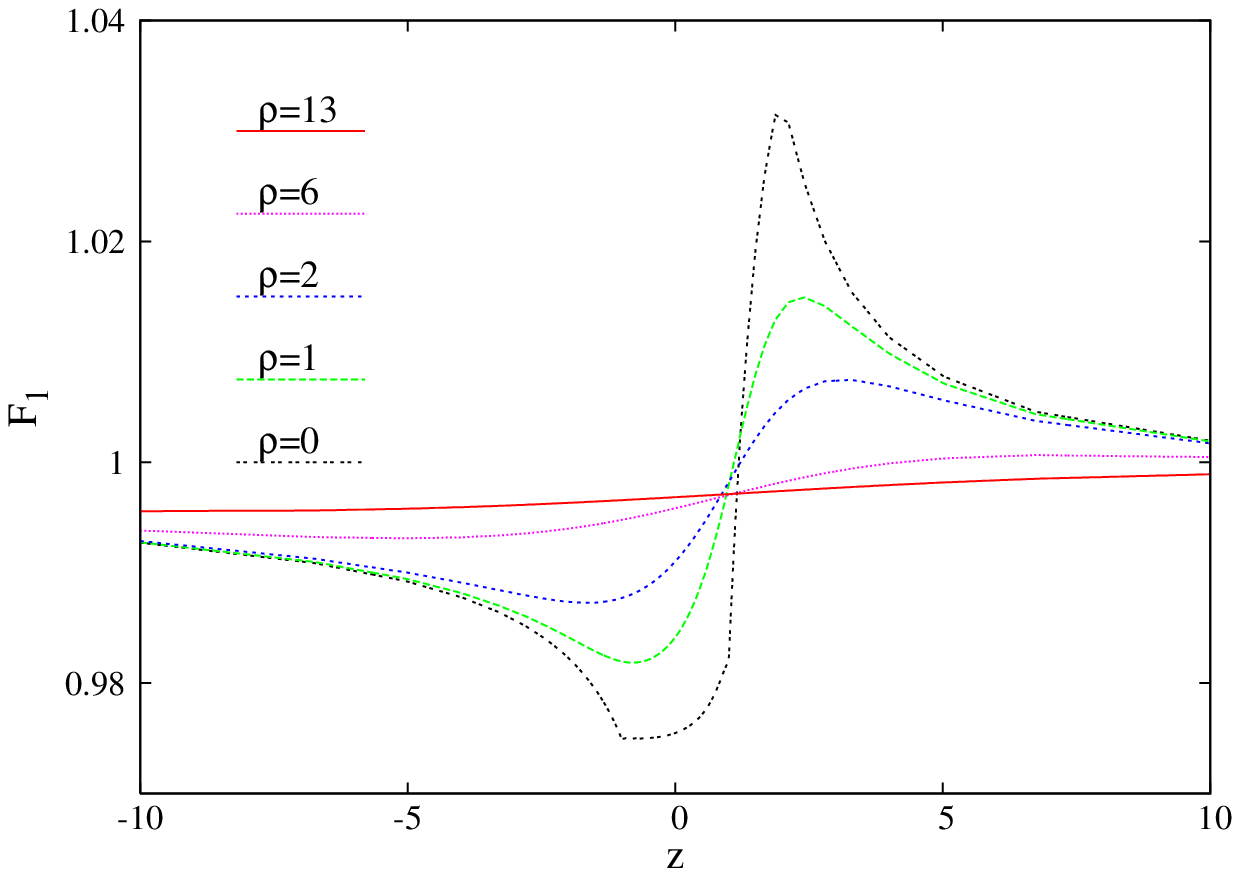,width=7.5cm}}
\put(7,6){\epsfig{file=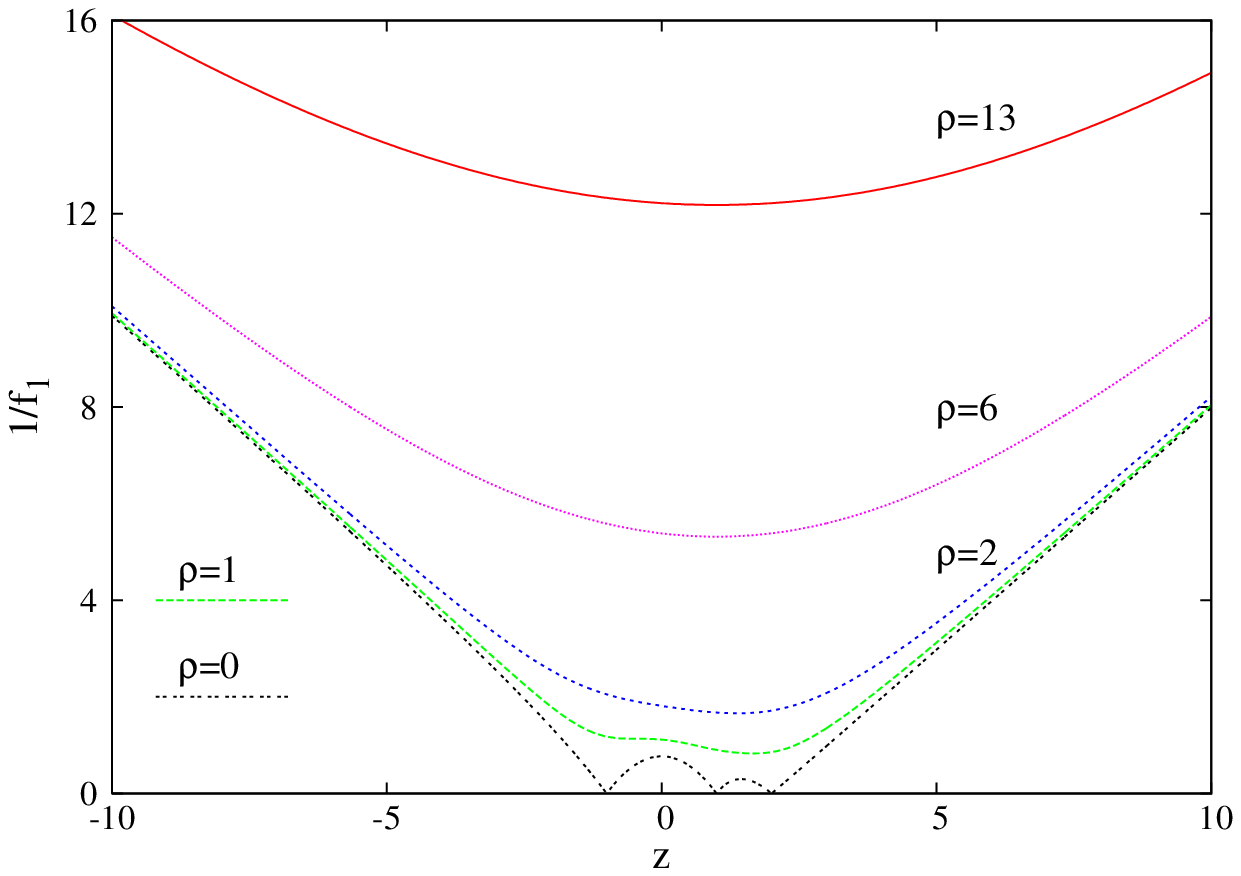,width=7.5cm}}
\put(-1,12){\epsfig{file=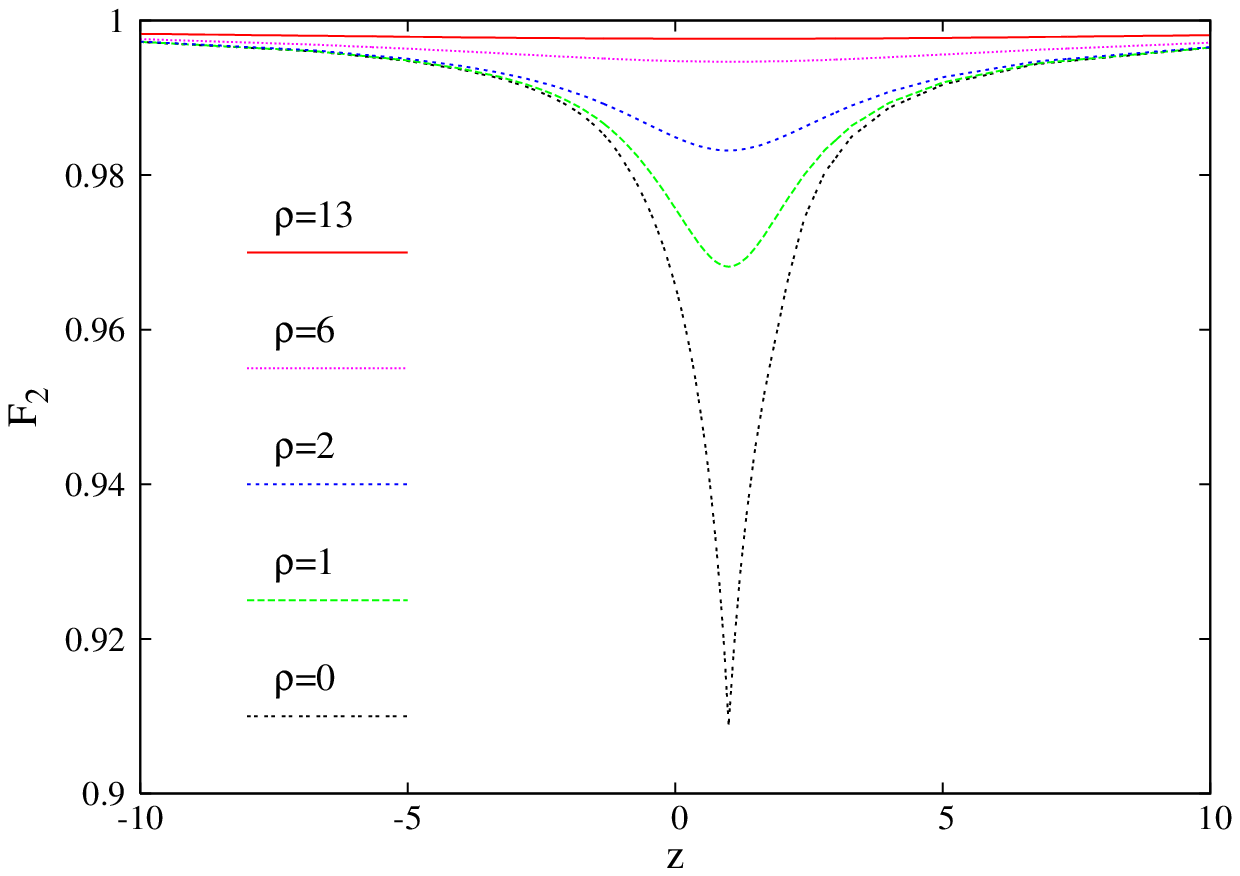,width=7.5cm}}
\put(7,12){\epsfig{file=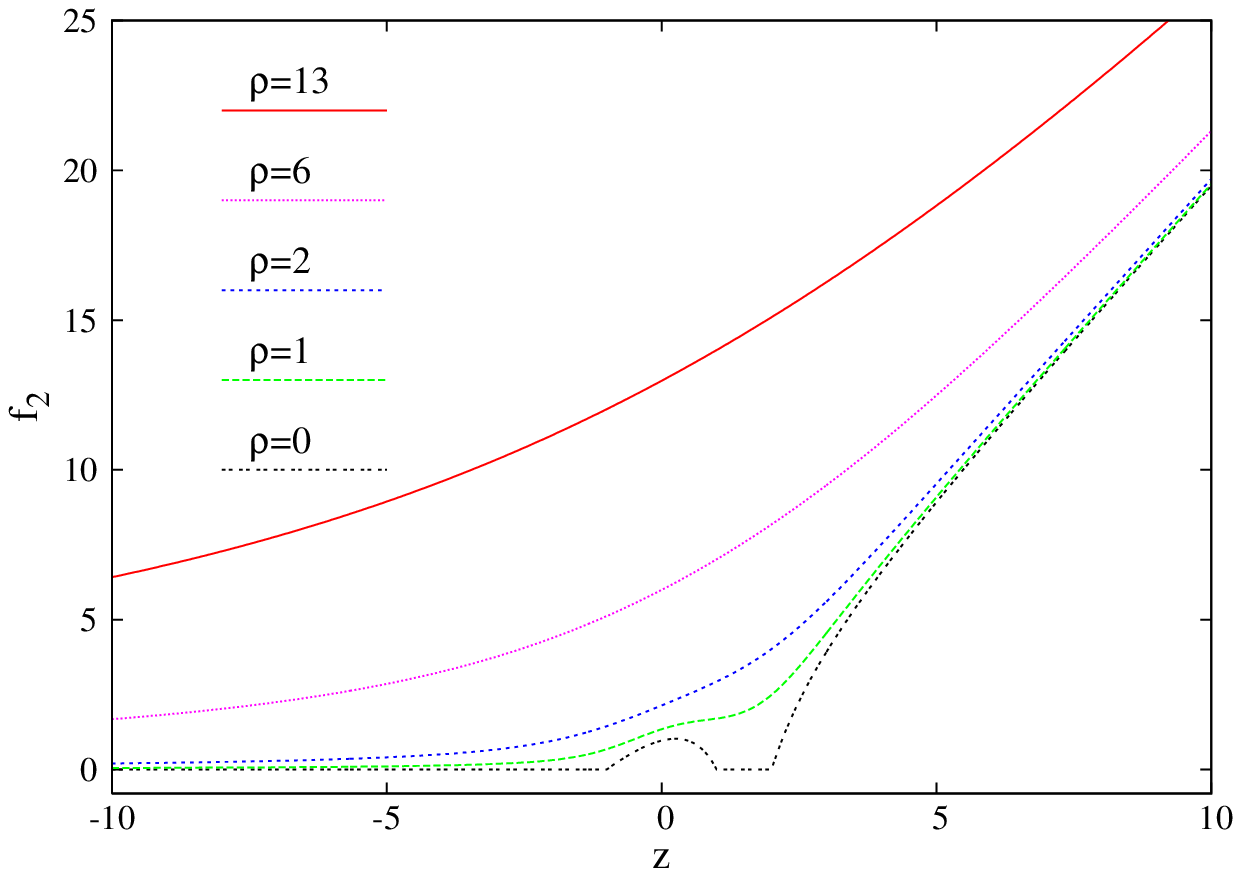,width=7.5cm}}
\put(-1,18){\epsfig{file=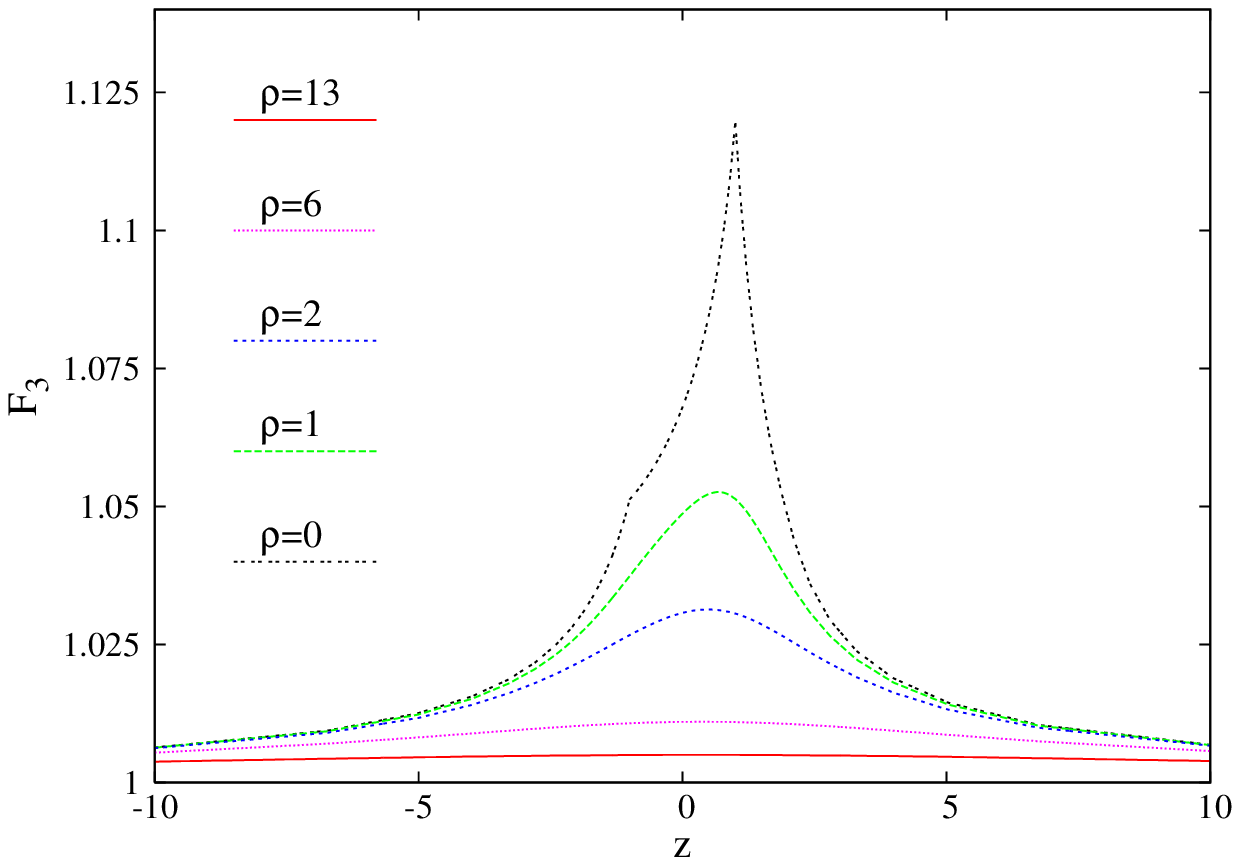,width=7.5cm}}
\put(7,18){\epsfig{file=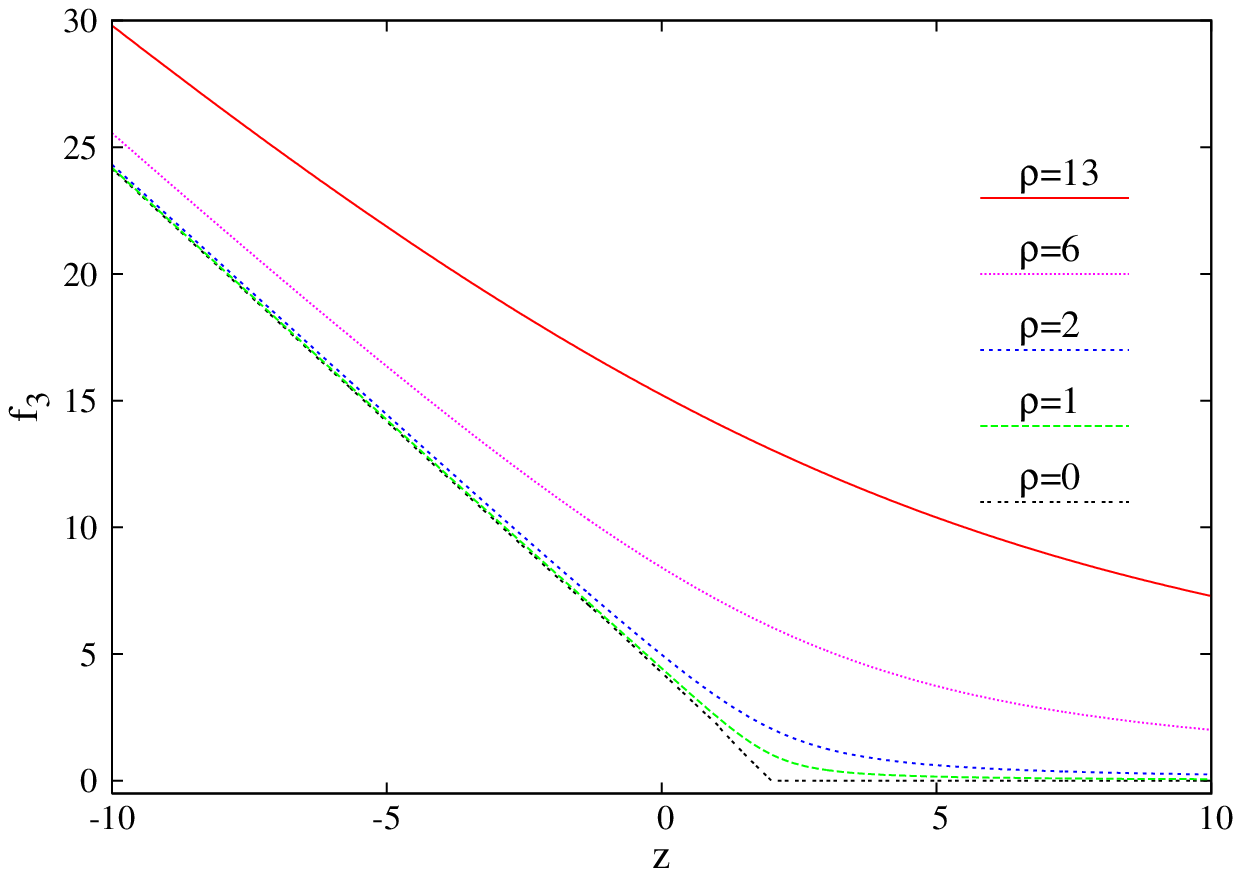,width=7.5cm}}
\label{Fi-2d}
\end{picture} 
\\
{\small {\bf Figure 5.}
The profiles of the functions $F_i$ used in numerics and of the metric functions $f_i$ 
are shown as functions of $z$ for several values of $\rho$
for a black ring solution with 
 $a=1$, $b=2$, $\alpha'=0.01$.
} 


The crucial point here is that the divergent behaviour of 
the functions $f_i$ has already been subtracted by the background
functions $f_i^{0}$.
We have verified that the Kretschmann scalar stays finite everywhere, in particular at $\rho=0$. 

A number of basic features of these EGB black ring solutions are analogous
to those of the static black rings of Einstein gravity.
In particular,
all solutions have a conical excess $\delta$ on the finite $\psi-$rod
(for the choice of $\Delta \psi=2\pi$ at infinity).
Moreover, on the horizon 
the circumference of the $S^1$  
is maximum for $z=-a$ and minimum for $z=a$. 

The isometric embedding of the horizon
is shown in Figure 6 
for a family of EGB black ring solutions
with fixed parameters $a=1$ and $\alpha'=0.015$ and several values of $b$,
chosen in the interval $1.1 \le b \le 8$.

\begin{figure}[ht]
\hbox to\linewidth{\hss%
        \resizebox{9cm}{7cm}{\epsfysize=8cm\epsffile{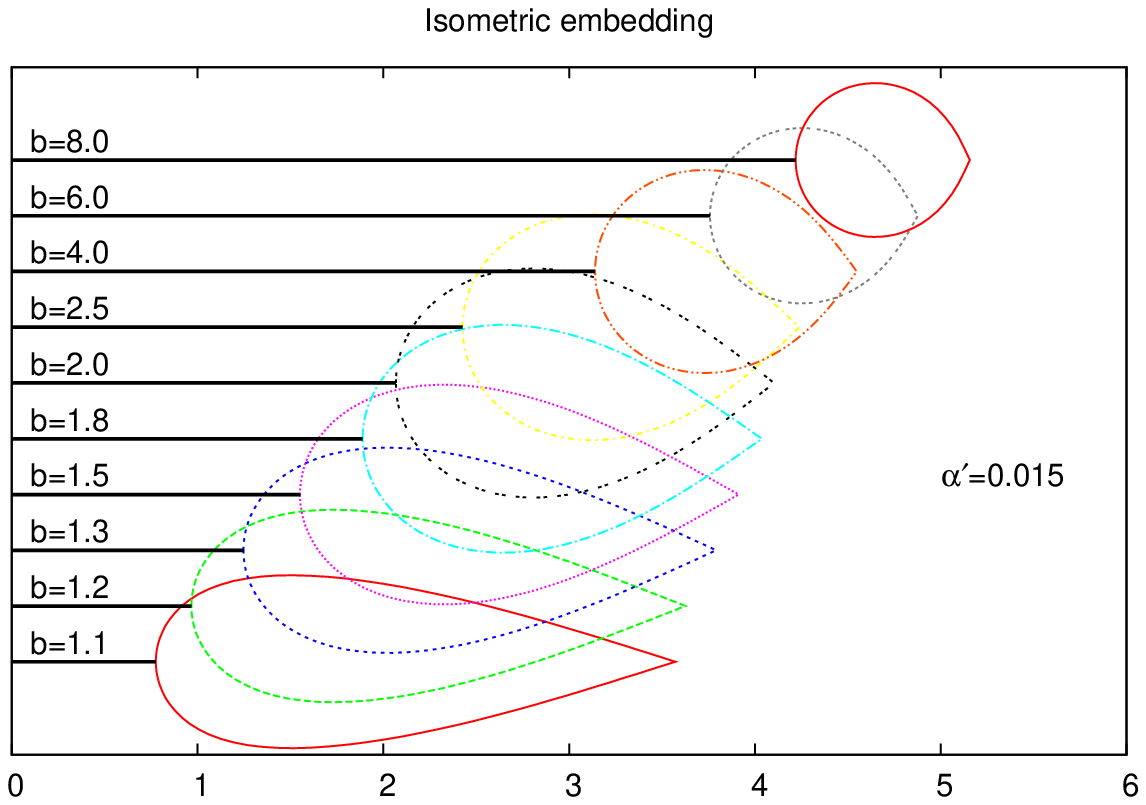}}
 \hss}
\label{fig11}
 \vspace{0.3cm}
{\small {\bf Figure 6.}
The embedding of the EGB black ring horizon
for a family of solutions
with $a=1$, $\alpha'=0.015$ and several values of the parameter $b$.
Note, that the conical singularity has been moved
to the outside of the ring, stretching from its surface to infinity.}
\end{figure}

We note, that for this embedding the conical singularity
was moved from the finite $\psi$-rod to the exterior of the ring,
extending from the outer ring circumference (at the
maximal radius $R_{\rm max}$) to infinity\footnote{Note that for the choice $\Delta\phi = 2\pi$ at infinity the
embedding would be pseudo-Euclidean.}.
Therefore the horizon is smooth and round
at the inner circumference of the ring, 
while it exhibits a cone-like edge at 
the outer circumference at
the maximal radius $R_{\rm max}$.
As the parameter $b$ is increased,
the proper distance from the center
of the ring to the horizon increases.
At the same time the conical excess decreases. 
Consequently, 
the shape of the horizon becomes more and more spherical
with increasing $b$.

Thus, for a given $\alpha'$, one might hope  that the conical excess could be completely
removed, by going to large enough values of $b$.
This hope, however, is dashed, when the parameter space is
fully explored.

Whereas for Einstein black rings
the ratio $c=a/b$ exploits the full range
$ 0 \le c \le 1$,
EGB black rings are restricted to the $\hat \alpha'$-dependent
range
\begin{equation}
0< 
c_{\rm min}(\hat \alpha') \le c \le c_{\rm max} (\hat \alpha')
< 1 .
\end{equation}
For a given $\hat \alpha'=\alpha'/a$, these minimal and maximal values
of $c$ can be read from Figure 7.
{This figure, which is one of the central results of this paper},
exhibits the maximal value of the scaled
coupling $\hat \alpha'$, up to which
EGB black rings can be obtained for a fixed
value of the parameter ratio $c=a/b$.
Therefore it delimits the domain of existence of EGB black rings.
The coupling $\hat \alpha'$ approaches its maximum value
approximately in the middle of the interval, $i.e.$ for $a/b \simeq 1/2$.

As $\hat \alpha'\to \hat \alpha'_{\rm max}$, 
the numerical process fails to converge, 
although no singular behaviour is found there.
This result is not a surprise given the black ring -- black string connection.
Heuristically, these EGB black ring solutions 
may be thought of as being obtained
by taking a piece of the EGB black string and forming a circle.
Thus they would inherit the $r_h-\alpha'$ 
constraint (\ref{UBS-min}) from there.
The technical reason which causes the solutions to cease to exist
at $\hat \alpha'_{\rm max}$ is discussed in Appendix B.3.
It involves an analytic explanation of this fact
based on a computation
performed in a special coordinate system introduced there.
Similar to the black string case, the argument in the Appendix
uses an analysis of the field equations at the event horizon.

\begin{figure}[h!]
\hbox to\linewidth{\hss%
        \resizebox{8cm}{6cm}{\epsfysize=8cm\epsffile{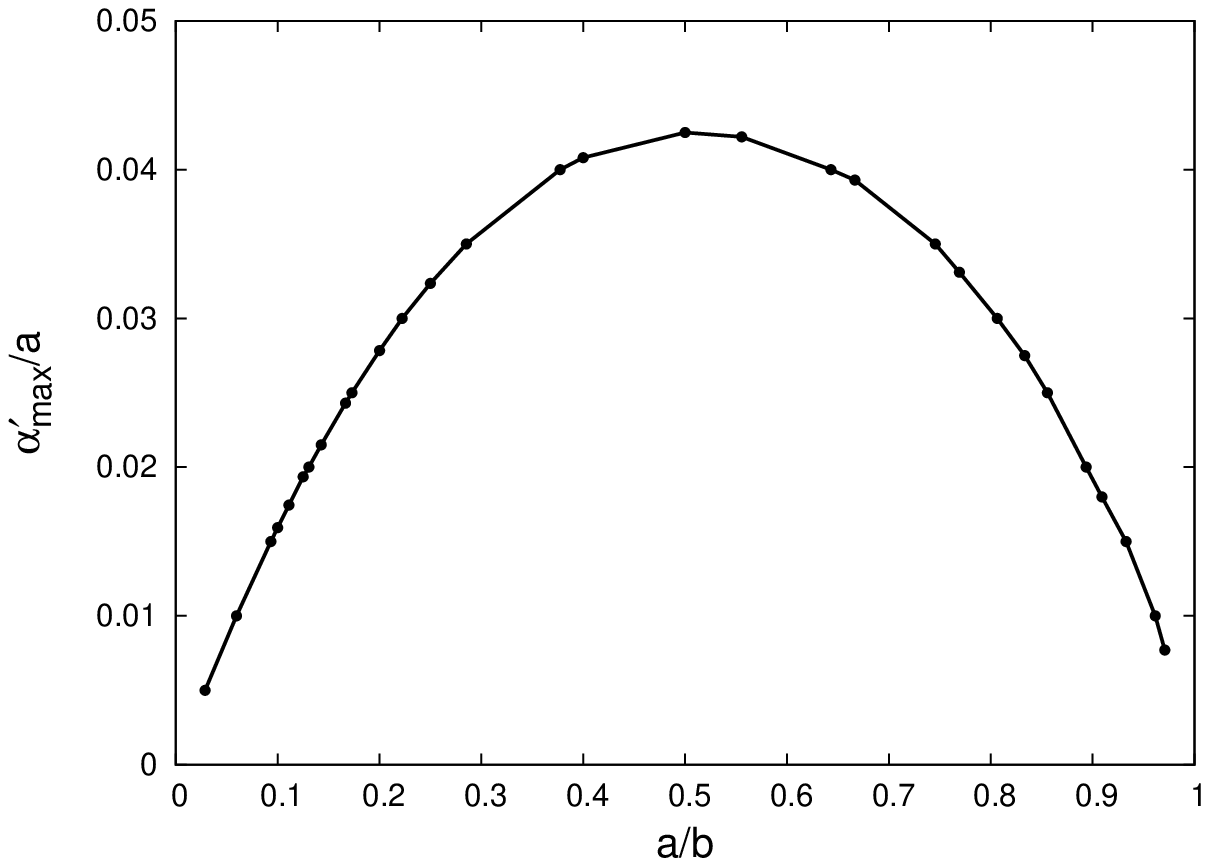}}
 \hss}
\label{fig13}
\vspace{0.2cm}
{\small {\bf Figure 7.}
The domain of existence of EGB black rings is delimited by
the maximal value of $\hat \alpha = \alpha'/a$ versus the
   parameter ratio $c=a/b$.
}
\end{figure}

In Figure 8 
a number of physically relevant properties of the EGB black rings
are exhibited as functions of the scaled coupling
$\hat \alpha'$ for a family of values of the ratio $b/a$.
Also included are the curves, showing the values of 
the respective properties on the
boundary of the domain of existence of the EGB black rings.

The first quantity shown is 
the relative conical excess $\bar \delta$.
Only for $\alpha'=0$, it covers the full range
$-1 \le \bar \delta \le 0$.
Thus it always differs from zero
in the domain of existence of EGB black rings,
as pointed out above.

The scaled mass $\hat M=M/a$ is constant for the Einstein black rings,
since it depends only on $a$. Therefore the EGB mass curves
start for all values of $b/a$ from this Einstein 

\setlength{\unitlength}{1cm}
\begin{picture}(15,21)
\put(-1,0) {\epsfig{file=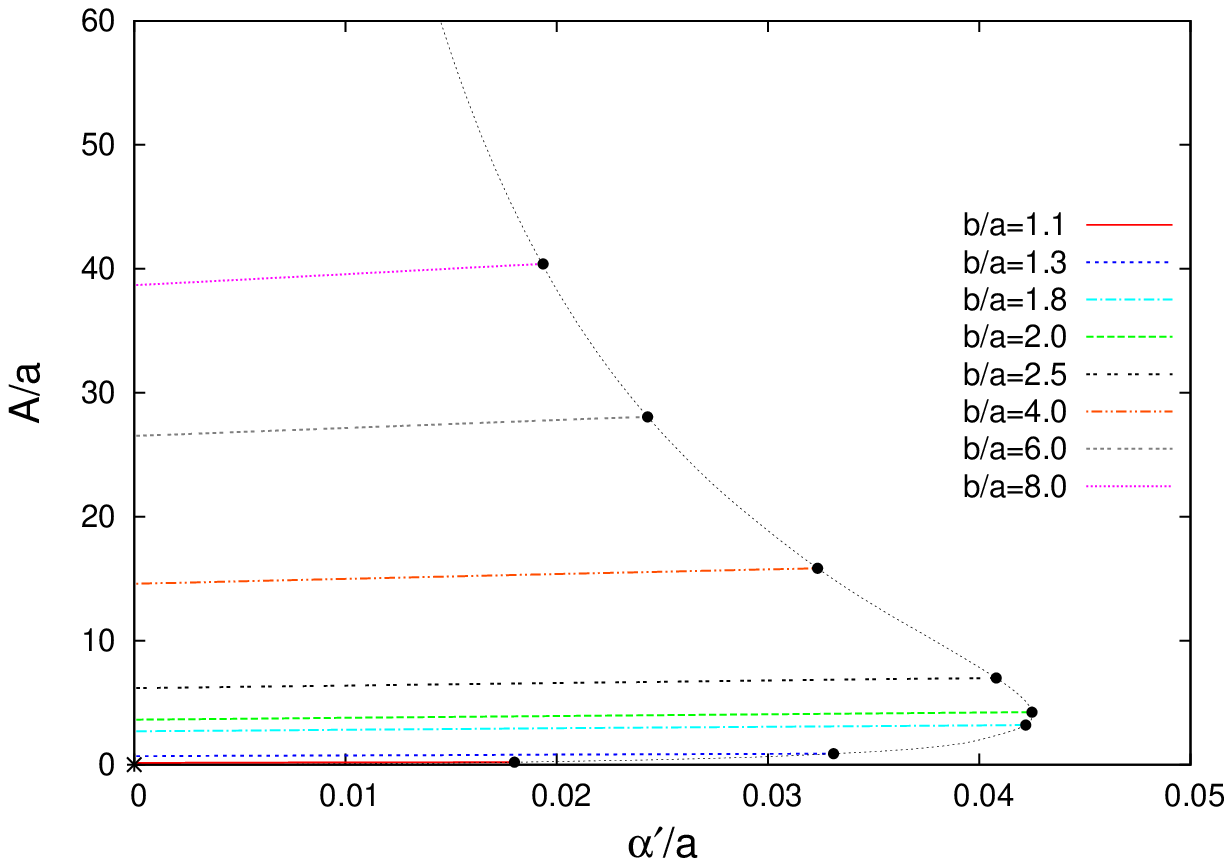,width=7.8cm}}
\put(7,0)  {\epsfig{file=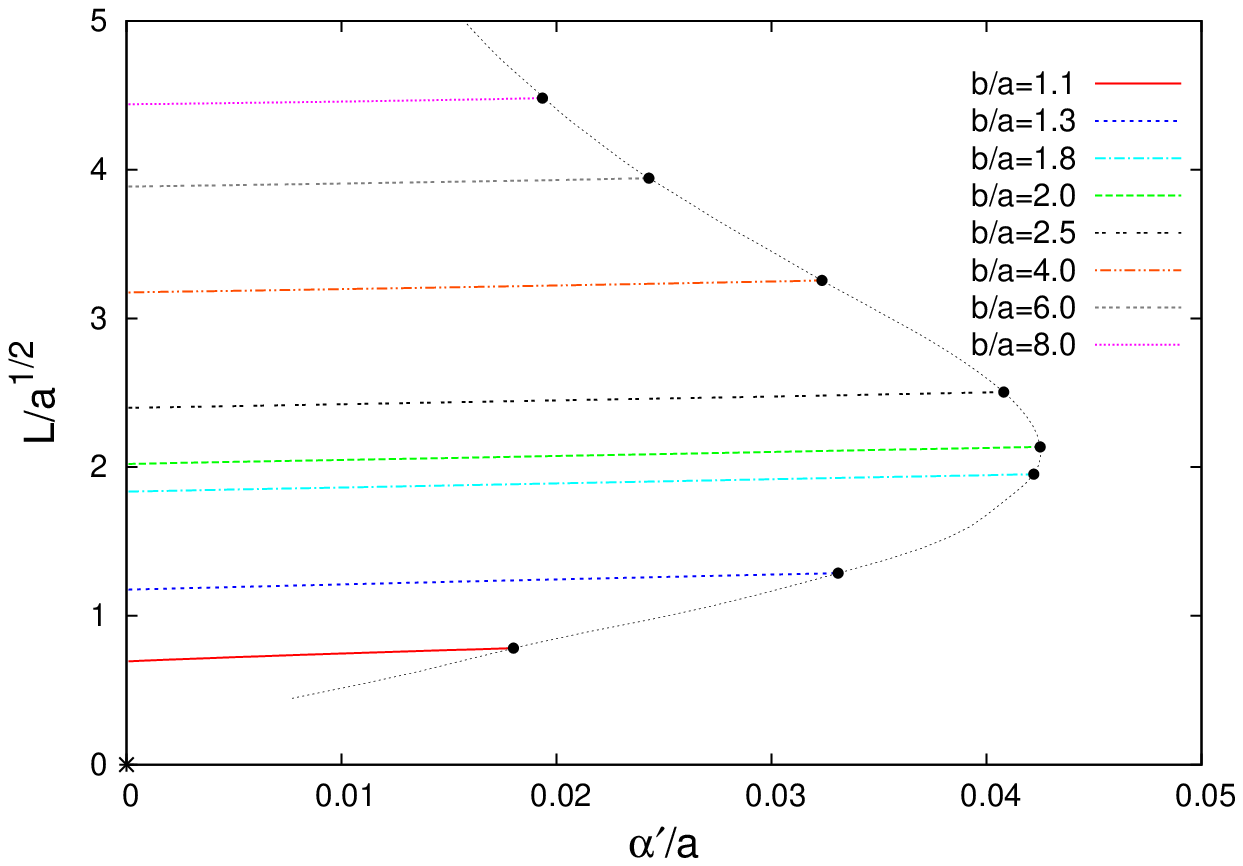,width=7.8cm}}
\put(-1,6) {\epsfig{file=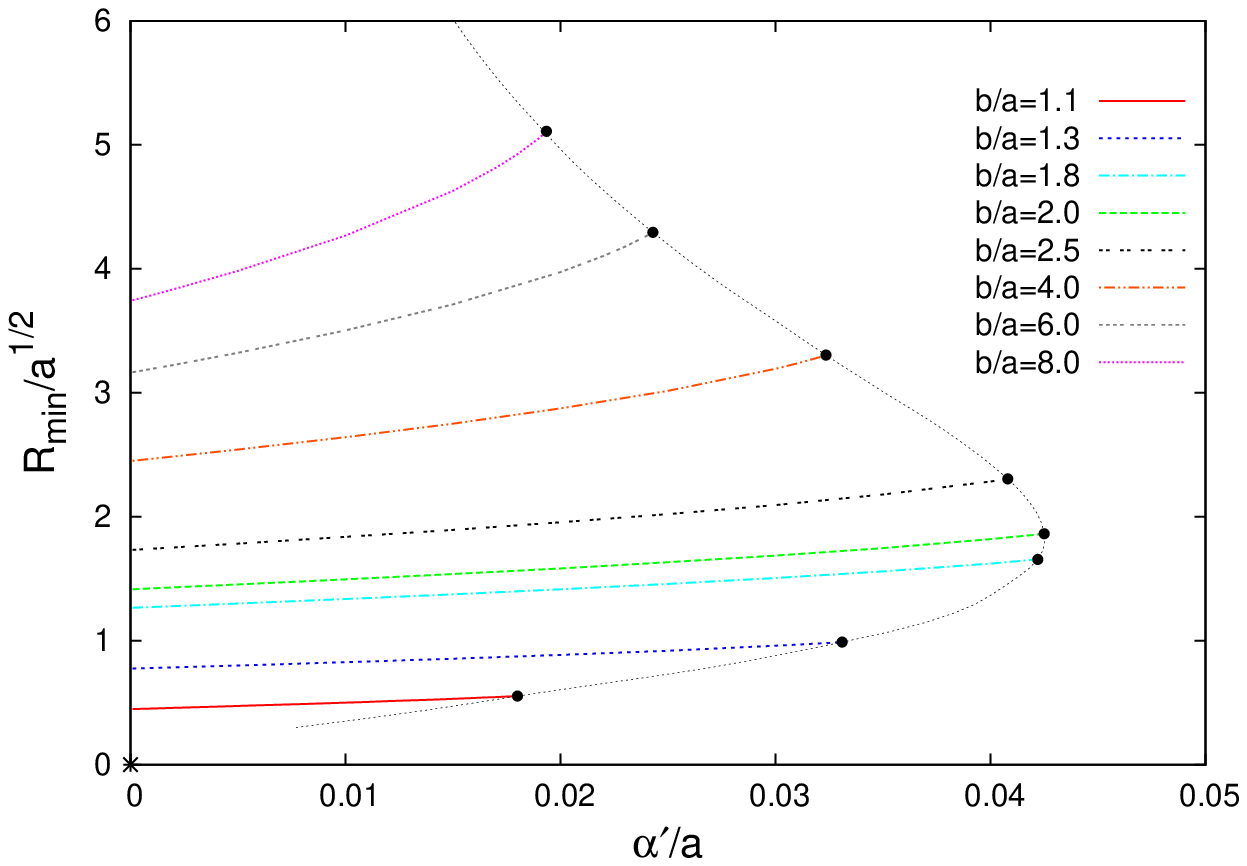,width=7.8cm}}
\put(7,6)  {\epsfig{file=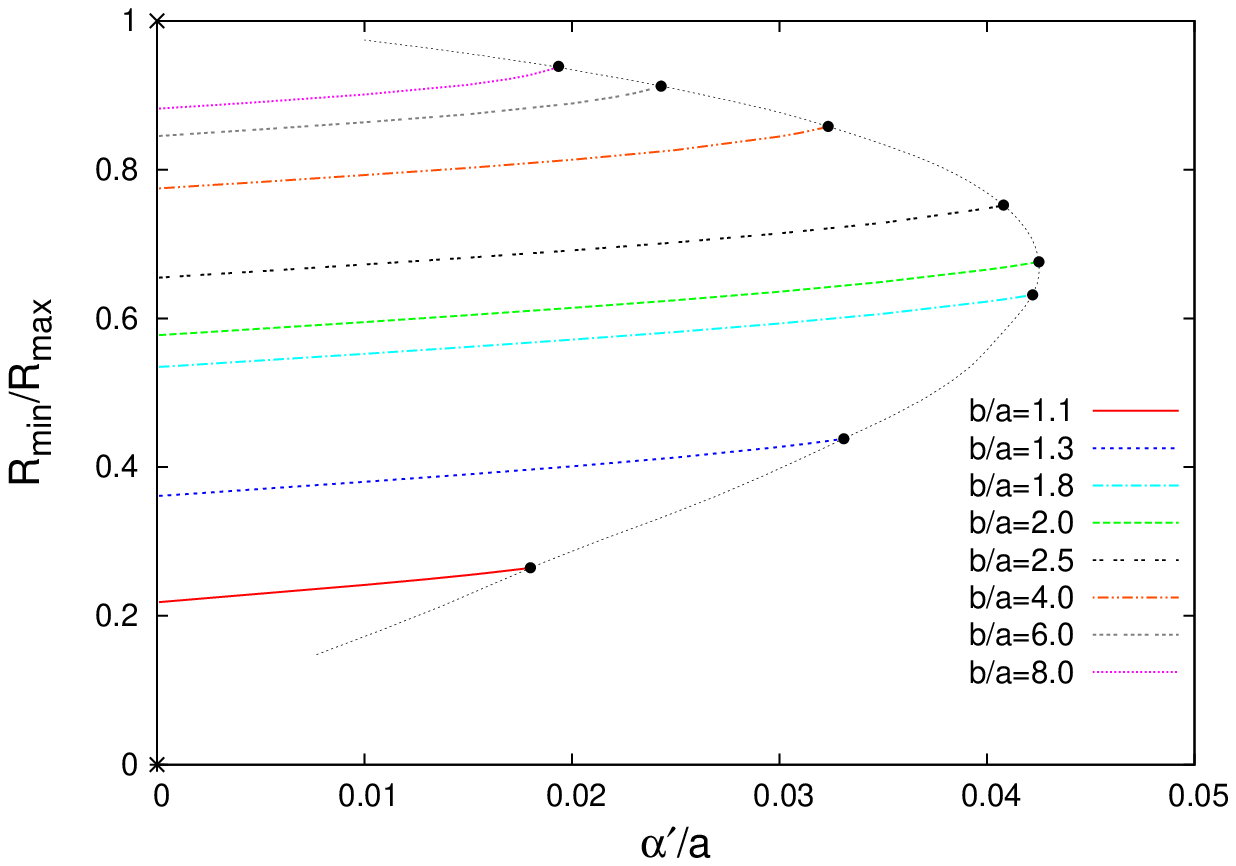,width=7.8cm}}
\put(-1,12){\epsfig{file=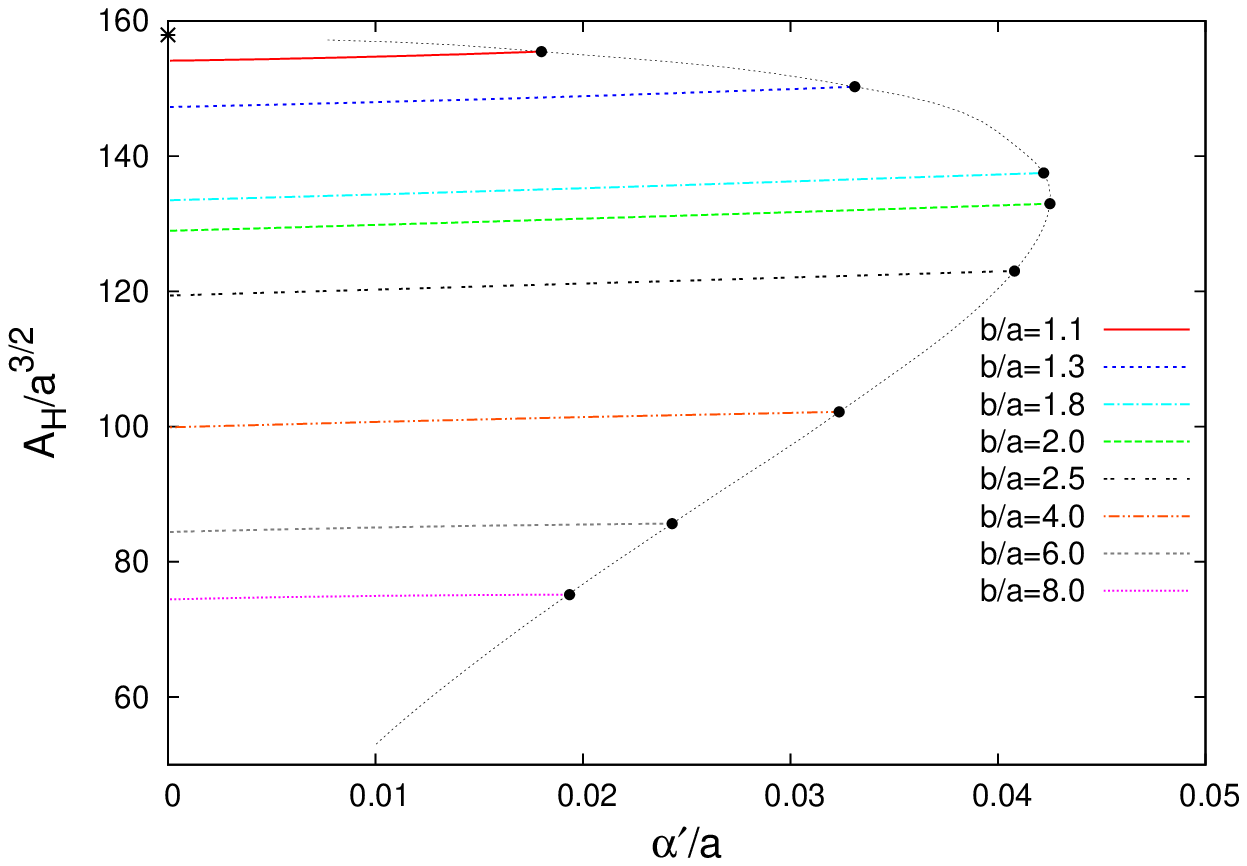,width=7.8cm}}
\put(7,12) {\epsfig{file=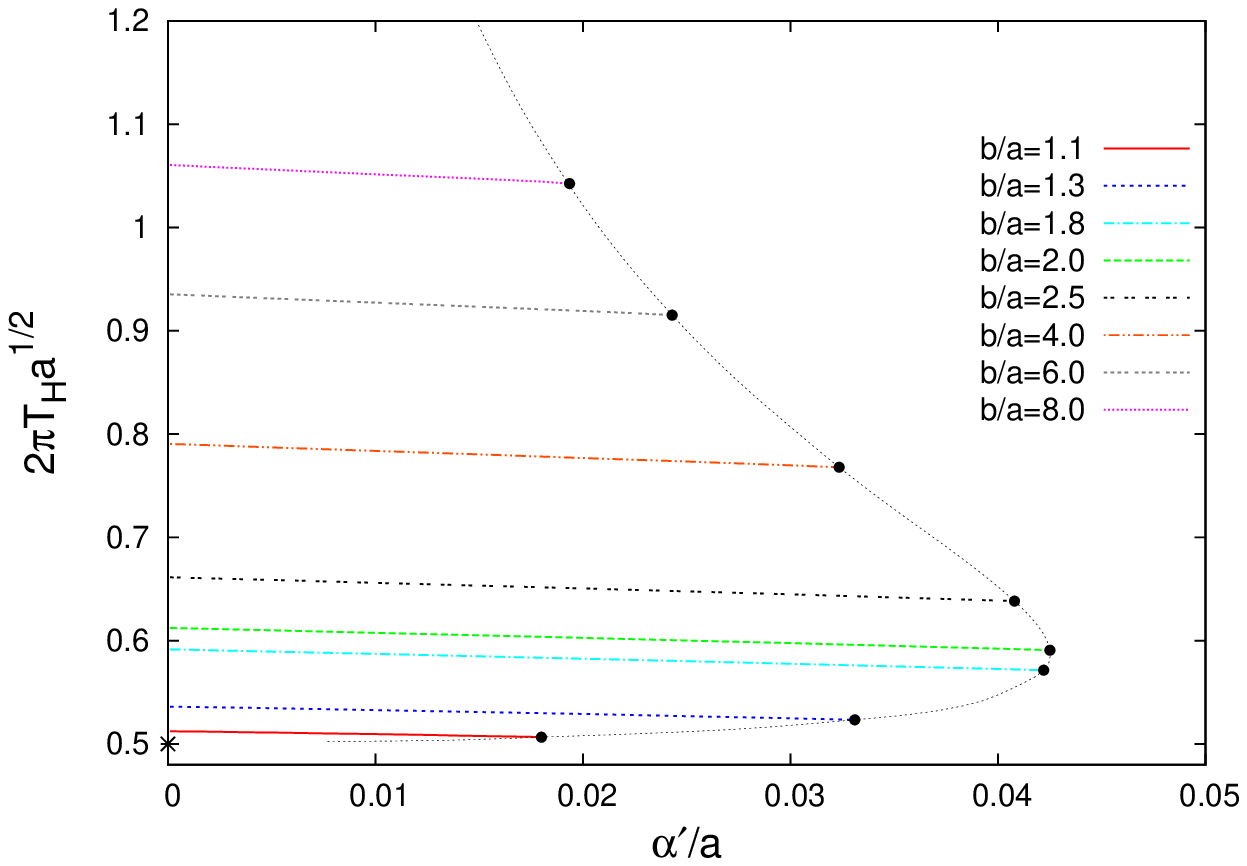,width=7.8cm}}
\put(-1,18){\epsfig{file=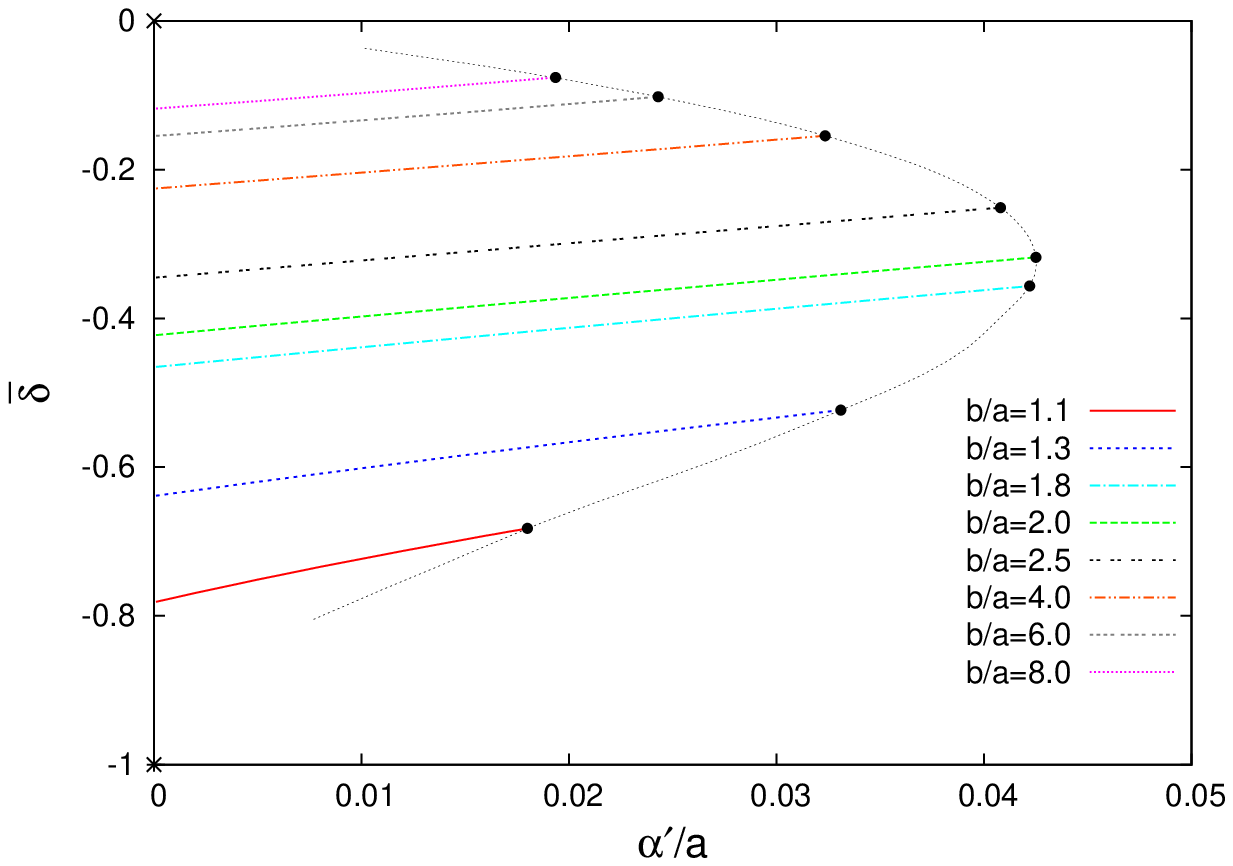,width=7.8cm}}
\put(7,18) {\epsfig{file=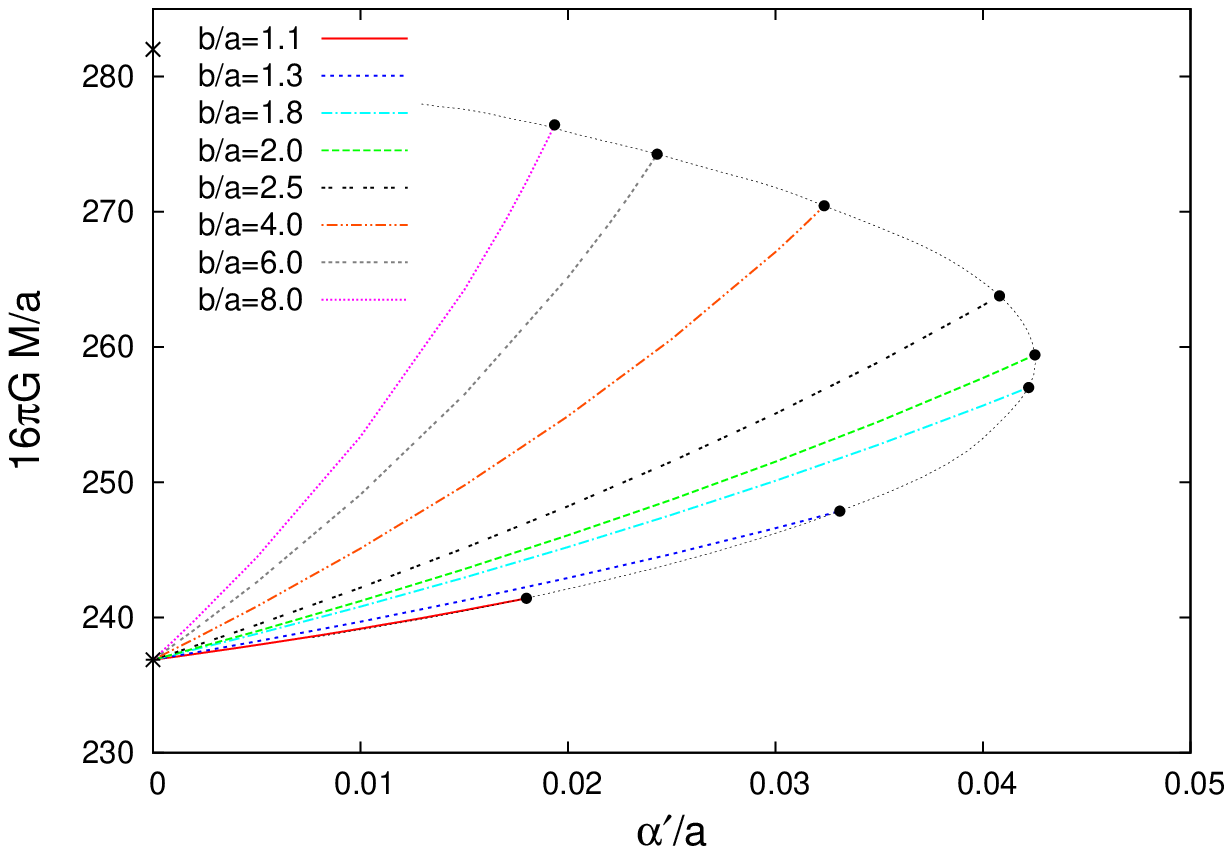,width=7.8cm}}
\label{var-alpha}
\end{picture} 
\\
{\small {\bf Figure 8.}
Several physically relevant (scaled) 
quantities are shown versus $\hat \alpha'$
for several fixed values of $b/a$.
The asterisks and crosses indicate the limits $b/a \to 1$, respectively
 $b/a \to \infty$
} 
\\
value and
increase monotonically until they reach 
the boundary of the domain of existence at the 
respective maximal value of $\alpha'/a$.

The maximal value of the mass,
reached on the boundary for $b/a\rightarrow \infty$,
is finite and less than 20\% above the Einstein value.
Recalling the Smarr-like formula (\ref{smarr}) for the mass, we conclude
that this increase of the mass is 
basically due to the integral $I_{\alpha'}$.

The scaled event horizon area $\hat A_H=A_H/a^{3/2}$ increases
slightly with the coupling $\alpha'/a$.
Its boundary line starts from the finite maximal value of the
Schwarzschild black hole at $b/a=1$.
Extrapolating the solutions on the boundary
to the limit $b/a \rightarrow \infty$,
we observe that the values of $\hat A_H$ of the solutions on the
boundary tend to zero 
as $\sqrt{\alpha'/a}$.

The scaled temperature $\hat T_H=a^{1/2} T_H$ on the other hand
exhibits a slight decrease with $\alpha'/a$.
(However, the solutions are far away from extremality.)
The boundary line for the temperature
starts from the finite minimal value of the
Schwarzschild black hole at $b/a=1$.
Extrapolation this boundary line to the limit $b/a=\infty$
shows, that the values of $\hat T_H$ on the boundary
tend to infinity as 
$1/\sqrt{\alpha'/a}$.
Consequently, the product of $\hat A_H$ and $\hat T_H$,
entering the Smarr-like formula (\ref{smarr}),
stays finite in the limit.

The inner and outer radii of the ring, $R_{\rm min}$ and $R_{\rm max}$, 
exhibit a more pronounced dependence on $\hat \alpha'$,
in particular for larger values of the ratio $b/a$.
The figure exhibits besides the scaled minimal radius $\hat R_{\rm min}$
also their ratio $R_{\rm min}/R_{\rm max}$.
The boundary line of the ratio starts from the Schwarzschild black hole
value $R_{\rm min}/R_{\rm max}=0$ and ends at the
maximally reachable value of $R_{\rm min}/R_{\rm max}=1$.

Figure 8 also shows the scaled proper length $\hat L$
of the $\psi$-rod
and the scaled thermodynamical parameter $\hat A$.
For these quantities we observe only a small dependence 
on
$\hat \alpha'$, increasing with $\hat \alpha'$ only by a few procent
for fixed $b/a$.
{A similar picture has been found for the quantities $A_1$ and $Area_1$, 
as given by (\ref{I0-c}), (\ref{comp2}), respectively.}

Finally, we would like to  return to the intriguing observation,
that the conical excess $\delta$ decreases with
increasing the  GB  coefficient $\hat \alpha'$.
While this result has been unexpected (for us),
it can be heuristically understood as follows.
In the presence of curvature-squared terms,
the modified Einstein equations leads to an effective stress tensor that involves the gravitational field
 \begin{eqnarray}
G_{\mu\nu}=-\alpha' H_{\mu\nu}=T_{\mu\nu}.
 \end{eqnarray}
Therefore, from some point of view, the quantity $\alpha' H_{t}^t(=-G_t^t)$ corresponds to a
local  `effective energy density'.
However, this effective stress tensor, 
thought of as a kind of matter distribution,
in principle may violate the weak energy 
condition\footnote{This mechanism
has been exploited to construct wormhole solutions in $d=5$ EGB theory,
see the discussion in \cite{Garraffo:2008hu} and the references there.}.
This is indeed the case, since as one can see in Figure 9, 
near the horizon this quantity
takes negative values in some region of the $\rho-z$ plane.
The picture is however quite complicated
and depends in a nontrivial way on the value 
of the input parameters $a/b$ and $\hat \alpha'$.
For fixed $\hat \alpha'$
and a value of the ratio $a/b$ close to one,
the region with a negative `effective energy density'
is localized around the topology changing point at $\rho=0,~z=b$.
This region expands when the ratio $a/b$ is decreased.
For sufficiently large values of $b$
a more complicated picture emerges,
with the occurrence of another region 
where $\alpha' H_{t}^t<0$, 
which is localized around the 
{horizon\footnote{We have noticed a similar picture for 
EGB black strings. For all solutions there we have found $H_{t}^t<0$ for a region
near the horizon.
However, the picture is different for the EGB black holes, in which case one can prove that
 $ H_{t}^t $ is always a positive quantity.
 A deeper understanding of this  difference  is still missing.}.
 }


\begin{figure}[ht!]
\setlength{\unitlength}{1cm}
\begin{picture}(15,20.0)
\put(-1,16.5){\epsfig{file=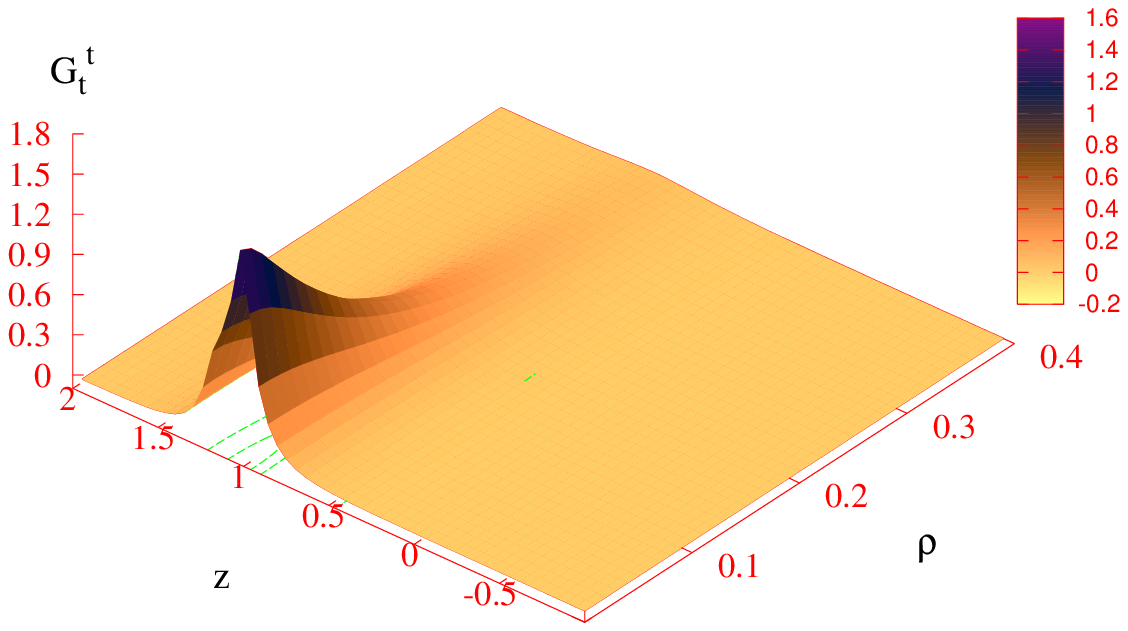,width=9.5cm}}
\put(9,17.){\epsfig{file=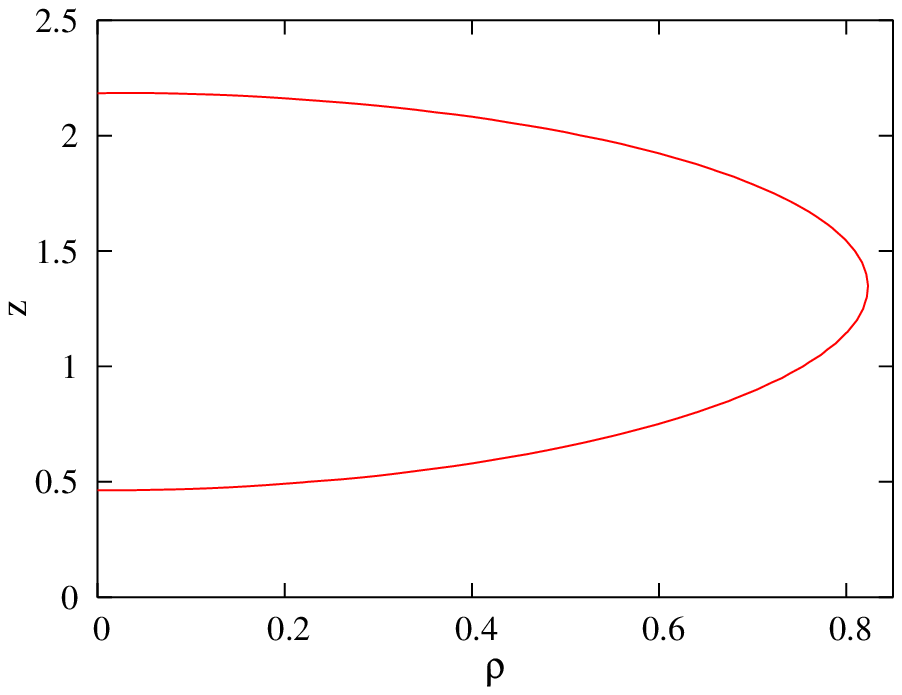,width=8.5cm}}
\put(-1,11){\epsfig{file=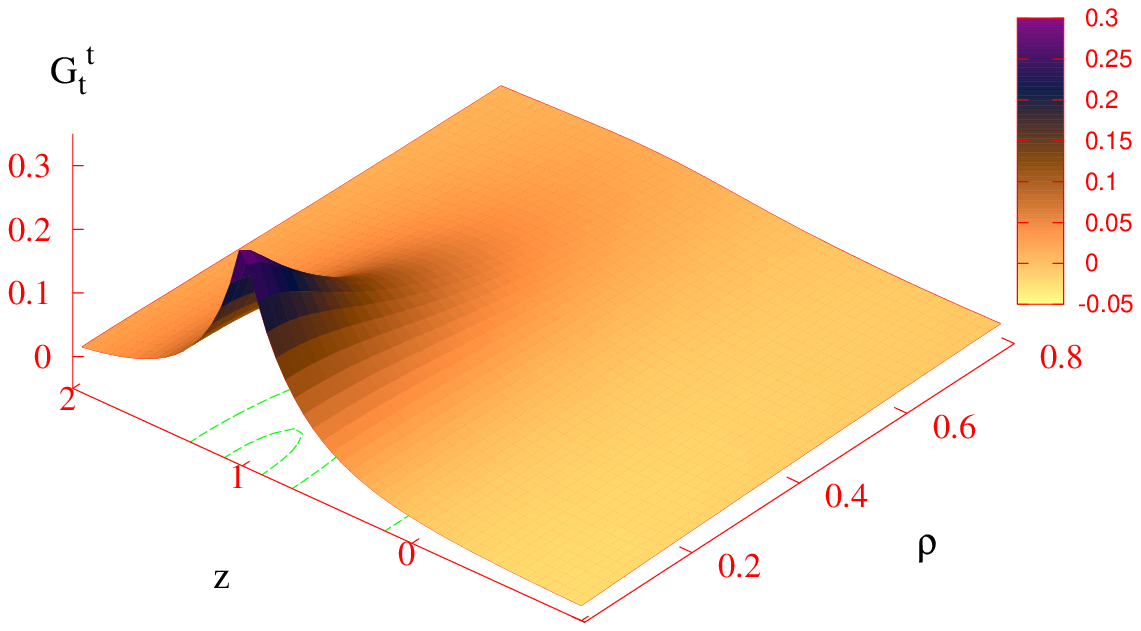,width=9.5cm}}
\put(9,11.5){\epsfig{file=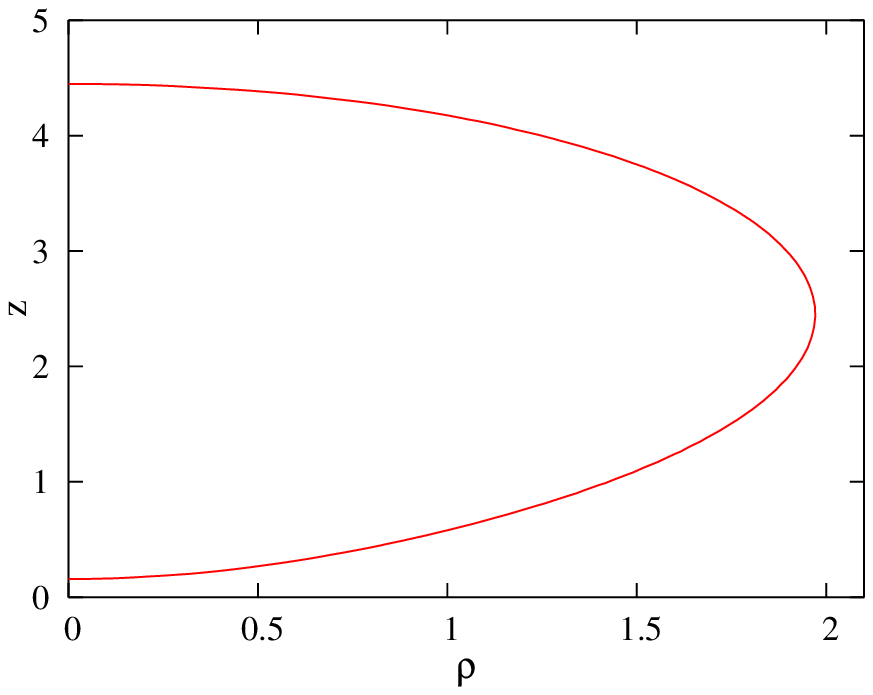,width=8.5cm}}
\put(-1,5.5) {\epsfig{file=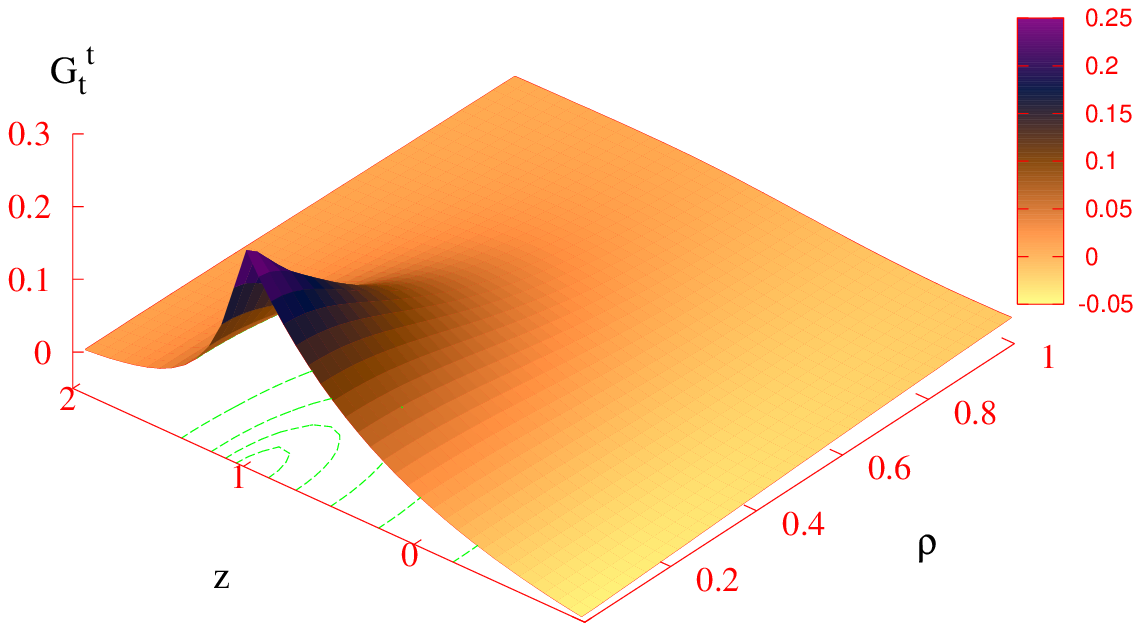,width=9.5cm}}
\put(9,6.) {\epsfig{file=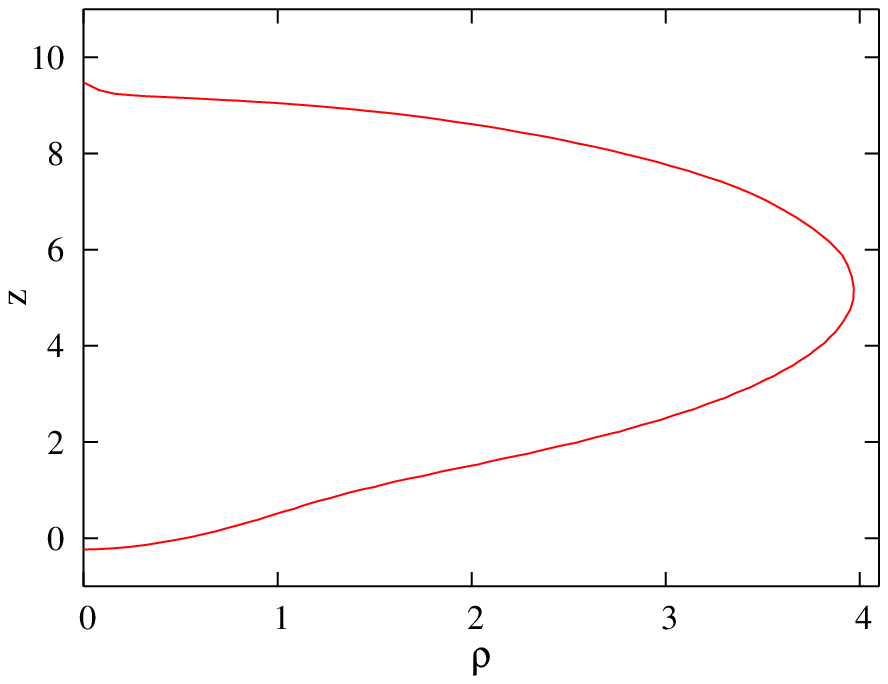,width=8.5cm}}
\put(-1,0) {\epsfig{file=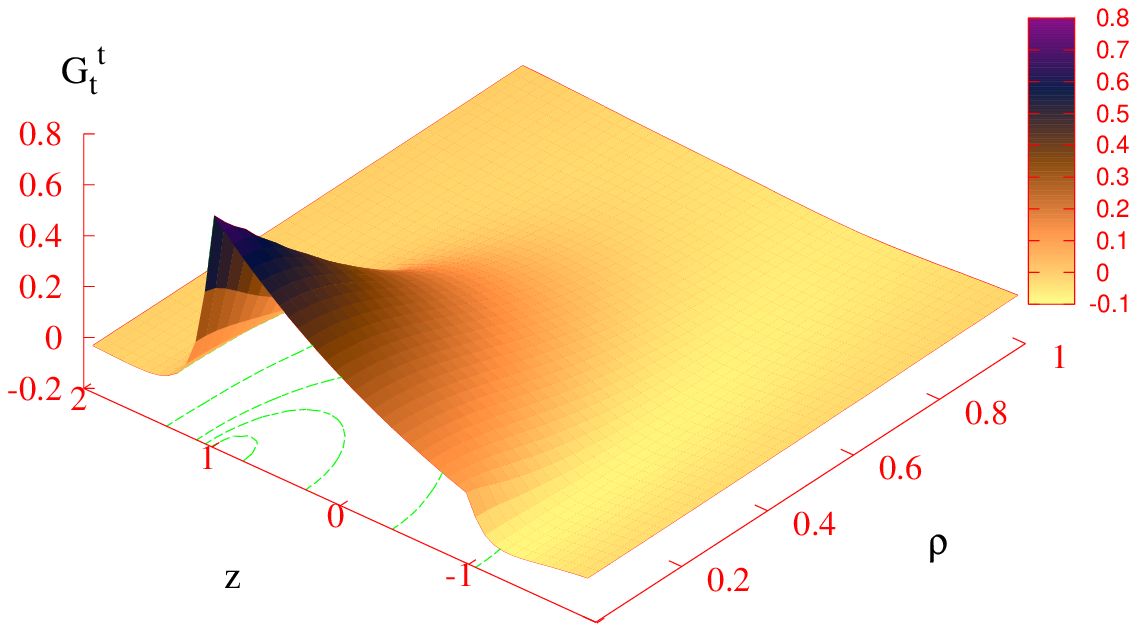,width=9.5cm}}
\put(9,0.5) {\epsfig{file=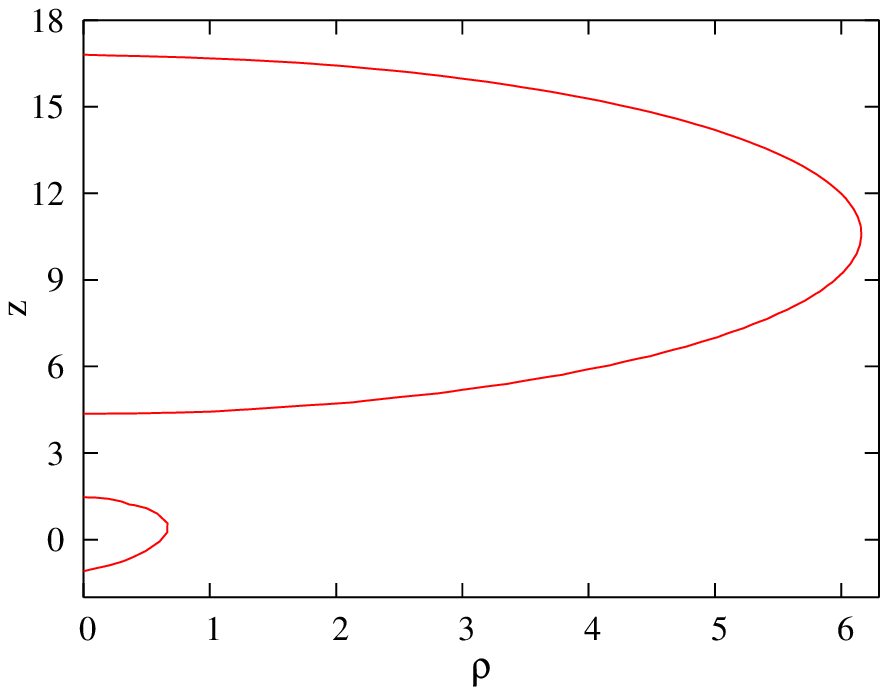,width=8.5cm}}
\end{picture} 
\\
{\small {\bf Figure 9.}
Left column: The $G^t_t$ component of the Einstein tensor near the
horizon versus $\rho$ and  $z$ for $\alpha'=0.015$,
$a=1$ and $b=1.2$, $2.0$, $4.0$ and $8.0$ from top to botton.
Right column: The curve $G^t_t= 0$ versus $\rho$ and  $z$
for the same set of parameters.
} 
\end{figure}

We conclude that, for a  black ring,
the contribution of the Gauss-Bonnet provides a repelling force
in addition to that supplied by the conical excess.
While this, in principle, might have given hope to construct balanced
static black rings in EGB theory,
these could unfortunately not be realized,
since the solutions cease to exist before
the limit $\delta=0$ is reached.

\subsubsection{The large $b$ limit and 
the maximal coupling $\alpha'_{\rm s,max}$}

Let us now address the large $b$ limit and try to understand
the character of the solutions obtained along the
boundary of the domain of existence,
when the ratio $a/b$ tends to zero along with the coupling $\hat \alpha'$.

In Figure 10 we show the ratio $\mu=16 \pi GM/6\pi A_H T_H$
versus the ratios
$R_{\rm min}/R_{\rm max}$
and $a/b$ for several values of $\alpha'$.
The dots indicate solutions on the boundary of the domain of existence.
They represent solutions with the maximal and minimal
values for the ratio $b/a$ for
a fixed value of the coupling $\hat \alpha'$.

\begin{figure}[h!]
\hbox to\linewidth{\hss%
        \resizebox{18cm}{6cm}{\epsfysize=5cm\epsffile{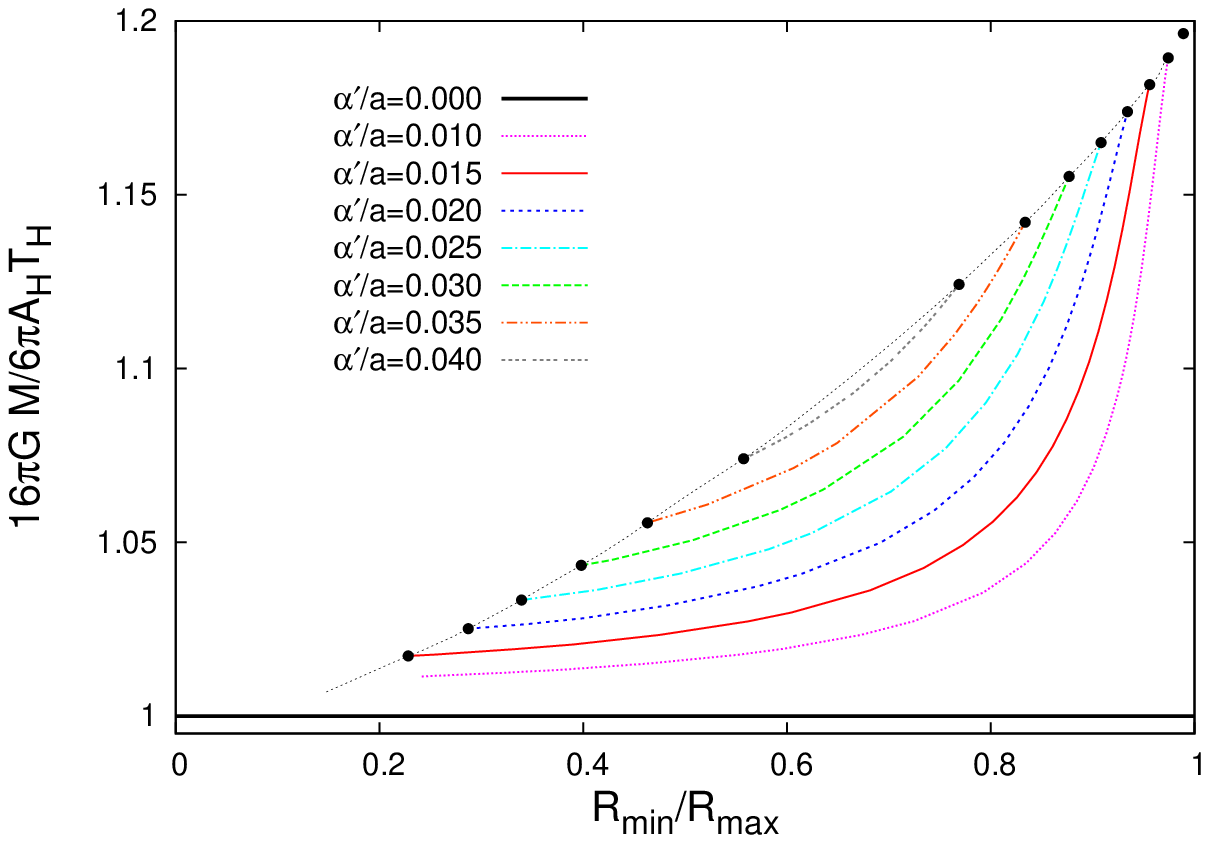}
                              \epsfysize=5cm\epsffile{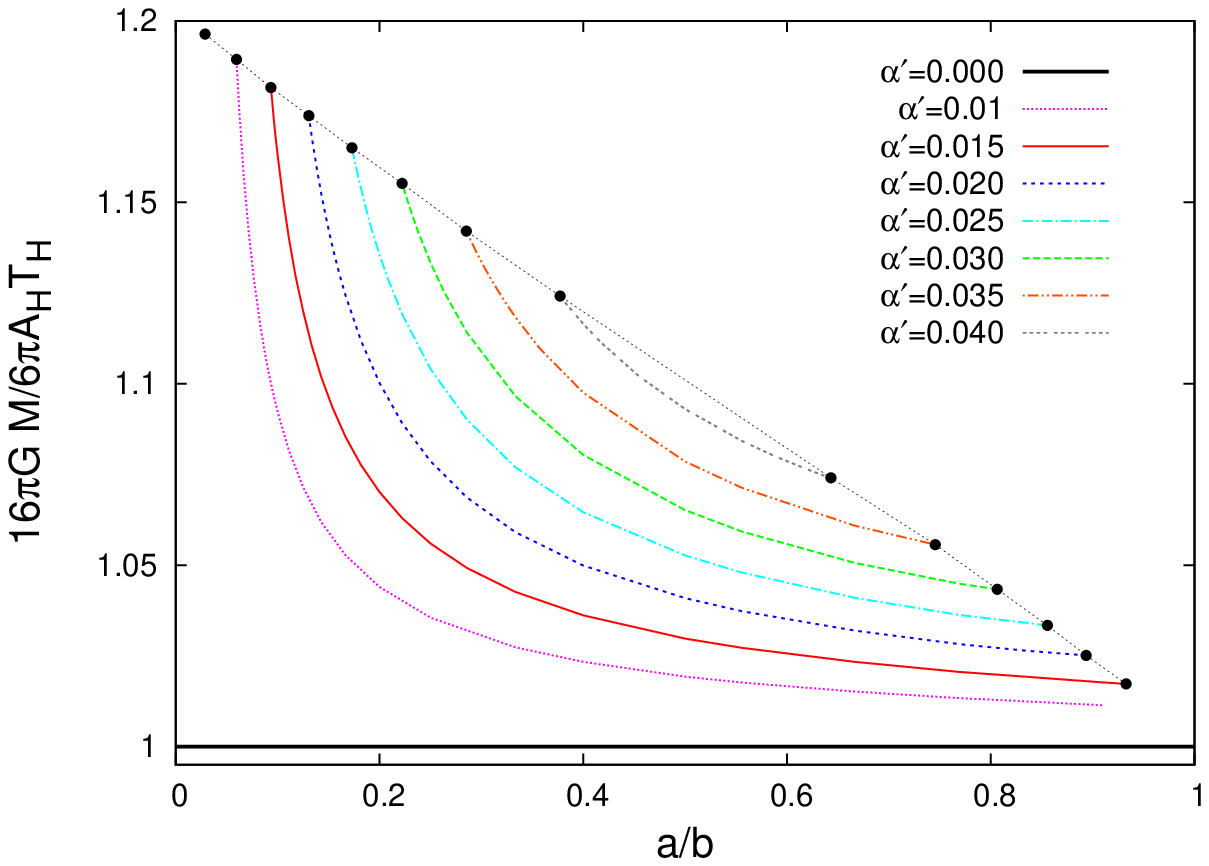}}
 \hss}
\label{fig12}
\vspace{0.3cm}
{\small {\bf Figure 10.}
The dimensionless ratio $16\pi GM/6\pi A_H T_H$ is shown versus the ratios
   $R_{min}/R_{max}$ and
 $a/b$, respectively, for several values of $\hat \alpha'$.}
\end{figure}

We note that with decreasing $\hat \alpha'$ the maximal value of $a/b$
increases. 
Here we expect to reach the five dimensional Schwarzschild
solution as $\hat \alpha'$  tends to zero and $b$ tends to $a$. In this limit
the ratio $\mu$ assumes the value $\mu=1$.

On the other hand, the minimal value of $a/b$ decreases with decreasing
$\hat \alpha'$, and tends to zero as $\hat \alpha'$ tends to zero, 
whereas the ratio $\mu$ tends to a finite value in this limit.
In the following we show that in this limit
the (suitably scaled) EGB uniform black string 
at its maximal value of $\alpha'$ is approached.

{For comparison with the black rings, we have performed a systematic study
of the EGB uniform black strings. Here we have used a coordinate system which is different from
 (\ref{UBS}), being related to a limit of the 
metric ansatz introduced in Appendix B. The line element of the  black strings
 is parametrized as
\begin{equation}
ds^2 = -f(r) dt^2 + m(r)\left(dr^2 +r^2 d\Omega_2^2\right)+ l(r) d\phi^2
\ .
\label{ds_string}
\end{equation}
(Note that the $d=4$ reduction along the $\phi$-direction 
gives black holes in isotropic coordinates.)
}

As noted above, the mass and tension 
of the black strings can be expressed in terms of 
the coefficients entering the asymptotic expansion of the 
$g_{tt}$ and $g_{\phi\phi}$ components of the metric.
Thus, from 
$
-g_{tt} \approx 1-\frac{c_t}{r} \ , \ 
 g_{\phi\phi} \approx 1+\frac{c_\phi}{r} \ , 
 $
we obtain
\begin{equation}
M_s/\Delta \phi = M_\ast +\frac{1}{2}{\cal T} \ , \ \ \ \
{\cal T} = \frac{1}{4\pi G}(c_t-2 c_\phi) \ , 
\label{M_string}
\end{equation}
with
\begin{equation}
M_\ast = \frac{1}{4\pi G} \frac{3}{2} c_t \  .
\label{M_ast}
\end{equation}
The Hawking temperature and horizon area of a black string are obtained as
\begin{equation}
T_{\rm H}^{s} = \frac{1}{2\pi r_0} \left. \sqrt{\frac{\hat{f}}{m}}\right|_{r_0} 
\ , \ \ \ 
A_{\rm H}^{s} = 4 \pi r_0^2 \Delta\phi\left. m\sqrt{l}\right|_{r_0} \  ,
\label{TA_string}
\end{equation}
where $r_0$ and $\Delta \phi$ denote the isotropic horizon radius
and the length of the compact coordinate, respectively,
and the function $\hat{f}(r)= f(r)/(1-r_0/r)^2$ is finite at the horizon. 
For fixed values of $r_0$, uniform black string solutions
exist for all $\alpha'_s/r_0^2 < \alpha'_{s,{\rm max}}/r_0^2 \approx 1.44$.
 
Let us now connect the uniform black strings with the black rings 
in the limit $a/b \to 0$.
From the results for the black rings (e.g.~Figure 7)
we observe that the product $\alpha' b_{\rm max}(\alpha')/a^2$
assumes a finite value when  $\alpha'/a \to 0$ and $b_{\rm max}/a \to \infty$,
which corresponds precisely to a fraction of the
scaled maximal string coupling $\alpha'_s/r_0^2$,
i.e.  
$\alpha' b_{\rm max}(\alpha')/a^2 \to \alpha'_{s,{\rm max}}/8 r_0^2$.

We therefore introduce the scaling parameter 
$\lambda = r_0 \sqrt{8b}/a$ such that
the scaled coupling $\bar{\alpha}' = \lambda^2 \alpha' = 8b r_0^2\alpha'/a^2$
tends to $\alpha'_{s,{\rm max}}$ for $a/b \to 0$.
The scaled parameters 
$$\bar{a}= \lambda^2 a = r_0^2 \frac{8b}{a} \ , \ \ \ \ 
  \bar{b}=\lambda^2 b = r_0^2 \frac{8b^2}{a^2} 
$$
then tend to infinity as $b/a \to \infty$ for fixed $a$.   
However, the ratio $\bar{a}/\sqrt{2\bar{b}} = 2 r_0$ remains finite in this
limit.

In order to obtain the uniform black string limit we also need to
scale the angle variable $\phi$ for the black rings,
since the function $f_3$ also diverges in the limit. 
Introducing
$\bar{\phi}= \sqrt{2\bar{b}} \phi = 4 r_0 \phi b/a$ 
and integrating
then yields $\Delta\bar{\phi} = 8 \pi r_0 b/a$,
which is the equivalent of the
asymptotic length of the compact dimension.
With this expression 
we thus find for the black rings 
the scaled Hawking temperature
$\bar{T}_{\rm H}$,
the scaled area per asymptotic length 
$\bar{A}_{\rm H}/\Delta\bar{\phi}$,
and the scaled mass per asymptotic length
$\bar{M}/\Delta\bar{\phi}$,
\begin{eqnarray}
\bar{T}_{\rm H} & = & T_{\rm H}/\lambda 
= \frac{a}{\sqrt{8b}} \frac{1}{r_0} T_{\rm H}
\nonumber\\
\bar{A}_{\rm H}/\Delta\bar{\phi}  & = & \lambda^3 A_{\rm H}/\Delta\bar{\phi}
=\frac{r_0^2}{\pi}\sqrt{\frac{8b}{a}} A_{\rm H}/a^{3/2}\ , 
\nonumber\\ 
\bar{M}/\Delta\bar{\phi}  & = & \lambda^2 M/\Delta\bar{\phi}
=\frac{r_0}{\pi} M/a \ , 
\nonumber
\end{eqnarray}

In Figure 11 we show 
the inverse of the dimensionless Hawking temperature
$\widetilde{T}_{\rm H} =\bar{T}_{\rm H} r_0$ 
and the dimensionless area 
$\widetilde{A}_{\rm H} =\bar{A}_{\rm H}/\Delta\bar{\phi}r_0^2$
of the black rings on the boundary of their domain of existence
versus the scaled coupling constant $\tilde {\alpha}' =\bar{\alpha}'/r_0^2$.
Also shown are the corresponding dimensionless
$\widetilde{T}_{\rm H}^s= {T}_{\rm H}^s r_0$ and
$\widetilde{A}_{\rm H}^s = {A}_{\rm H}^s/\Delta {\phi}r_0^2$ 
of the uniform black string solutions
versus the scaled coupling constant $\tilde {\alpha}'_s=\bar{\alpha}'_s/r_0^2$.
We observe 
that $\widetilde{T}_{\rm H}$ and $\widetilde{T}_{\rm H}^s$ 
assume the same limiting values,
when the maximal $\tilde {\alpha}'$ resp.~$\tilde {\alpha}'_s$
is approached.
The same holds for the dimensionless entropies
$\widetilde{A}_{\rm H}$ and $\widetilde{A}_{\rm H}^s$.
We note, that for the black rings the values for
$\bar{\alpha}'=\alpha'_{s,{\rm max}}$ are extrapolated.
The figure also demonstrates,
that the product $\widetilde{A}_{\rm H} \widetilde{T}_{\rm H}$
has only a very slight $\alpha'$-dependence,
i.e.~for finite 
$\alpha'$ it differs only slightly from its pure Einstein value 
(which is $4$).

\begin{figure}[h!]
\hbox to\linewidth{\hss%
	\resizebox{18cm}{6cm}{\epsfysize=5cm\epsffile{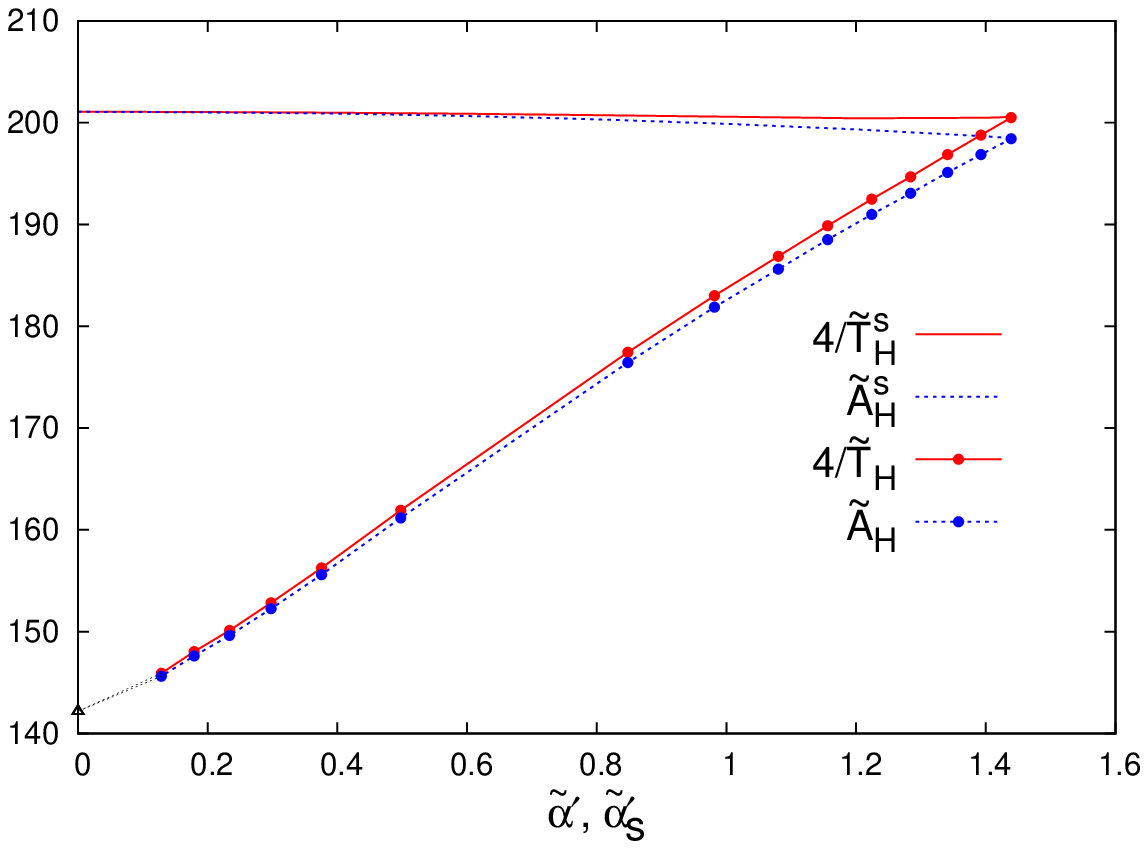}
	                      \epsfysize=5cm\epsffile{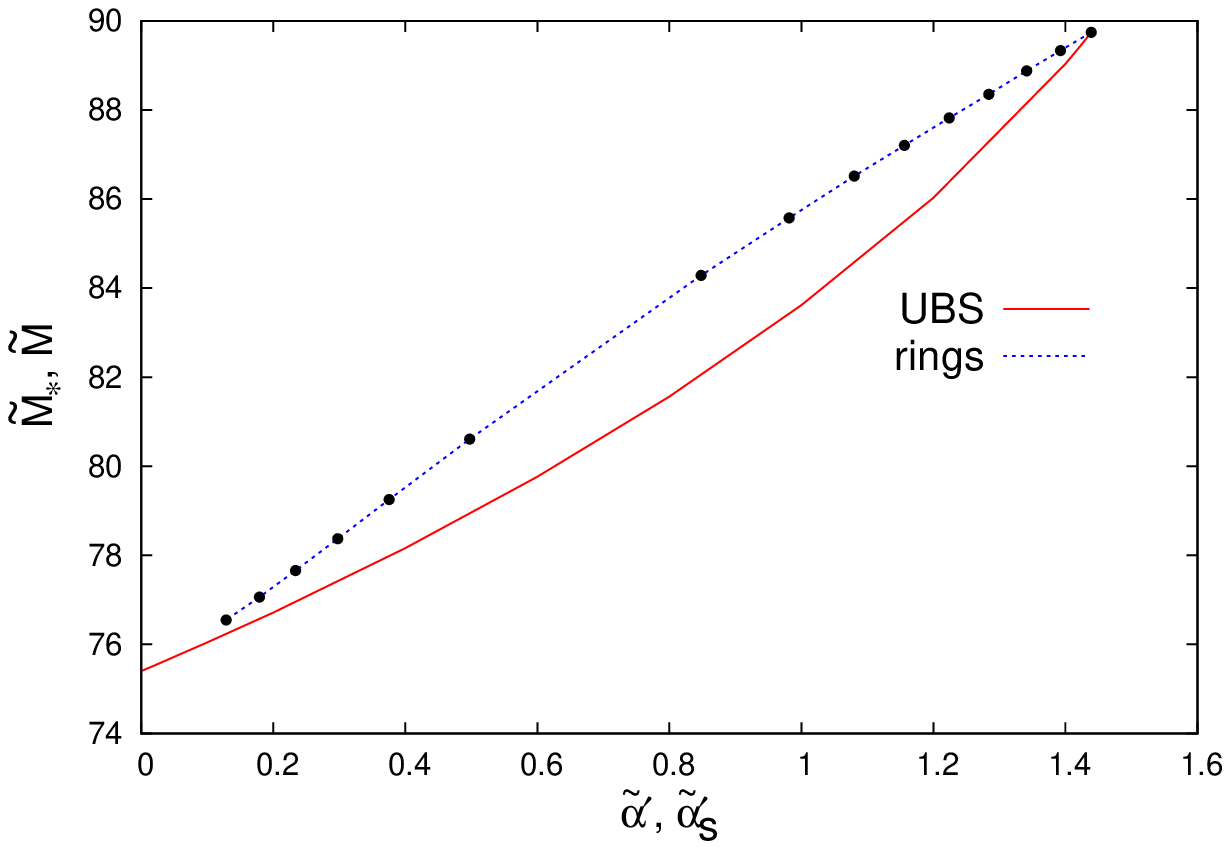}}
 \hss}
\vspace{0.3cm}
{\small {\bf Figure 11.}
   Left:
   The inverse of the dimensionless Hawking temperature
   $\widetilde{T}_{\rm H}= {T}_{\rm H} r_0$
   and the dimensionless area 
   $\widetilde{A}_{\rm H}=\bar{A}_{\rm H}/\Delta\bar{\phi}r_0^2$ 
   of the black rings on the boundary of their domain of existence
   and the dimensionless Hawking temperature
   $\widetilde{T}_{\rm H}^s= {T}_{\rm H}^s r_0$
   as well as the dimensionless area
   $\widetilde{A}_{\rm H}^s=A_{\rm H}^s/\Delta {\phi}r_0^2$ 
   of the uniform black strings
   are shown versus
   $\tilde{\alpha}'=\bar{\alpha}'/r_0^2$ and
   $\tilde{\alpha}'_s=\alpha'_{s}/r_0^2$, respectively.
   Right:
   The same for the dimensionless mass
   $\widetilde{M}=\bar{M}/\Delta\bar{\phi}r_0$ 
   of the black rings and the dimensionless mass
   $\widetilde{M}_\ast=M_\ast/r_0$ of the black strings.
   }
\end{figure}

To compare the scaled mass of the black rings with the mass of the
uniform black strings we have to keep in mind that the former are derived
for an asymptotically flat space-time and therefore have no contribution
from the tension. 
Consequently, the relevant quantity of the 
uniform black strings to compare with is $M_\ast$ in Eq.~(\ref{M_ast}).
In Figure 11 we also show the dimensionless masses
$\widetilde{M}=\bar{M}/\Delta\bar{\phi}r_0$ and
$\widetilde{M}_\ast=M_\ast/r_0$ as functions of $\bar{\alpha}'/r_0^2$
and $\bar{\alpha}'_s/r_0^2$, respectively.
We observe that both masses assume the same values in the limit
where the couplings  $\alpha'_{s}$ and $\bar{\alpha}'$
tends to the maximal value $\alpha'_{s,max}$.  

\subsubsection{The phase diagram}

\begin{figure}[ht]
\hbox to\linewidth{\hss%
        \resizebox{18cm}{6cm}{\epsfysize=5cm\epsffile{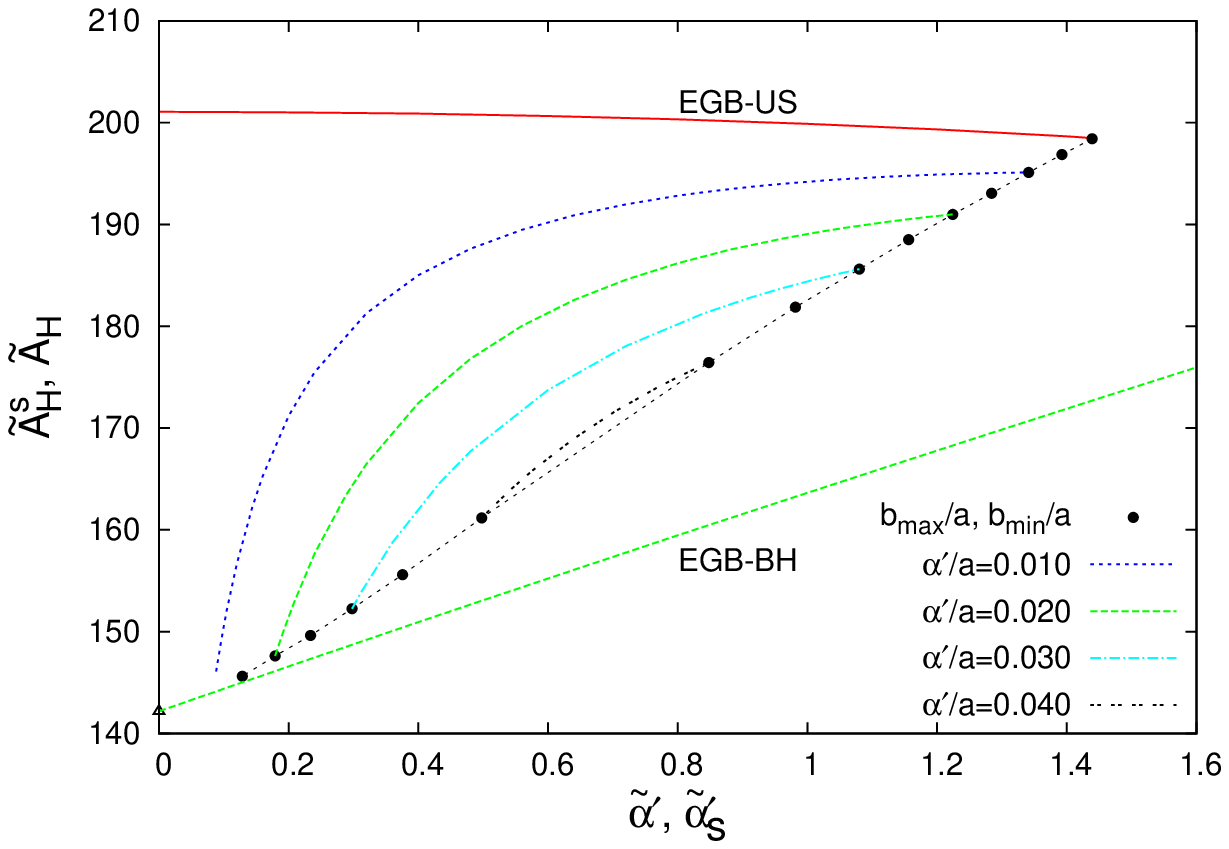}
                              \epsfysize=5cm\epsffile{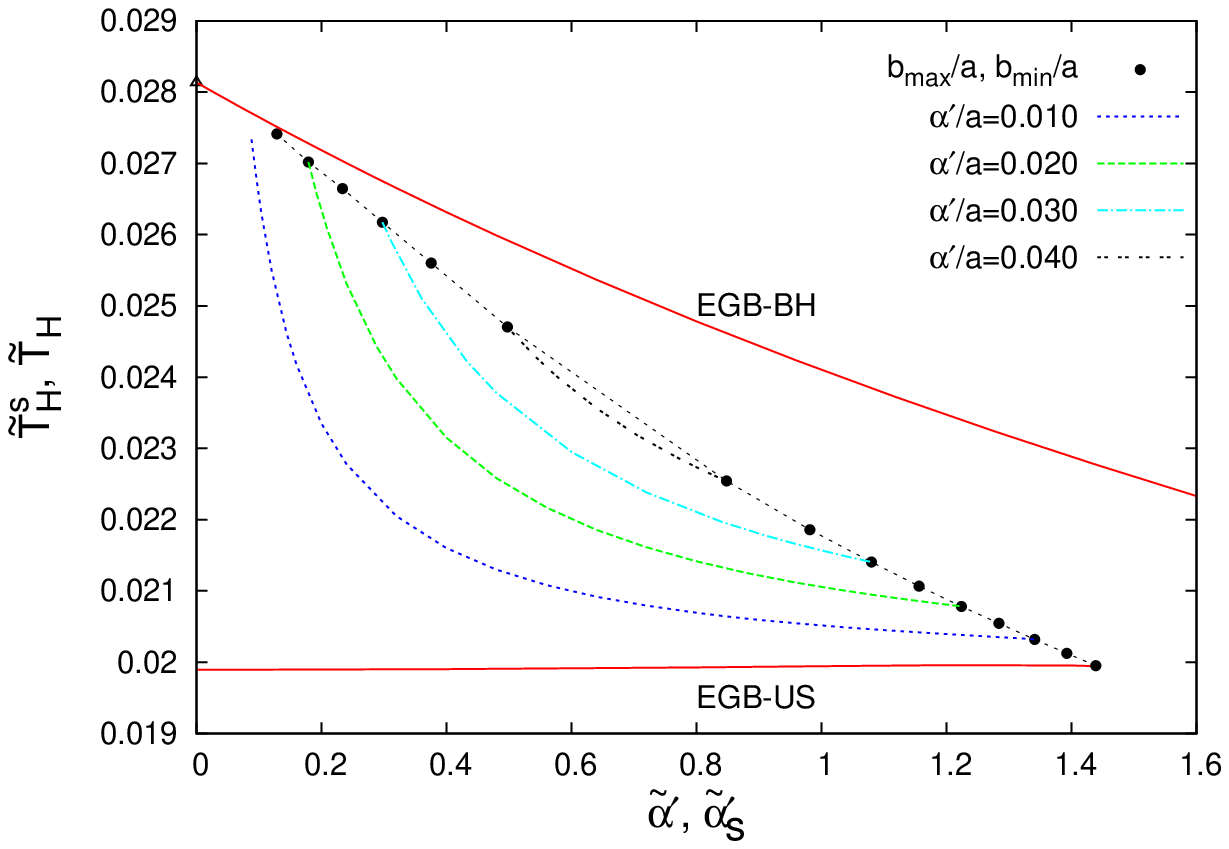}}
 \hss}
\hbox to\linewidth{\hss%
        \resizebox{18cm}{6cm}{\epsfysize=5cm\epsffile{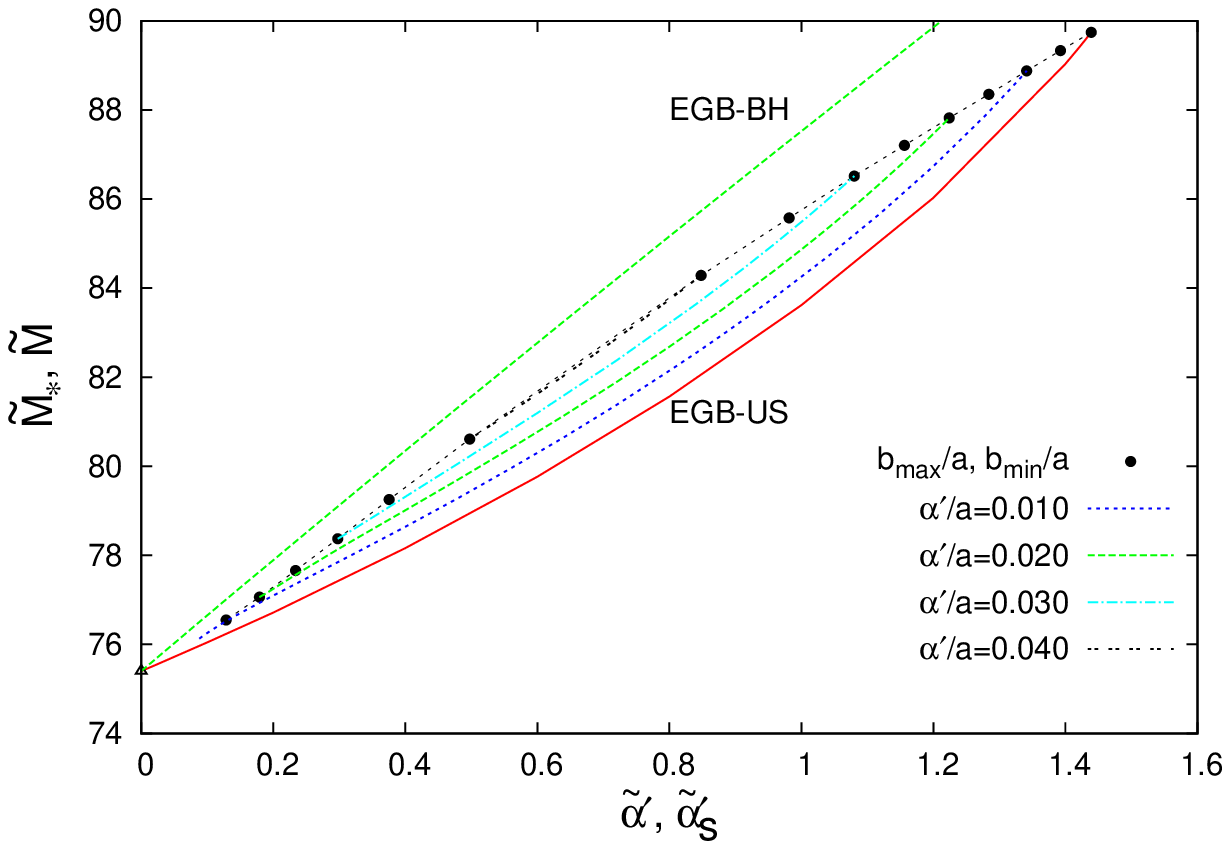}
                              \epsfysize=5cm\epsffile{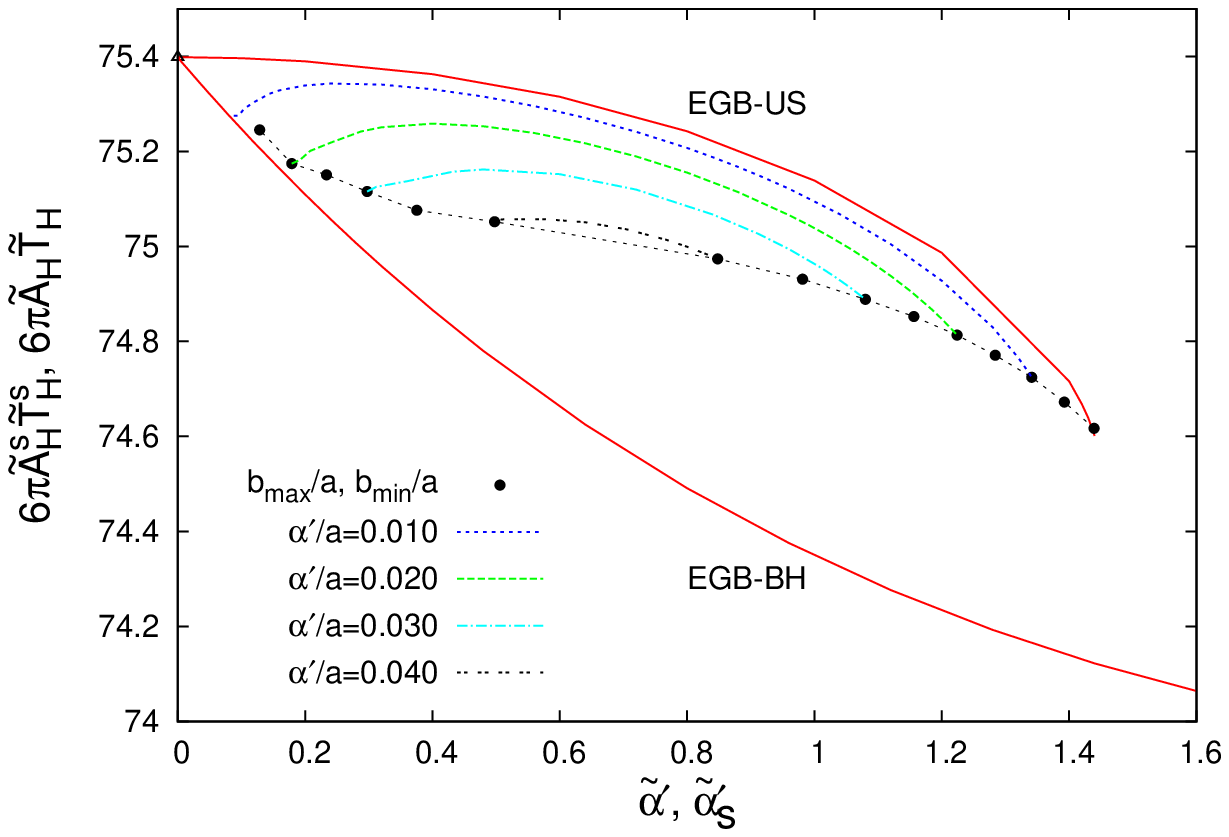}}
 \hss}
\vspace{0.3cm}
{\small {\bf Figure 12.}
   The dimensionless area
   $\widetilde{A}_{\rm H}$,
   temperature $\widetilde{T}_{\rm H}$,
   mass $\widetilde{M}_{\rm H}$, and
   the product $\widetilde{A}_{\rm H} \widetilde{T}_{\rm H}$
   of the EGB black rings and Schwarzschild black holes 
   versus
   $\tilde{\alpha}'=\bar{\alpha}'/r_0^2$
   as well as the respective dimensionless quantities
   $\widetilde{A}_{\rm H}^s$ 
   $\widetilde{T}_{\rm H}^s$, $\widetilde{M}_{\rm H}^s$ and
   $\widetilde{A}_{\rm H}^s \widetilde{T}_{\rm H}^s$
   of the uniform black strings
   versus
   $\tilde{\alpha}'_s=\alpha'_{s}/r_0^2$.
   }
\end{figure}

Let us now compare the properties of the three types of static solutions,
the EGB Schwarzschild black holes, the EGB black rings
and the EGB black strings.
To obtain a phase diagram for these solutions
we exhibit the dimensionless quantities
   $\widetilde{A}_{\rm H}$,
   $\widetilde{T}_{\rm H}$, and
   $\widetilde{M}_{\rm H}$
of the black holes and black rings
together with the corresponding black string quantities
$\widetilde{A}_{\rm H}^s$,
   $\widetilde{T}_{\rm H}^s$, and
   $\widetilde{M}_{\rm H}^s$
in Figure 12.

We note, that the EGB Schwarzschild black holes
connect to the domain of existence of the EGB black rings
only in the limit of Einstein gravity.
Starting from this boundary point they form an infinite branch,
otherwise disconnected from the black rings' domain of existence,
with their scaled mass higher and their scaled area
lower than the values possible for black rings.

In terms of the scaled coupling $\tilde \alpha'$,
the domain of existence of the EGB black ring solutions
is bounded on one side by the set of black ring solutions with
the minimal (respectively maximal) values of $b/a$,
while on the other side 
it is bounded by the set of EGB black string solutions.
The two boundary lines merge and end in a cusp
at the maximal scaled coupling $\tilde \alpha'$.

The presence of cusps in the sets of solutions
typically signals a change of stability.
The question of stability of the sets of solutions
studied here is, however, beyond our current understanding.
It would certainly be interesting
to see whether the analogue of the
Gregory-Laflamme instability \cite{Gregory:1993vy} is present for
EGB back strings and leads to
associated sets of nonuniform EGB back strings.
These latter would then be expected to
be reflected in the emergence of  new black rings,
which are nonuniform along the $S^1$.
The presence of such solutions would 
give rise to a much more intricate EGB phase diagram
(as would be presence of black Saturns, di-rings, etc.).

Moreover, it would be interesting to consider
the phase diagram for dimensions higher than five.
Construction of the uniform black strings would be
straightforward, and one might observe
a dimension dependence of the maximal GB coupling.
Most intriguing would, however, be to find a possibility
to obtain the corresponding sets of higher dimensional black rings.

\section{Further remarks. Conclusions}

In four spacetime dimensions, the Einstein-Hilbert action augmented by a
cosmological constant term is the unique geometrical action 
(thus depending only
on the metric and the curvature tensor) 
leading to field equations which involve
at most second order derivatives of the metric.  However, this is no longer true
if the dimensionality of spacetime is greater than four, in which case, 
from a purely geometrical perspective, there is no
compelling reason to consider only the Einstein-Hilbert
action.

For $d=5$, the most general theory of gravity leading to
second order field equations for the metric is the 
 EGB theory, which contains quadratic powers of 
the curvature.  
Although the solutions of this model  
have been been studied for a long time,
most of the literature considered only spherically symmetric configurations.
Very few EGB exact solutions are known and none of them
is axially symmetric.

In this paper we have argued that the approach used to construct
$d=5$ axially symmetric configurations 
in Einstein gravity within the Weyl formalism, 
can be used also for EGB solutions. 
This opens the possibility to generalize to EGB theory the known solutions with a 
nonspherical topology of the horizon.

Although it would clearly be 
preferable to have analytic solutions (if at all possible), we have made
some progress in this direction by solving numerically the EGB equations. 
The main purpose of this work was to present 
a systematic analysis of the static black rings in EGB theory. 
The results of our investigation show that, for a given value of the GB coupling constant $\alpha'$,
the properties of the solutions are similar to their Einstein gravity counterparts.
For example, all solutions we have found suffer from conical singularities and such solutions
are presumably unstable.
(The apparent unavoidable existence of conical singularities
plague all known asymptotically flat static  solutions 
with a nonspherical topology of the horizon (including here also multi-black objects)).
Interestingly, the absolute value of the conical excess decreases with the GB coupling constant $\alpha'$,
but the solutions stop to exist for some  $\alpha'_{max}$,  before 
approaching a regular configuration.

The techniques  proposed in this paper can easily be extended to
other types of $d=5$ static black objects in EGB theory 
($e.g.$ multi-black holes with $S^3$ topology of the horizon, 
black Saturns, di-rings).
Another possible direction to approach in the future 
is the generalization of the results in Section 2  to the case of 
$d-2$ commuting orthogonal Killing vector fields, with $d>5$.
However, these solutions will not be globally asymptotically flat.
One way to construct $d>5$ solutions approaching at infinity the Minkowski background
 and possesing  a nonspherical topology of the horizon
would be to extend the approach in the recent work 
\cite{Kleihaus:2009wh} by including a GB term in the action.

Let us close this work by briefly 
mentioning the issue of charged static black rings in $d=5$ EGB-Maxwell theory.
One may hope that the inclusion of some matter fields will cure the conical singularity
plaguing the vacuum solutions.
The natural candidate here is a gauge field, 
in the simplest case an electromagnetic field.
Soon after the discovery of the vacuum solution,  
an exact solution describing a U(1) electrically charged static black ring was   
found by several different authors \cite{Ida:2003wv}, \cite{Kunduri:2004da}.
However,  in Einstein-Maxwell theory,  the presence of an electric charge alone was found insufficient to
stabilize a static black ring and prevent it from collapsing, since conical
singularities were unavoidable also in the charged case.

{\it A priori}, it is not obvious that this result holds also for EGB-Maxwell theory.
Therefore we have considered a generalization of the EGB black ring solutions discussed in Section 4
 by including a Maxwell term in the action
(\ref{action}). 
Since the Harrison-type generation techniques used the construct electrically 
charged solutions in Einstein-Maxwell theory
do not hold in the presence of a GB term, 
one is constrained again to approach  this problem numerically.
Restricting ourselves to a purely electric U(1) potential,
$A=V(\rho,z)dt$, we have performed for the
charged case a similar computation to that described in Section 4.
The boundary conditions satisfied by the metric functions are the same as before, while for the electric potential
we have imposed
$V(0,z)=\Phi$ (with $\Phi$ a constant) on the horizon and $\partial_\rho V(\rho,z)|_{\rho=0}=0$ for the rest of the $z-$axis.
At infinity, the electric potential vanishes, the electric charge being read from the asymptotic expansion of $V(\rho,z)$.
The solutions were found starting with EGB configurations and slowly increasing the value for the electric potential
on the horizon.

Although we did not explore yet systematically the full set of parameters, for all solutions we have found,
the absolute value of the conical excess {\it increases} with  $\Phi$.
Thus we conclude that the presence of an electric charge alone 
in very unlikely to stabilize the EGB black rings.
Nonetheless, we expect that, similar to the Einstein-Maxwell theory,  
the conical singularities would be eliminated by submerging
an EGB charged static black ring  into a  background gauge field. 
A drawback of this construction is that, due to the
backreaction of the background electromagnetic field, the black ring will no
longer be asymptotically flat.

However, the  solutions in Section 3 may be viewed as 
an intermediate step towards the construction 
of a rotating {\it balanced} black ring in EGB theory.
In principle, the approach in this work can straightforwardly be generalized to the case of 
spinning solutions \cite{EGBrot}.  
The only obstacle we can see at this moment 
is the tremendous complexity of the EGB 
equations in the presence of rotation.

\section*{Acknowledgements}
B.K. gratefully acknowledges support by the DFG. 
The work of E.R. was supported by a fellowship from the Alexander von Humboldt Foundation.


\appendix

\section{The components of $G_{\mu\nu}$ and $H_{\mu\nu}$}

For the metric ansatz (\ref{metric-canonical}),
the essential nonvanishing components of the Einstein tensor $G_\mu^\nu$
are:
\begin{eqnarray}
\label{Gi}
&&G_\rho^\rho=e^{-2\nu }
\bigg(
-\dot \nu (\dot U_1+\dot U_2+\dot U_3)+
\nu' ( U_1'+ U_2'+  U_3')+
\dot U_1^2+\dot U_2^2+\dot U_3^2
\end{eqnarray}
\begin{eqnarray}
&&{~~~~~~~~~~~}+(\nabla U_1)\cdot (\nabla U_2) 
+(\nabla U_1)\cdot (\nabla U_3) 
+(\nabla U_2)\cdot (\nabla U_3) 
+\ddot U_1+\ddot U_2+\ddot U_3
\bigg),
\end{eqnarray}
\begin{eqnarray}
&&G_z^z=e^{-2\nu }
\bigg(
-\nu' ( U_1'+ U_2'+  U_3')
+\dot \nu (\dot U_1+\dot U_2+\dot U_3)+
 U_1'^2+  U_2'^2+  U_3'^2
\\
\nonumber
&&{~~~~~~~~~~~}+(\nabla U_1)\cdot (\nabla U_2) 
+(\nabla U_1)\cdot (\nabla U_3) 
+(\nabla U_2)\cdot (\nabla U_3)
+  U_1''+ U_2''+ U_3''
\bigg),
\end{eqnarray}
\begin{eqnarray}
&&G_z^\rho=e^{-2\nu }
\bigg(
\nu'(\dot U_1+\dot U_2+\dot U_3)+
\dot \nu ( U_1'+ U_2'+  U_3')
-U_1' \dot U_1-U_2' \dot U_2-U_3' \dot U_3
-\dot U_1'-\dot U_2'-\dot U_3'
\bigg),~~{~~~~}
\end{eqnarray}
\begin{eqnarray}
&&G_\psi^\psi=e^{-2\nu }
\bigg(
(\nabla U_1)^2+(\nabla U_3)^2+(\nabla U_1)\cdot (\nabla U_3)+
\nabla^2\nu +\nabla^2U_1 +\nabla^2U_3
\bigg),
\end{eqnarray}
\begin{eqnarray}
&&G_\varphi^\varphi=e^{-2\nu }
\bigg(
(\nabla U_1)^2+(\nabla U_2)^2+(\nabla U_1)\cdot (\nabla U_2)+
\nabla^2\nu +\nabla^2U_1 +\nabla^2U_2
\bigg),
\end{eqnarray}
\begin{eqnarray}
&&G_t^t=e^{-2\nu }
\bigg(
(\nabla U_2)^2+(\nabla U_3)^2+(\nabla U_2)\cdot (\nabla U_3)+
\nabla^2\nu +\nabla^2U_2 +\nabla^2U_3
\bigg),
\end{eqnarray}

The essential nonvanishing components of the  tensor $H_\mu^\nu$
are:
\begin{eqnarray}
\label{Hi}
&&H_\rho^\rho=4e^{-4\nu }
\bigg(
-\nu'(\dot U_1\dot U_3 U_2'+\dot U_2\dot U_3 U_1'+\dot U_2\dot U_1 U_3')
+\dot \nu (\dot U_3  U_1' U_2'+\dot U_2  U_1' U_3'+\dot U_1 U_2' U_3')
\\
\nonumber
&&{~~~~~~~~~~~~}
-\dot U_1 \dot U_2 \dot U_3 (\dot U_1+\dot U_2+\dot U_3)
+3(\dot \nu \dot U_1 \dot U_2 \dot U_3 -\nu' U_1'U_2'U_3' )
-(\dot U_3^2U_1'U_2'+\dot U_2^2U_1'U_3'+\dot U_1^2U_1'U_3')
\\
\nonumber
&&{~~~~~~~~~~~~}-\ddot U_1 (\nabla U_2)\cdot (\nabla U_3)
-\ddot U_2 (\nabla U_1)\cdot (\nabla U_3)
-\ddot U_3 (\nabla U_2)\cdot (\nabla U_1)
\bigg),
\end{eqnarray}
\begin{eqnarray}
&&H_z^z=4e^{-4\nu }
\bigg(
  \nu'(\dot U_1\dot U_3 U_2'+\dot U_2\dot U_3 U_1'+\dot U_2\dot U_1 U_3')
-\dot \nu (\dot U_3  U_1' U_2'+\dot U_2  U_1' U_3'+\dot U_1 U_2' U_3')
\\
\nonumber
&&
{~~~~~~~~~~~~}
-  U_1' U_2'  U_3' ( U_1'+ U_2'+  U_3')
+3( \nu'   U_1'   U_2'  U_3'  -\dot  \nu  \dot U_1 \dot U_2 \dot U_3  )
-(  U_3'^2\dot U_1 \dot U_2 +  U_2'^2\dot U_1 \dot U_3 + U_1'^2\dot U_2 \dot U_3 )
\\
\nonumber
&&{~~~~~~~~~~~~}- U_1'' (\nabla U_2)\cdot (\nabla U_3)
-  U_2'' (\nabla U_1)\cdot (\nabla U_3)
-  U_3''(\nabla U_2)\cdot (\nabla U_1)
\bigg),
\end{eqnarray}
\begin{eqnarray}
&&H_z^\rho=4e^{-4\nu }
\bigg(
-\nu'(\dot U_3U_1'U_2'+\dot U_2U_1'U_3'+\dot U_1U_2'U_3')
-\dot \nu (\dot U_1 \dot U_3 U_2'+\dot U_3 \dot U_2 U_1'+\dot U_1 \dot U_2 U_3')~~~~~~~{~~~~}
\\
&&
\nonumber
{~~~~~~~~}-3(\nu' \dot U_1 \dot U_2 \dot U_3+\dot \nu U_1'U_2'U_3')+(U_1'\dot U_1+\dot U_1')(\nabla U_2)\cdot (\nabla U_3)
\\
&&
\nonumber
{~~~~~~~~}+(U_2'\dot U_2+\dot U_2')(\nabla U_1)\cdot (\nabla U_3)
+(U_3'\dot U_3+\dot U_3')(\nabla U_2)\cdot (\nabla U_1)
\bigg)
\end{eqnarray}
\begin{eqnarray}
\nonumber
&&H_\psi^\psi=4e^{-4\nu }
\bigg(
2 (\nabla \nu)^2 (\nabla U_1) \cdot (\nabla U_3)
+\nu' 
\left(
-(U_1'+U_3')(2\dot U_1\dot U_3+U_1'U_3')+\dot U_3^2U_1'+\dot U_1^2U_3'
\right)
\\
\label{Hpsipsi}
&&
{~~~~~~~~}+
\dot \nu 
\left(
-(\dot U_1 +\dot U_3 )(2 U_1'  U_3'+\dot U_1 \dot U_3 )+  U_3'^2\dot U_1 + U_1'^2 \dot U_3 
\right)
-(U_1' \dot U_3-U_3' \dot U_1)^2~~~~~~~~~~~~~~~~~~~{~~~~~~~~~~}
\\
\nonumber
&&
{~~~~~~~~}-2(U_3'\dot \nu+ \dot U_3 \nu'-\dot U_3 U_3')\dot U_1'
-2(U_1'\dot \nu+ \dot U_1 \nu'-\dot U_1 U_1')\dot U_3'
+2 \dot U_1' \dot U_3'
\\
\nonumber
&&
{~~~~~~~~}
+(\dot \nu \dot U_3 -\nu' U_3')(U_1''-\ddot U_1)
+(\dot \nu \dot U_1 -\nu' U_1')(U_3''-\ddot U_3)
\\
\nonumber
&&
{~~~~~~~~}
- \ddot U_1 U_3'^2-U_1'' \dot U_3^2 
- \ddot U_3 U_1'^2-U_3'' \dot U_1 
-\ddot U_1 U_3''-U_1'' \ddot U_3
- (\nabla U_1) \cdot (\nabla U_3)
\nabla^2\nu
\bigg),
\end{eqnarray}
\begin{eqnarray}
&&H_\varphi^\varphi=4e^{-4\nu }
\bigg(
2 (\nabla \nu)^2 (\nabla U_2) \cdot (\nabla U_1)
+\nu' 
\left(
-(U_1'+U_2')(2\dot U_2\dot U_1+U_2'U_1')+\dot U_1^2U_2'+\dot U_2^2U_1'
\right)~~~~~~~~~~~~~~{~~}
\\
\nonumber
&&
{~~~~~~~~}+
\dot \nu 
\left(
-(\dot U_1 +\dot U_2 )(2 U_2'  U_1'+\dot U_2 \dot U_1 )+  U_1'^2\dot U_2 + U_2'^2 \dot U_1 
\right)
-(U_2' \dot U_1-U_1' \dot U_2)^2
\\
\nonumber
&&
{~~~~~~~~}-2(U_1'\dot \nu+ \dot U_1 \nu'-\dot U_1 U_1')\dot U_2'-2(U_2'\dot \nu+ \dot U_2 \nu'-\dot U_2 U_2')\dot U_1'
+2 \dot U_2' \dot U_1'
\\
\nonumber
&&
{~~~~~~~~}
+(\dot \nu \dot U_1 -\nu' U_1')(U_2''-\ddot U_2)
+(\dot \nu \dot U_2 -\nu'  U_2')(U_1''-\ddot U_1)
- \ddot U_2 U_1'^2-U_2'' \dot U_1^2 
- \ddot U_1 U_2'^2-U_1'' \dot U_2^2
\\
\nonumber
&&
{~~~~~~~~}-\ddot U_2 U_1''-U_2'' \ddot U_1
- (\nabla U_1) \cdot (\nabla U_2)
\nabla^2\nu
\bigg),
\end{eqnarray}
\begin{eqnarray}
 &&H_t^t=4e^{-4\nu }
\bigg(
2 (\nabla \nu)^2 (\nabla U_3) \cdot (\nabla U_2)
+\nu' 
\left(
-(U_2'+U_3')(2\dot U_3\dot U_2+U_3'U_2')+\dot U_2^2U_3'+\dot U_3^2U_2'
\right)~~~~~~~~~~{~~}
\\
\nonumber
&&
{~~~~~~~~}+
\dot \nu 
\left(
-(\dot U_2 +\dot U_3 )(2 U_3'  U_2'+\dot U_3 \dot U_2 )+  U_2'^2\dot U_3 + U_3'^2 \dot U_2 
\right)
-(U_3' \dot U_2-U_2' \dot U_3)^2
\\
\nonumber
&&
{~~~~~~~~}-2(U_2'\dot \nu+ \dot U_2 \nu'-\dot U_2 U_2')\dot U_3'-2(U_3'\dot \nu+ \dot U_3 \nu'-\dot U_3 U_3')\dot U_2'
+2 \dot U_3' \dot U_2'
\\
\nonumber
&&
{~~~~~~~~}+(\dot \nu \dot U_2-\nu' U_2')(U_3''-\ddot U_3)
+(\dot \nu \dot U_3-\nu' U_3')(U_2''-\ddot U_2)~
-\ddot U_3 U_2'^2- U_3'' \dot U_2^2 
\\
\nonumber
&&
{~~~~~~~~}- \ddot U_2 U_3'^2 - U_2'' \dot U_3^2 
-\ddot U_3 U_2''- \ddot U_2 U_3''
- (\nabla U_3) \cdot (\nabla U_2)
\nabla^2\nu
\bigg),
 \end{eqnarray}
where we define 
\begin{eqnarray}
\label{rel}
(\nabla U) \cdot (\nabla V)=\partial_\rho U \partial_\rho V+  \partial_z U \partial_z V,~~~
\nabla^2 U=\partial_\rho^2U+ \partial_z^2 U,
\end{eqnarray} 
The  expressions of these tensors  in terms of the functions $f_i$ is straightforward.
 
In practice we have solved the following combination of the EGB equations
\begin{eqnarray}
&&E_\rho^\rho+E_z^z+E_\psi^\psi+E_\varphi^\varphi-2E_t^t=0,
~~~~E_\rho^\rho+E_z^z-2E_\psi^\psi+E_\varphi^\varphi+E_t^t=0,
\\
&&E_\rho^\rho+E_z^z+E_\psi^\psi-2E_\varphi^\varphi+E_t^t=0,
~~~~E_\rho^\rho+E_z^z-\frac{1}{2}(E_\psi^\psi+E_\varphi^\varphi+E_t^t)=0,
\end{eqnarray}
which diagonalizes the Einstein tensor w.r.t. $\nabla^2 U_1$, $\nabla^2 U_2$, $\nabla^2 U_3$
and respectively $\nabla^2\nu$.

\section{Details on the numerics}
\subsection{A new coordinate system}

Although we could construct\footnote{ The methods in this case were similar to those used in \cite{Kleihaus:2009wh} to construct
$d=6,7$ black holes with $S^2\times S^{d-4}$ topology of the horizon.}  EGB black ring 
solutions by employing the 
Weyl-type coordinates $(\rho,z)$, the metric ansatz (\ref{metric}) has a number of disadvantages.
For example, it has proven difficult to extract with enough accuracy the value of the mass parameter $M$
from the asymptotic form of $f_0$
and also to study solutions with $a/b\to 0 $ or $a/b \to 1$.

To solve  numerically  the EGB equations, we have found 
it more convenient to introduce
the new coordinates $r,\theta$ and reparametrize the metric (\ref{metric}) as
\begin{eqnarray}
& &
ds^2=-\hat{f}_0(r,\theta)dt^2+\frac{1}{\hat{f}_1(r,\theta)}(dr^2+r^2 d\theta^2)
+\frac{\hat{f}_2(r,\theta)}{\hat{f}_3(r,\theta)}d\psi^2
+\hat{f}_3( r,\theta)d\varphi^2,
\label{remetric} \\
& & r_0 \leq r < \infty \ , \ \ \ \ 
0 \leq \theta \leq \frac{\pi}{2} \ ,
 \end{eqnarray}
where $(\rho,z)$ are related to $(r,\theta)$ by\footnote{Note that this coordinate transformation
reduces to (\ref{coord-tr}) for $r_0=0$.}
\begin{equation}
\rho= \frac{r^4-r_0^4}{2r^2} \sin 2\theta \ , \ \ \ \ 
z= \frac{r^4+r_0^4}{2r^2} \cos 2\theta \ , 
\end{equation}
with $r_0^2= a$.
In these coordinates the horizon is located at $r=r_0$, 
$0 \leq \theta \leq \frac{\pi}{2}$. The semi-finite and finite $\psi$-rods
are mapped to $r_0 \leq r < \infty$, $\theta=\pi/2$ and
$r_0 \leq r \leq r_b $, $\theta=0$, respectively,
while the semi-finite  $\vphi$-rod is on the interval
$r_b \leq r < \infty$, $\theta=0$.
Here $r_b=\sqrt{b+\sqrt{b^2-a^2}}$.\\

 \begin{figure}[ht]
\hbox to\linewidth{\hss%
        \resizebox{8cm}{6cm}{\includegraphics{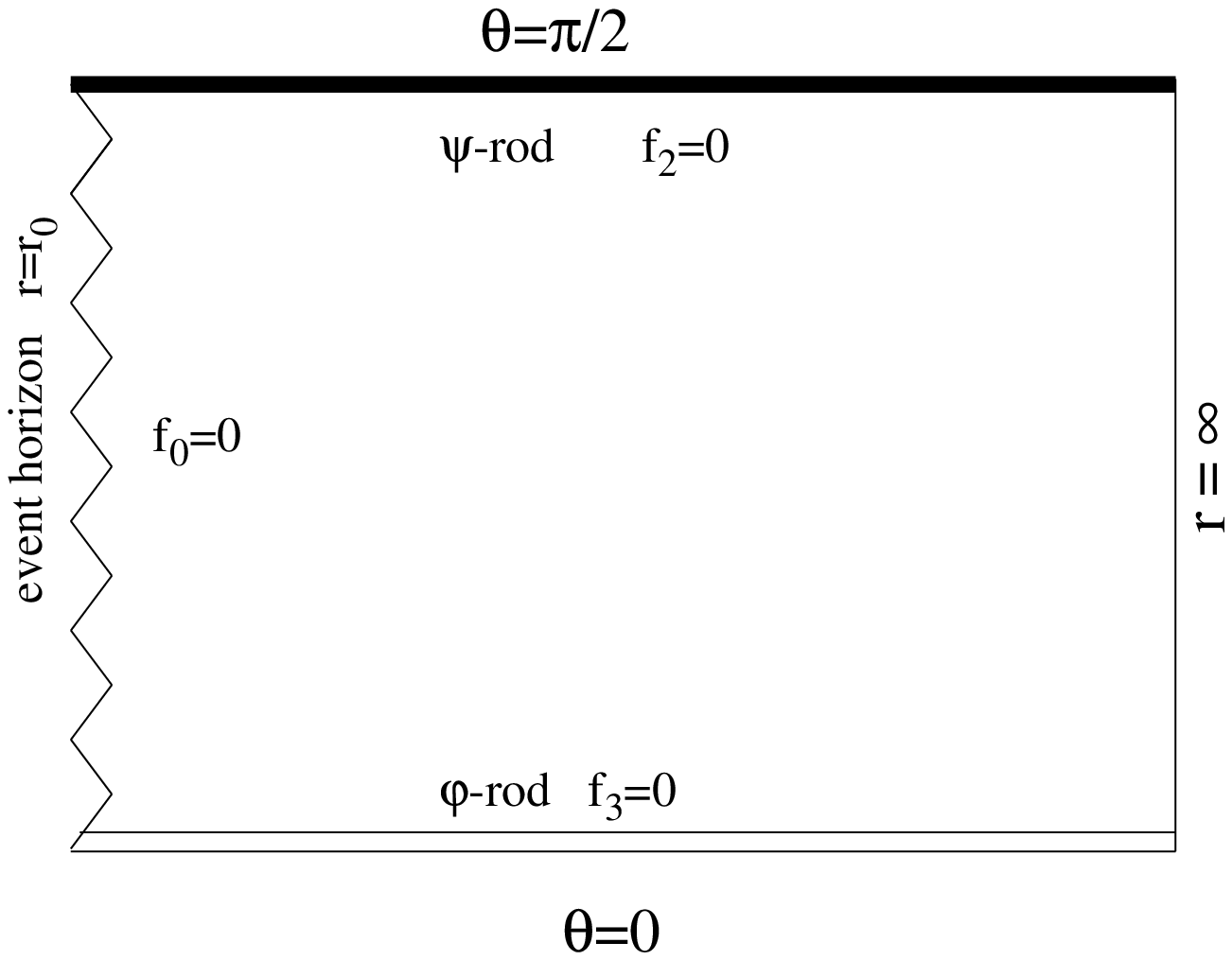}}
\hspace{5mm}%
        \resizebox{8cm}{6cm}{\includegraphics{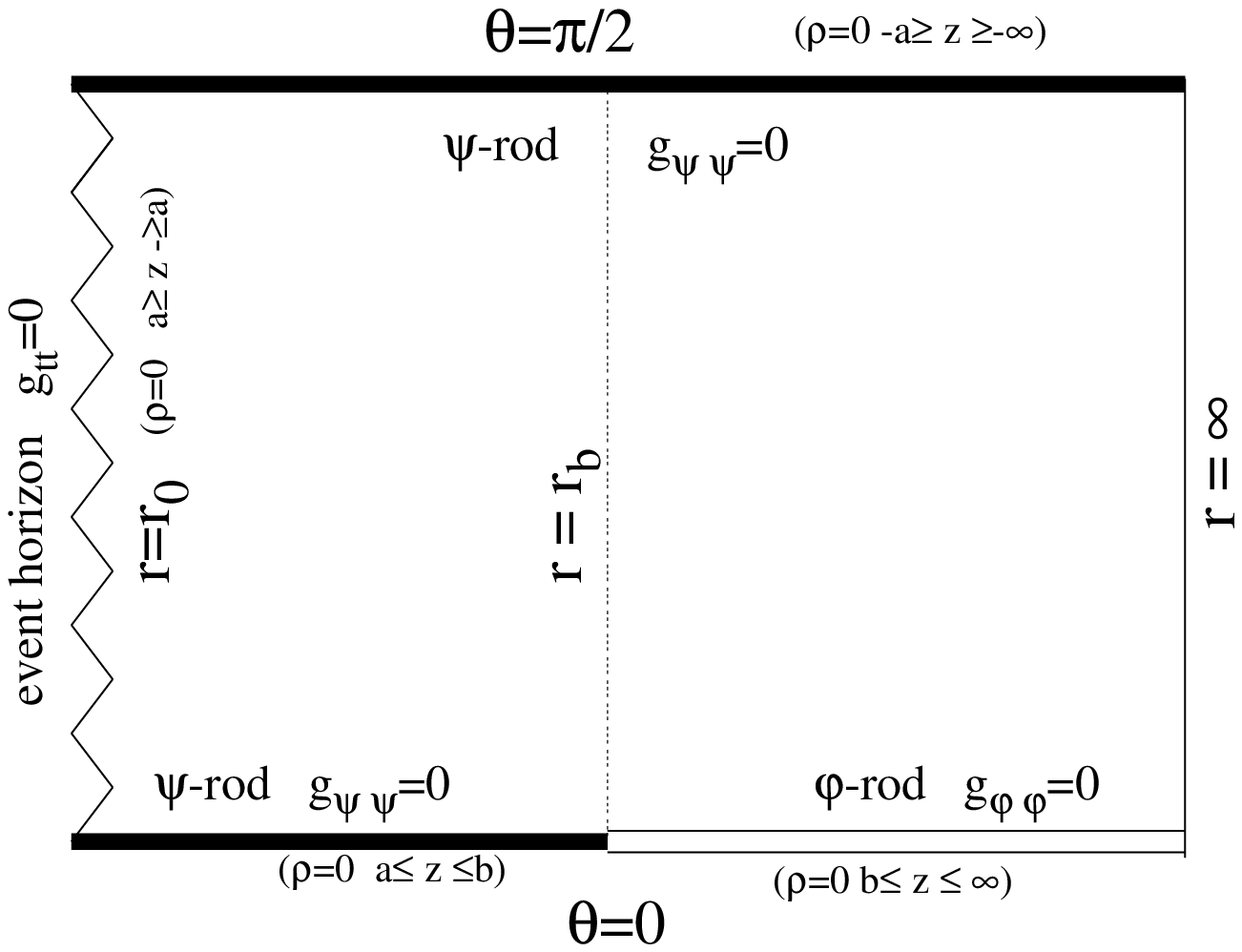}}
\hss}
\label{diagram}
\vspace{0.7cm}
 {\small {\bf Figure 13.}
The domain of integration for the
 coordinate system (\ref{remetric})   is shown for
 a EGB Schwarzschild black hole (left) and a static black ring (right).
 The relation with the Weyl-type coordinates in (\ref{metric})
is also presented.}
\end{figure}

Next we introduce background functions $\hat{f}_i^{0}$ 
\begin{eqnarray}
\label{bgf2}
\hat{f}_i=\hat{f}_i^{0}\hat{F}_i ,
\end{eqnarray}
analogous to Eq.~(\ref{ans1}). 
Note that in the coordinates $(r,\theta)$
the background functions $\hat{f}_0^{0},\hat{f}_2^{0}$ 
simplify to
\begin{equation}
\hat{f}_0^{0}  =  \left(\frac{r^2-r_0^2}{r^2+r_0^2}\right)^2 \ ,
\ \ \ \ \ 
\hat{f}_2^{0} = \left(\frac{r^2+r_0^2}{r}\right)^4 
                     \cos^2\theta \sin^2\theta \ ,
\nonumber
\end{equation}
and  $\hat{f}_1^{0}$ is regular and finite except 
at the intersection of the $\psi$-rod and the $\vphi$-rod.
For completeness we also include the background functions 
$\hat{f}_1^{0}$ and $\hat{f}_3^{0}$
\begin{equation}
\hat{f}_1^{0}
=  \frac{2 R_3}{r^2}
 \left(
 \frac{1+\left(\frac{r_0}{r_b}\right)^2}{1+\left(\frac{r_0}{r}\right)^2 }
                            \right)^2
\left[
\left(1+\left(\frac{r_0}{r_b}\right)^4\right)\left(1+\left(\frac{r_0}{r}\right)^4\right)
+4\left(\frac{r_0}{r_b}\right)^2
\left(\frac{R_3}{r^2}-\left(\frac{r_0}{r }\right)^2\cos 2\theta \right)
\right]^{-1}  \ ,
\nonumber
\end{equation}
\begin{equation}
\hat{f}_3^{0}=
\frac{1}{2}
\bigg(
2 R_3+ {r_b^2} \bigg(
1+\left(\frac{r_0}{r_b}\right)^4
-\left(\frac{r_0}{r_b}\right)^2\cos 2\theta 
\left[\left(\frac{r_0}{r}\right)^2+\left(\frac{r}{r_0}\right)^2\right]
\bigg)
\bigg)
 \ ,
\nonumber
\end{equation}
where
\begin{equation}
R_3= \frac{r^2}{2}\left[
\left(1+\left(\frac{r_b}{r}\right)^4
         -2 \cos 2\theta \left(\frac{r_b}{r}\right)^2\right)
\left(1+\left(\frac{r_0^2}{r r_b}\right)^4
         -2\cos 2\theta  \left(\frac{r_0^2}{r r_b}\right)^2\right)	 
\right]^{1/2} \ .
\nonumber
\end{equation}

The black hole limit corresponds to $r_b=r_a$, in which case one can easily see that the 
EGB Schwarzschild black hole written in isotropic coordinates is recovered.
The relation between the  coordinates in the new metric form (\ref{remetric})
and the Weyl ones in (\ref{metric})
is shown in Figure 13 for both black holes and black rings solutions.

The boundary conditions for the functions $\hat{F}_i$ follow from the 
expansions Eqs.~(\ref{nrod1}), (\ref{nrod2}) and the assumption of asymptotic flatness.
Thus $\hat{F}_i \to 1$ as $r \to \infty$ and 
the normal derivatives of all functions vanish on the other boundaries,
except $\hat{F}_1\hat{F}_3 = 1$ along
$\theta =0$, $r_b \leq r < \infty $.

\subsection{The numerical methods}

With these parametrisation 
we solve the resulting set of four coupled non-linear
elliptic partial differential equations numerically,
subject to the above boundary conditions.

First, one introduces  the new radial variable  
$x=1-r_0/r$ 
which maps the semi infinite region $[r_0,\infty)$ to the closed region $[0,1]$.
This leads to the following substitutions in the differential equations
\begin{eqnarray}
r \hat F_{,r}   \longrightarrow    (1-x) \hat F_{,x}
~~~
r^2 \hat F_{,rr}   \longrightarrow
(1-x)^2  \hat F_{,xx}
  - 2 (1-x) \hat F_{,x}
\end{eqnarray}
for any function $\hat F_i$.

The equations for $\hat{F}_i$ are then discretized on a non-equidistant grid in
$x$ and $\theta$. 
Typical grids used have sizes $90 \times 50$,
covering the integration region
$0\leq x \leq 1$ and $0\leq \theta \leq \pi/2$.  

All numerical calculations  
are performed by using the programs FIDISOL/CADSOL,
which uses a  Newton-Raphson method.  
 A detailed presentation of the this code is presented in  \cite{fidisol}.
This code requests the system of nonlinear partial differential equations to be written in the
form
$
P(x,\theta,u,u_{x},u_{\theta}
,u_{x \theta},u_{xx},u_{\theta \theta})=0,
$
(where $u$ denotes the set of unknown functions) subject to a set of boundary
conditions on a rectangular domain.
The user must deliver to FIDISOL/CADSOL the equations, the boundary conditions,
the Jacobian matrices
for the equations
and the boundary conditions, and some initial guess functions.
The numerical procedure works as follows:
for an approximate solution $u^{(1)}$,
$P(u^{(1)})$ does not vanish.
The next step is to consider an improved  solution
$u^{(2)}=u^{(1)}+\Delta u$, supposing that $P(u^{(1)}+\Delta u)=0$.
The expansion in the small parameter
$\Delta u$ gives in the first order
$
0=P(u^{(1)}+\Delta u) \approx
P(u^{(1)})+\frac{\partial P}{\partial u }(u^{(1)}) \Delta u 
+\frac{\partial P}{\partial u_x }(u^{(1)}) \Delta u_x
+ \dots
\ .
$
This equation can be used to determine the correction
 $\Delta u^{(1)}= \Delta u$.
 Repeating the calculations
iteratively ($u^{(3)}=u^{(3)}+\Delta u^{(2)}$ etc), the approximate solutions will converge,
provided the initial guess solution is close enough to the exact solution.
The iteration stops after $i$ steps if the Newton residual $P(u^{(i)})$
is smaller than a prescribed tolerance.
Therefore it is essential to have a good
first guess, to start the iteration procedure.

In each iteration
step a correction to the initial guess configuration is 
computed. The maximum of the relative defect decreases 
by a factor of $20$ from one iteration step to another.
However, for large values of $\alpha'$ convergence is slower.
In this case we re-iterate the solution until the defect is 
small enough (about $10^{-4}$). 
Note, that this defect concerns the discretized equations. 
The estimates of the relative
error of the solution (truncation error) are computed separately.
They are of the order $0.001$.
The errors also depend on the order of consistency of the method,
$i.e.$ on the order of the discretisation of derivatives. 
For the solutions in this paper,  this order was six.  
We have also monitored the quantities
\begin{eqnarray}
n_{(k)}= \left({\sum_i e_{(k)}^2(x_i,\theta_i)}\right)^{1/2},
\end{eqnarray}
(with $x_i,\theta_i$ a point of the mesh and $e_{(k)}$ a discretized equation), which provide 
an average error estimate.
For most of the solutions, we have found $n_{(k)}<10^{-11}$
(this holds also for the constraint equations  $E_r^\theta$ and $E_r^r-E_\theta^\theta $).
Most of the errors come from the region around the point $r=r_b,~\theta=0$,
where the distribution of the points in the mesh should be carefully chosen.
 
In this scheme, the input parameters are the positions of the rods fixed by 
$a$ and $b$, (resp. $r_0$ and $r_b$)
and the value $\alpha'$ of the GB coupling parameter.
To obtain EGB  black rings, 
one starts with the Einstein gravity solution 
as initial guess ($i.e.$ $\alpha'=0$ and $F_i=1$) and increases the value 
of $\alpha'$ slowly.
The iterations converge, and, in principle, repeating the procedure one obtains
in this way solutions for higher  $\alpha'$. 
For some of the configurations, we interpolate the resulting
 configurations and use them as a starting guess on a finer grid.

The mass $M$ of the solutions can be determined by extracting the
coefficient of the $1/r^2$ decay of the metric component $g_{tt}$ 
\begin{equation}
-g_{tt} \to 1 - c_0/r^2,
\end{equation}
with $c_0=16\pi GM/3V_3$.
Alternatively, 
the Smarr-like relation (\ref{smarr})
between mass, surface area, Hawking temperature, and the 
integral $I_{\alpha'}$ can be employed to determine the mass once
the other quantities are computed. 
Comparing this Smarr value for the mass with the mass evaluated
from the asymptotic decay we find excellent agreement, $i.~e.$ 
deviations occur only after 6 digits.

\subsection{The issue of the maximal value of $\alpha'$}

\begin{figure}[h]
\hbox to\linewidth{\hss%
        \resizebox{16cm}{6cm}{\epsfysize=4cm\epsffile{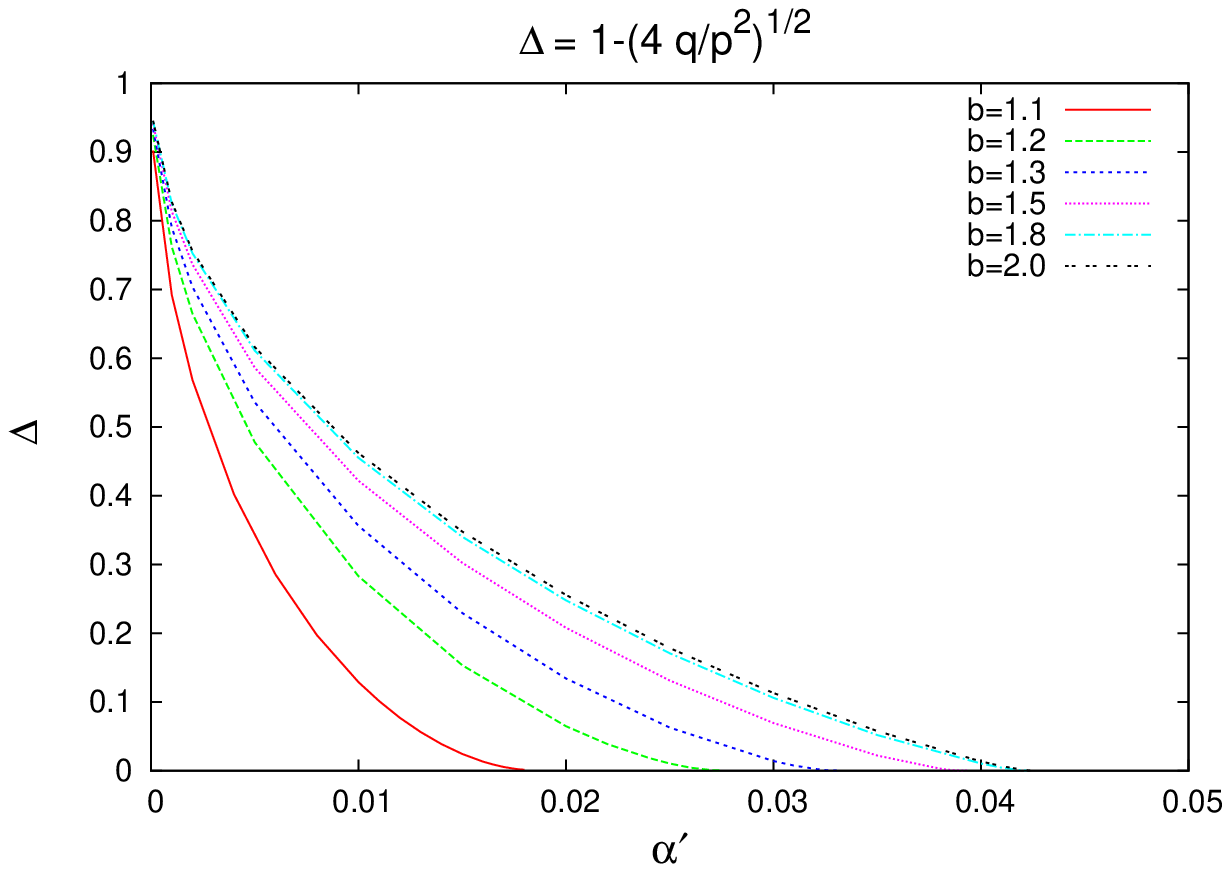}
                              \epsfysize=4cm\epsffile{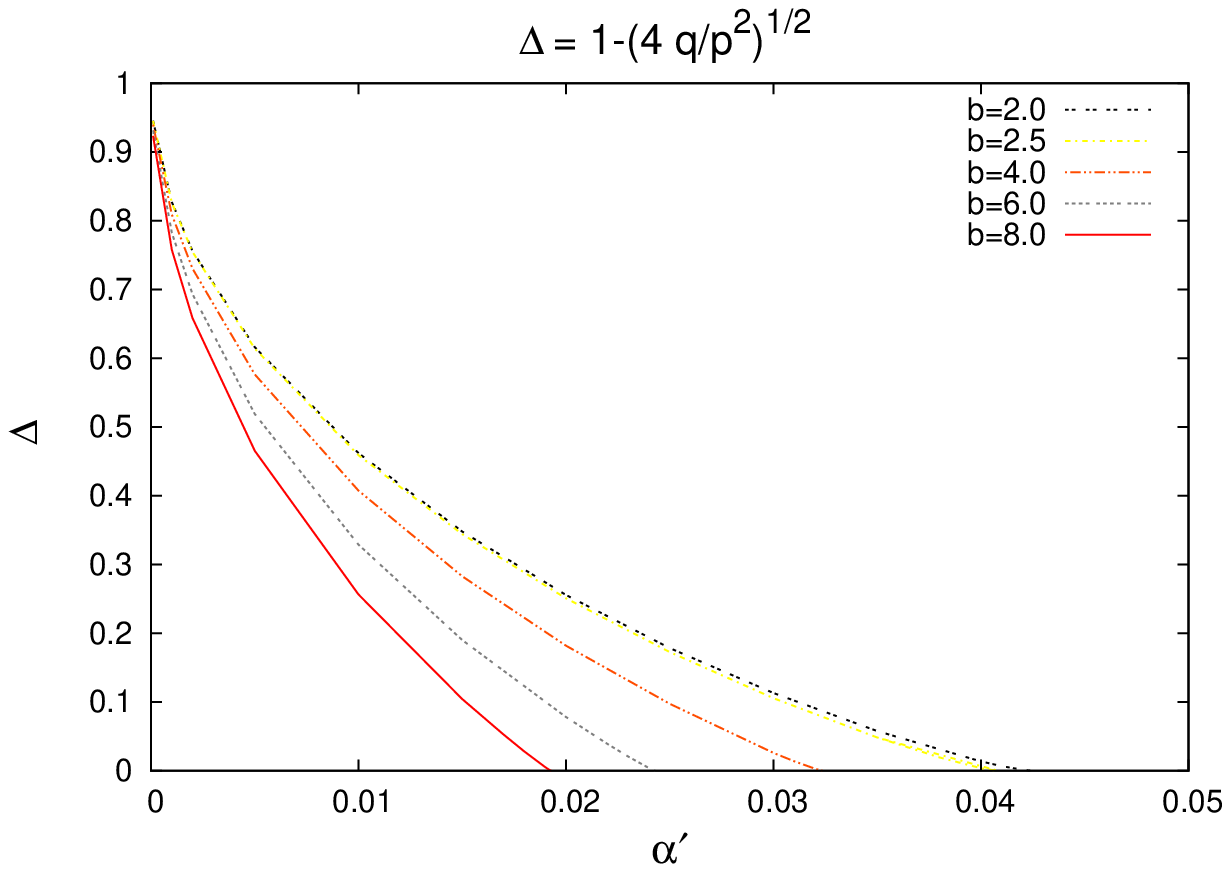}}
 \hss}
\vspace{0.3cm}
 {\small {\bf Figure 14.}
 The discriminant $\Delta$ is shown as a function $\alpha'$
 for several values of the ratio $b/a$.
 }
\end{figure}

When $\alpha'$ approaches its maximal value,  no singularity shows up.
In order to get more insight into this behaviour we follow
the approach of ref.~\cite{Kobayashi:2004hq} and study the expansion
of the solution near $(r=r_0, \theta =0)$. 

Defining $\eta =  r-r_0 $ we parametrise the solution as
\begin{eqnarray}
\hat{f}_0 & = & \eta^2 
\left[(h_{0}+h_{0,r} \eta + h_{0,rr} \eta^2)
      +(h_{0,\zz}+h_{0,r\zz} \eta + h_{0,rr\zz} \eta^2)\frac{\theta^2}{2} 
\right]      \ ,
\nonumber \\      
\hat{f}_1 & = & 
(h_{1}+h_{1,r} \eta + h_{1,rr} \eta^2)
      +(h_{1,\zz}+h_{1,r\zz} \eta + h_{1,rr\zz} \eta^2)\frac{\theta^2}{2} \ ,
\nonumber \\      
\hat{f}_2 & = & 
       (h_{2}+h_{2,r} \eta + h_{2,rr} \eta^2)\frac{\theta^2}{2} 
      +(h_{2,\zz}+h_{2,r\zz} \eta + h_{2,rr\zz} \eta^2)\frac{\theta^4}{24} \ ,
\nonumber \\      
\hat{f}_3 & = & 
(h_{3}+h_{3,r} \eta + h_{3,rr} \eta^2)
      +(h_{3,\zz}+h_{3,r\zz} \eta + h_{3,rr\zz} \eta^2)\frac{\theta^2}{2} \ ,
\nonumber      
\end{eqnarray}
where already the boundary conditions at $\theta=0$ have been taken into
account. Here the quantities $h_{i,\mu\nu \dots}$ denote constants
(in abuse of notation).

Substitution in the EGB equations yields for the first order 
expansion coefficients
\begin{eqnarray}
& &
h_{0,r}=  -h_0/r_0 \ , \ \ \ h_{0,r\zz}=-h_{0,\zz}/r_0 \ , \ \ \ 
h_{1,r}=-2 h_0/r_0 \ , \ \ \ h_{1,\zz}=-h_{0,\zz}\frac{h_1}{h_0} \ , 
\nonumber \\
& &
h_{1,r\zz}=-2 h_{0,\zz}/r_0\frac{h_1}{h_0} \ , \ \ \ 
h_{2,r}=0 \ , \ \ \ h_{3,r}=0 \ , \ \ \  h_{3,r\zz}=0
\nonumber
\end{eqnarray}
For the second order expansion coefficients $h_{i,rr}$ we find a system of 
quadratic equations. 
After some algebra we find for the coeffcients $h_{1,rr}$ 
an equations of the form 
\begin{equation} 
h_{1,rr}^2 + p h_{1,rr} +q = 0 \  .
\label{eqh1rr}
\end{equation}
The coefficients $h_{0,rr}$, $h_{2,rr}$ and $h_{3,rr}$ 
can be expressed in terms of $h_{1,rr}$.

The discriminant of Eq.~(\ref{eqh1rr}) is given by
\begin{eqnarray}
\frac{p^2}{4} - q 
   & = & 
\frac{
\left[ 2 \left( ( 8 h_2 h_{3,\zz} -  h_{2,\zz} h_3 ) h_0 
                + 2 h_2 h_3 h_{0,\zz} \right) 
\left( 2 \alpha' h_{3,\zz} - h_3 h_1 r_0^2 \right) \alpha'
 + h_0 h_2 h_3^2 h_1^2 r_0^4\right] h_1^4}
{64\left[ 2 \alpha' h_{3,\zz} - h_1 h_3 r_0^2\right]^2 {\alpha'}^2 h_0 h_2~.
 }~~~~~~~~~{~~}
\label{diss} 		
\end{eqnarray}
A real solution to  Eq.~(\ref{eqh1rr}) exists only
if $\frac{p^2}{4} - q \geq 0$.
We monitored the discriminant and observed that 
the solutions cease to exist
exactly when the discriminant becomes negative. This is demonstrated in 
Figure~14.


 \begin{small}
 
 \end{small}

\end{document}